**Theory of Cross Phenomena and Their Coefficients Beyond Onsager Theorem**

Zi-Kui Liu

Department of Materials Science and Engineering, The Pennsylvania State University,

University Park, PA 16803, USA

## Abstract

Cross phenomena, representing responses of a system to external stimuli, are ubiquitous from quantum to macro scales. The Onsager theorem is often used to describe them, stating that the coefficient matrix of cross phenomena connecting the driving forces and the fluxes of internal processes is symmetric. Here we show that this matrix is intrinsically diagonal when the driving forces are chosen from the gradients of potentials that drive the fluxes of their respective conjugate molar quantities in the combined law of thermodynamics including the contributions from internal processes. Various cross phenomena are discussed in terms of the present theory.

## Impact Statement

Through flux equations based on the combined law of thermodynamics, theory of cross phenomena is developed and applied to thermoelectricity, thermodiffusion, diffusion, electromigration, electrocaloric and electromechanical effects, and thermal expansion.

# Table of Contents

# 1 Introduction

In a recent overview article [1], the author discussed fundamental thermodynamics, thermodynamic modeling, and the applications of computational thermodynamics. Some kinetic aspects were briefly mentioned at the end of the overview including diffusion and Seebeck coefficients and general discussion on off-diagonal transport coefficients in connection with the Onsager reciprocal relationships (ORRs) [2]. The present paper aims to expand those discussions and present a broader perspective on thermodynamic basis of kinetic coefficients in the framework of the materials science and engineering (MSE) discipline. The hierarchical relationships among MSE discipline's fundamental components, including chemistry and thermodynamics, processing and kinetics, structure and crystallography, and defect and property of materials, can be represented by the off-central circles shown in Figure 1 proposed by the author [3] coinciding with the time when he coined the term "Materials Genome®" [4]. Figure 1(i) and (ii) represent the engineering and scientific views of the MSE discipline, respectively, and demonstrate the core importance of thermodynamics and kinetics in the discipline. The off-center feature of the circles implies the author's perspectives on both the broadness and coherency of material research in terms of the distinctive features of individual components and their integrations in shaping the strength of the MSE discipline.

*Figure 1: Fundamental components of materials science and engineering with (i) engineering focus and (ii) science focus* [3].

Following the discussions in the overview article [1], the present article focuses on the fundamentals of cross phenomena, such as transports of mass and electron due to heat conduction, and the predictions of their coefficients in connection with the ORRs [2]. Instead of the commonly used phenomenological kinetic equations proposed by Onsager and broadly used in the community, the present discussion starts from the combined first and second law of thermodynamics applicable to nonequilibrium systems, beyond its equilibrium form established by Gibbs [5], by including the entropy productions due to irreversible internal kinetic processes inside the system. It will be articulated that under the linear proportionality approximation of kinetic processes, the ORRs are naturally fulfilled with the off-diagonal coefficients being zero. Consequently, it follows that the coefficient of a cross phenomenon is the product of the kinetic coefficient of the transport process and the derivative of the potential driving the transport process with respect to the potential that triggers the cross phenomenon. The derivatives between potentials play a central role in materials functionalities and can be predicted from free energy containing independent internal state variables. In the present paper, free energy is used to describe all energies except the internal energy represented by the combined first and second law of thermodynamics, which is sometimes simplified as combined law of thermodynamics. While the Gibbs energy [6] and Helmholtz energy [7] are used without "free", as suggested by the International Union of Pure and Applied Chemistry.

The present article is organized as follows. In Section 2, the combined first and second law of thermodynamics applicable to both equilibrium and non-equilibrium systems is discussed by retaining the entropy productions due to irreversible internal processes in a system. In Section 3, the flux of an internal process is introduced for the change of a molar quantity based on the combined law that it is driven solely by the gradient of its conjugate potential, demonstrating that cross phenomena manifest through the derivatives between potentials. Various cross phenomena studied in the literature are systematically discussed in Section 4 in terms of the present theory, including thermoelectricity, thermodiffusion, uphill diffusion, electromigration, electrocaloric and electromechanical effects, and thermal expansion. In Section 5, our zentropy theory built on statistical mechanics and quantum mechanics in terms of first-principles calculations based on density functional theory (DFT) [8,9] is reviewed including fundamentals and applications in predicting cross phenomena in relation to critical points and associated property anomaly through Maxwell relations, including superconductivity. Finally, a summary and outlooks are presented.

## 2 Combined first and second law of thermodynamics for nonequilibrium systems

Thermodynamics is a science concerning the state of a system, whether it is stable, metastable, or unstable, when interacting with its surroundings. The first law of thermodynamics specifies that the exchange of heat, work, and chemical energies between a system and its surrounding must be represented by the change of the internal energy of the system, i.e. the energy is conserved. It does not prescribe whether the system is in internal equilibrium or not, though equilibrium is usually assumed in most textbooks. Furthermore, in most textbooks, the chemical energy is introduced much later, causing significant confusion on the concept of chemical potential. The second law represents an inequality, i.e. any irreversible internal processes (*ip*) in a system must generate entropy, resulting in positive entropy production written as $d_{ip}S > 0$. Consequently, the total entropy change of a system can be obtained as follows [1,10,11],

$$dS = \frac{dQ}{T} + \sum_i S_i dN_i + d_{ip}S \qquad \textit{Eq. 1}$$

where $dQ$ and $dN_i$ are the exchanges of heat and moles of component $i$ from the surroundings to the system (negative when from the system to the surroundings), $T$ is the temperature, and $S_i$ is the partial entropy of component *i* defined as

$$S_i = \left(\frac{\partial S}{\partial N_i}\right)_{dQ=0, d_{ip}S=0, dN_{j\neq i}=0} \qquad \textit{Eq. 2}$$

The general form of the combined law of thermodynamics can thus be written as follows aided by Eq. 1 [1,10,11]

$$\begin{aligned} dU &= dQ + dW + \sum_i U_i dN_i = TdS + dW + \sum_i \mu_i dN_i - Td_{ip}S \\ &= \sum_a Y^a dX^a - Td_{ip}S \end{aligned} \qquad \textit{Eq. 3}$$

where $dU$ is the change of the internal energy in the system, $dW$ is the exchange of work from the surroundings to the system, including mechanical, electric, and magnetic work, $U_i$ and $\mu_i = U_i - TS_i$ are the partial internal energy and chemical potential of component $i$. $Y^a$ and $X^a$ represent the pairs of conjugate variables with $Y^a$ for potentials, such as temperature, stress/pressure with negative sign [10,11], chemical potential, electrical and magnetic fields, and $X^a$ for molar quantities, such as entropy, strain/volume, moles of components, and electrical and magnetic displacements. It is important to note that the change of the internal energy does not depend on internal processes because this change concerns only the exchanges between the surroundings and the system as shown by the first part of Eq. 3, i.e. for an isolated system with $dQ = dW = dN_i = 0$, the internal energy of the system is constant with $dU = 0$ independent of whether the system is at equilibrium or not. This can also be understood because $dS$ in Eq. 3 contains the contribution from $d_{ip}S$ as shown by Eq. 1.

All kinetics processes in a nonequilibrium system are irreversible and contribute to the total entropy change as shown by $d_{ip}S$ in Eq. 1. Under the *linear proportionality approximation*, the last term in Eq. 3 can be written as

$$T d_{ip}S = \sum_j D_j d\xi_j \qquad \textit{Eq. 4}$$

where $d\xi_j$ represents the $j^{th}$ internal process for the change of the internal variable $\xi_j$, $D_j$ is the driving force for the $j^{th}$ internal process, and the summation goes over all irreversible internal processes inside the system. In reality, internal processes do not obey linear proportionality, but in principle, one can always select a small enough $d\xi_j$ so the higher-order terms are less important in defining Eq. 4 and perform integrations along the pathways to obtain overall behaviors. It is also important to differentiate externally controlled variables, i.e. all $X^a$ in Eq. 3, *vs* internal variables, i.e. all $\xi_j$ in Eq. 4, as emphasized by Hillert [10]. Furthermore, to study the stability of an internal process or a system such as instability and critical points discussed in Sections 4.10 and 5.3, one would need to include higher order terms beyond linear proportionality.

As an example, let us consider the diffusion of component $i$ as an internal process, i.e. $d\xi = dc_i$ with $c_i = N_i/V$ being the concentration of component $i$ in terms of moles per volume, and the driving force is the decrease of the chemical potential of the component, i.e. $D = -\Delta\mu_i$. This will be denoted by $D = -\Delta Y_\xi$ or $D_j = -\Delta Y_{\xi_j}$ with $Y_{\xi_j}$ being the conjugate potential of $\xi_j$ in the rest of the manuscript. When an internal process consists of changes of several molar quantities *simultaneously*, such as chemical reactions involving several components, the entropy production of the internal process can be written as the sum of entropy production of the change of each molar quantity as follows

$$D d\xi = -\sum_i \Delta\mu_i dc_i \qquad \textit{Eq. 5}$$

It should be noted that there are internal variables that are related to microstructure such as grain size [12] and phase morphologies [13], which will not be discussed further in the present article because it is not directly related to cross phenomena. Another significant challenge is to quantify the entropy in a system because it contains the contributions from internal processes as shown in Eq. 1 and is multiscale in nature [1,14], and these contributions are also important in equilibrium systems due to thermal fluctuations at finite temperatures above zero K [15–18] and will be discussed in Section 5.

For equilibrium systems, there are no irreversible internal processes, i.e. $Td_{ip}S = 0$, and one obtains the combined law derived by Gibbs [5] as follows

$$dU = \sum_a Y^a dX^a \qquad \textit{Eq. 6}$$

All $dX^a$ in Eq. 6 refer to exchanges from the surroundings to the system or negative exchanges from the system to the surroundings. The internal energy of the system is determined by the amount of each $X^a$ which is controlled from the surroundings, i.e. $X^a$ are independent variables of $U(X^a)$. All $X^a$ are called natural variables of the internal energy as they are defined from the combined law of thermodynamics [10]. All quantities in the system are also dependent on all $X^a$, including all internal variables, i.e. $\xi_j(X^a)$. The values of $\xi_j(X^a)$ can be determined by solving the following equilibrium equations for all conceivable internal processes in the system except thermal fluctuations, which will be further discussed later in the present paper

$$D_j d\xi_j = 0 \qquad \textit{Eq. 7}$$

As mentioned above, for nonequilibrium systems, $dS$ in Eq. 3 contains the contributions from irreversible internal processes as shown by Eq. 1 with $Td_{ip}S > 0$, and $\xi_j$ thus become independent variables of the system. All properties of the system are the functions of $\xi_j$, including the internal energy, i.e. $U\left(X^a, \xi_j\right)$, so are all the potentials and driving forces of internal processes, which are the partial derivative of the internal energy to the molar quantity as follows

$$Y^a\left(X^a, X^b, \xi_j\right) = \left(\frac{\partial U}{\partial X^a}\right)_{X^b \neq X^a, \xi_j} \qquad \textit{Eq. 8}$$

$$D_j\left(X^a, X^b, \xi_j, \xi_k\right) = \left(\frac{\partial U}{\partial \xi_j}\right)_{X^a, X^b, \xi_{k \neq j}} \qquad \textit{Eq. 9}$$

where $X^b$ denotes all other molar quantities except $X^a$, and $\xi_k$ represents all independent internal variables in the system except $\xi_j$. Eq. 8 represents a critically important relationship that the values of $Y^a$ are not only affected by its conjugate molar quantity $X^a$, but also by all other non-conjugate molar quantities $X^b$ and all independent internal variables $\xi_j$. Eq. 9 seems in contradiction with the statement just made that the internal energy is independent of internal variables, but it is not because when the entropy of the system is kept constant, there must be either exchange of heat or mass between the surroundings and the system as stipulated by Eq. 1 if there are internal processes in the system, i.e. $d_{ip}S > 0$, which results in the change of the internal energy of the system. In the rest of this paper, the variables kept constant for partial derivatives are those remaining natural variables and independent internal variables and will not be listed for the sake of brevity of formulas unless necessary.

Furthermore, the system can be constrained from the surroundings through the control of one or more potentials, $Y^c$, so that it becomes the independent variables of the system and the natural variable of a new free energy defined as follows along with its combined law

$$\Phi = U - \sum Y^c X^c \qquad \textit{Eq. 10}$$

$$d\Phi = \sum\nolimits_{a \neq c} Y^a dX^a - \sum\nolimits_{c \neq a} X^c dY^c - T d_{ip} S \qquad \textit{Eq. 11}$$

Eq. 8 and Eq. 9 thus become

$$Y^a\left(X^a, X^b, Y^c, \xi_j\right) = \left(\frac{\partial \Phi}{\partial X^a}\right)_{X^b \neq (X^a, X^c), Y^c, \xi_j} \qquad \textit{Eq. 12}$$

$$D_j\left(X^a, X^b, Y^c, \xi_j, \xi_k\right) = \left(\frac{\partial \Phi}{\partial \xi_j}\right)_{(X^a, X^b) \neq X^c, Y^c, \xi_{k \neq j}} \qquad \textit{Eq. 13}$$

This represents the core concept of cross phenomena such as thermoelectrics and thermodiffusion with $Y^c$ being temperature and pressure and $\Phi$ being the Gibbs energy and will be discussed in the rest of the paper. It is noted that in both thermoelectrics and thermodiffusion, the systems are inhomogeneous and can be divided into sufficiently smaller regions or subsystems that may be considered to be homogeneous to apply Eq. 12 and Eq. 13 [19]. Various constraints from the surroundings were presented in a recent publication [20], including the microcanonical, canonical, grand canonical, isothermal–isobaric (Gibbs), isoenthalpic-isobaric, and partial grand Gibbs ensembles, plus the isoentropic-isobaric ensemble, which is very hard to realize experimentally due to the contributions to entropy from internal processes as shown by Eq. 1.

Let us introduce the hydrostatic, mechanical, and electrical works for the interest of the present work as follows

$$dW_h = -PdV \qquad \textit{Eq. 14}$$

$$dW_m = \sigma d\varepsilon \qquad \textit{Eq. 15}$$

$$dW_e = Ed\theta \qquad \textit{Eq. 16}$$

where $P$ and $V$ are pressure and volume, $\varepsilon$ and $\sigma$ are the strain and stress components of mechanical energy, and $\theta$ and $E$ are the components of the electrical displacement and electrical field, both with their tensor indexes omitted for the sake of brevity of formulas. $\theta$ is used here because $D_j$ is designated for the driving force of internal process. The negative sign in Eq. 14 is because negative volume change implies the increase of the system energy, while the positive sign in Eq. 15 and Eq. 16 is because both a compressive strain and an electrical displacement in the system are denoted by positive $\varepsilon$ and $\sigma$ and increase the system energy [21,22], though other conventions are also used in the literature [23].

# 3 Kinetic equations

## 3.1 Entropy production and flux through conjugate variables

From the above discussion, the rate of the entropy production per volume due to the $j^{th}$ internal process in a sufficiently small region with a thickness of $\Delta$z and area of $A$ can be written as

$$\frac{Td_{ip}\dot{S}_j}{V} = \frac{D_j d\dot{\xi}_j}{A\Delta z} = -\frac{d\dot{\xi}_j}{A}\frac{\Delta Y_{\xi_j}}{\Delta z} = -\frac{d\dot{\xi}_j}{A}\nabla Y_{\xi_j} \qquad \textit{Eq. 17}$$

with $D_j = -\Delta Y_{\xi_j}$ and $Y_{\xi_j}$ being the conjugate potential of the internal variable of $\xi_j$ as discussed above and $\nabla Y_{\xi_j}$ the gradient of $Y_{\xi_j}$. The flux of $\xi_j$, $J_{\xi_j}$, can then be defined as follows

$$J_{\xi_j} = \frac{d\dot{\xi}_j}{A} = -L_{\xi_j}\nabla Y_{\xi_j} \qquad \textit{Eq. 18}$$

where $L_{\xi_j}$ is the kinetic coefficient for the change of $\xi_j$. Eq. 17 can be re-written as

$$\frac{Td_{ip}\dot{S}_j}{V} = L_{\xi_j}\left(\nabla Y_{\xi_j}\right)^2 \qquad \textit{Eq. 19}$$

Eq. 18 can be further expanded to independent variables of $\nabla Y_{\xi_j}$ as follows using Eq. 12

$$\begin{aligned} J_{\xi_j} &= -L_{\xi_j}\nabla Y_{\xi_j} \\ &= -L_{\xi_j}\left(\frac{\partial Y_{\xi_j}}{\partial \xi_j}\nabla \xi_j + \sum\nolimits_{\xi_k \neq \xi_j}\frac{\partial Y_{\xi_j}}{\partial Y_{\xi_k}}\nabla Y_{\xi_k} + \sum\nolimits_{\xi_l \neq \xi_j, \xi_k}\frac{\partial Y_{\xi_j}}{\partial X_{\xi_l}}\nabla X_{\xi_l} \right. \\ &\left. + \sum\nolimits_{\xi_n \neq \xi_j, \xi_k, \xi_l}\frac{\partial Y_{\xi_j}}{\partial \xi_n}\nabla \xi_n\right) \end{aligned} \qquad \textit{Eq. 20}$$

where the 1st term is the gradient of the conjugate internal molar quantity, the 2nd and 3rd terms denote the independent internal potentials and molar quantities controlled from the surroundings, respectively, and the 4th term represents the contributions from the remaining independent internal molar quantities due to their redistributions in the system. It is noted that $L_{\xi_j}$ would also depend on all independent variables in Eq. 20, i.e. $\xi_j$, $Y_{\xi_k}$, $X_{\xi_l}$, and $\xi_n$. The fluxes as a function of time are evaluated from both the initial and boundary conditions of the system depicted by Eq. 12. The 2nd and 3rd terms represent the cross phenomena from externally controlled variables, while the 4th term reflects the cross phenomena due to interactions among internal processes in the system.

The significance of the combined law of thermodynamics, Eq. 3, and the flux equation, Eq. 18 and Eq. 20, is as follows

1. The flow of a molar quantity is ***solely*** driven by the gradient of its conjugate potential with a characteristic kinetic coefficient under the linear proportionality approximation, derived from the combined law of thermodynamics ***without phenomenological*** considerations.
2. Both the potential gradient and the characteristic kinetic coefficient are functions of all independent variables in an internal process, resulting in the cross-phenomena shown by Eq. 20 to be discussed in detail in the rest of the paper.
3. The product of the molar quantity and its conjugate potential results in the entropy production due to the internal process that contributes to the energy of the system as one term in the combined law of thermodynamics, enabling its applicability to non-equilibrium systems.

### 3.2 Onsager reciprocal relation, its limitations, and improvements

Onsager [2,24] proposed the ***phenomenological*** flux equation as follows

$$J_{\xi_j}^{Onsager} = -\sum_{\xi_k} L_{\xi_j \xi_k} \nabla Y_{\xi_k} \qquad \textit{Eq. 21}$$

where $L_{\xi_j \xi_k}$ is the kinetic coefficient for the flux of $\xi_j$ due to $\nabla Y_{\xi_k}$ with the summation going through gradients of all potentials. Eq. 21 implies that all driving forces contribute to the flux of $\xi_j$, in an *apparent* accordance with experimental observations of cross-phenomena. However, it is a postulation, neither from fundamental physics as pointed out by Hillert [10] nor the first and second laws of thermodynamics as mentioned by Balluffi et al. [25]. Following the phenomenological definition, i.e. Eq. 21, the rate of entropy production is

$$\frac{T d_{ip}\dot{S}}{V} = -J_{\xi_j}^{Onsager} \nabla Y_{\xi_j} = \nabla Y_{\xi_j} \sum_{\xi_k} L_{\xi_j \xi_k} \nabla Y_{\xi_k} \qquad \textit{Eq. 22}$$

Onsager's fundamental theorem [2,10,24,26], commonly referred to as ORRs, states that the phenomenological coefficient matrix in Eq. 21 is symmetric for independent fluxes and driving forces, i.e.

$$L_{\xi_j \xi_k} = L_{\xi_k \xi_j} \qquad \textit{Eq. 23}$$

However, since a symmetric matrix can be diagonalized through the standard principal component analysis, there must exist a set of unique and independent driving forces with $L_{\xi_j \xi_k} = 0$ for $\xi_j \neq \xi_k$, i.e. eigenvectors of the coefficient matrix that are mutually orthogonal. This is exactly what Eq. 18 represents. It is noted that there are two additional fundamental issues related to Eq. 21 and Eq. 22:

1. When $\nabla Y_{\xi_j} = 0$, Eq. 21 shows that $J_{\xi_j}^{Onsager}$ is not equal to zero unless all other $\nabla Y_{\xi_k}$ or all off-diagonal $L_{\xi_j \xi_k}$ values are zero, with the latter equivalent to Eq. 18. This means that even though the conjugate driving force for the flow of $\xi_j$ is zero, there could be a flux of $\xi_j$ due to non-conjugate driving forces based on Eq. 21, which is incorrect for a steady-state system to be discussed in the present work.

2. Eq. 22 indicates that there will be no entropy production if $\nabla Y_{\xi_j} = 0$ even when $J_{\xi_j}^{Onsager}$ is not zero. This means that the flow of $\xi_j$ does not result in entropy, while the second law of thermodynamics stipulates that any irreversible process must result in a positive entropy production, including the flow of $\xi_j$.

Both issues are resolved if the flux is represented by Eq. 18 derived from the combined law of thermodynamics, while Eq. 20 represents the cross phenomena that Onsager's phenomenological Eq. 21 was intended to describe. Eq. 18 shows that when there is no gradient of a potential locally, its conjugate molar quantity will not change. While Eq. 20 depicts that the vanishing potential gradient can be realized by a set of balanced gradients such as those in steady-state systems. It is self-evident that Eq. 18 and Eq. 21 are equivalent with $L_{\xi_j\xi_k} = L_{\xi_k\xi_j} = 0$ for $\xi_j \neq \xi_k$ so the ORR, Eq. 23, is naturally fulfilled. It should be emphasized that $\nabla\xi_j$ and $\nabla\xi_k$ in Eq. 20 are not the true driving force for internal processes based on Eq. 3 and Eq. 4, one should not attempt to apply the ORR to Eq. 20 as recently commented by the author [27].

It is noted that Coleman and Truesdell [28] also pointed out the self-contradictory in ORRs and mentioned that "unless 'forces' and 'fluxes' are defined by some property more specific than mere occurrence in the expression for production of entropy, there is no content in the statement that the matrix of phenomenological coefficients is or is not symmetric". Ågren [29] recently revisited the ORRs and suggested that it can be interpreted as there is a frame of reference where all the transport processes are independent, through illustrations with isobarothermal diffusion and the Kirkendall effect, which is discussed in Section 4.6.

In the rest of the present paper, following cross phenomena will be discussed as examples to illustrate the applications of Eq. 20:

1. Thermoelectricity for transport of electron or hole $(c_e\ or\ c_h)$, i.e. $\nabla\xi_j = \nabla c_e\ or\ \nabla c_h$, with respect to the gradient of the non-conjugate externally controlled potential, temperature, i.e. $\nabla Y_{\xi_j} = \nabla T$, under the condition that there is no diffusion of other components.
2. Thermodiffusion for transport of a component $(c_i)$, i.e. $\nabla\xi_j = \nabla c_i$, with respect to the gradient of the non-conjugate externally controlled potential, temperature, i.e. $\nabla Y_{\xi_j} = \nabla T$, for a closed system. This situation can be complex if the diffusion of other components takes place at the same time, i.e. $\nabla c_{k\neq i} \neq 0$, which is related to the next example.
3. Isothermal interdiffusion with $\nabla\xi_j = \nabla c_i$ for all components for a closed system without other externally controlled potentials. It is self-evident that the interdiffusion can happen simultaneously in thermodiffusion.
4. Electromigration for transport of component $(c_i)$ and electrocaloric effect for heat conduction $(S)$, i.e. $\nabla\xi_j = \nabla c_i$ and $\nabla\xi_j = \nabla S$ with respect to the gradient of the non-conjugate externally controlled potential, electrical field, i.e. $\nabla Y_{\xi_j} = \nabla E$.
5. Electromechanical effect for transport of electron $(c_e)$, i.e. $\nabla\xi_j = \nabla c_e$, with respect to the non-conjugate externally controlled potential, stress, i.e. $\nabla Y_{\xi_j} = \nabla\sigma$.

# 4 Kinetic coefficients and cross-phenomenon coefficients

There are four common types of transport phenomena in the MSE discipline: heat, electron, mass, and fluid, commonly represented by the Fourier's, Ohm's, Fick's, and Darcy's laws based on experimental observations, with $\xi_j = S, c_e, c_i$ and $V$, and $\nabla Y_{\xi_j} = T, E, \mu_i,$ and $-P$, respectively. In the Fourier's, Ohm's, and Darcy's laws, Eq. 18 is used with the gradient of conjugate potentials as the driving force, so their linear proportionality coefficients represent the kinetic coefficients, i.e. $L_{\xi_j}$ in Eq. 18. On the other hand, the concentration gradients, i.e. $\nabla c_i$, are used in Fick's law in terms of Eq. 20. Consequently, the linear proportionality coefficient is the product of the kinetic coefficient and thermodynamic factor, commonly referred as diffusion coefficient. It should be emphasized that when Eq. 20 is used, all terms should be included, and Fick's law for diffusion of component $i$ in a temperature gradient can be written in general as one of the following forms under fixed *external* pressure or volume, respectively,

$$J_i = -L_i \nabla \mu_i = -L_i \left( \sum_j \frac{\partial \mu_i}{\partial c_j} \nabla c_j + \frac{\partial \mu_i}{\partial V} \nabla V + \frac{\partial \mu_i}{\partial T} \nabla T \right)$$
$$= -L_i \left( \sum_j \Phi_{ij} \nabla c_j - P_i \nabla V - S_i \nabla T \right) \qquad \textit{Eq. 24}$$

$$J_i = -L_i \nabla \mu_i = -L_i \left( \sum_j \frac{\partial \mu_i}{\partial c_j} \nabla c_j + \frac{\partial \mu_i}{\partial P} \nabla P + \frac{\partial \mu_i}{\partial T} \nabla T \right)$$
$$= -L_i \left( \sum_j \Phi_{ij} \nabla c_j + V_i \nabla P - S_i \nabla T \right) \qquad \textit{Eq. 25}$$

where the summation goes through all independent diffusion components including $i$, $\Phi_{ij} = \frac{\partial \mu_i}{\partial c_j}$ is the thermodynamic factor between components $i$ and $j$, and $P_i = \frac{\partial \mathrm{P}}{\partial c_i}$, $S_i = \frac{\partial \mathrm{S}}{\partial c_i}$, and $V_i = \frac{\partial \mathrm{V}}{\partial c_i}$ are the partial pressure, partial entropy, and partial volume of component $i$, respectively. $\Phi_{ij}$ can be further written as follows [10]

$$\Phi_{ij} = \frac{\partial \mu_i}{\partial c_j} = \frac{\partial(\mu_i^0 + RTlna_i)}{\partial c_j} = \frac{RT}{a_i}\frac{\partial a_i}{\partial c_j} \quad \textit{Eq. 26}$$

where $\mu_i^0$ and $a_i$ are the reference chemical potential and the activity of component $i$, and $R$ is the gas constant. $P_i$ and $V_i$ can be represented in terms of the Helmholtz energy, $F = U - TS$, and Gibbs energy, $G = F + PV$, under constant temperatures as follows

$$\frac{\partial \mu_i}{\partial V} = \frac{\partial^2 F}{\partial V \partial c_i} = -\frac{\partial P}{\partial c_i} = -P_i \quad \textit{Eq. 27}$$

$$\frac{\partial \mu_i}{\partial P} = \frac{\partial^2 G}{\partial P \partial c_i} = \frac{\partial V}{\partial c_i} = V_i \quad \textit{Eq. 28}$$

The expressions for $S_i$ in Eq. 24 and Eq. 25 need more discussion. Under the *externally* fixed pressure for Eq. 24 or *externally* fixed volume for Eq. 25, the system exhibits a volume gradient or a pressure gradient *in the system*, respectively. Therefore, $S_i$ in Eq. 24 and Eq. 25 are represented in terms of the Helmholtz energy, $F = U - TS$, or Gibbs energy, $G = F + PV$, as follows

$$\left(\frac{\partial \mu_i}{\partial T}\right)_P = \left(\frac{\partial^2 G}{\partial T \partial c_i}\right)_P = -\left(\frac{\partial S}{\partial c_i}\right)_P = -(S_i)_P \quad \textit{Eq. 29}$$

$$\left(\frac{\partial \mu_i}{\partial T}\right)_V = \left(\frac{\partial^2 F}{\partial T \partial c_i}\right)_V = -\left(\frac{\partial S}{\partial c_i}\right)_V = -(S_i)_V \quad \textit{Eq. 30}$$

Hillert [10] showed that for a stable system, the derivative of a potential to its conjugate molar quantity under constant volume is larger than its value under constant pressure. While no such conclusion for derivatives between two potentials, i.e. Eq. 29 and Eq. 30, has been reported in the literature and is worth further investigations.

Let us have some general discussions using Eq. 24 and Eq. 25. Since $S_i$ is positive, both equations indicate that the flux of each component is in the same direction of temperature gradient, i.e. migrating from a low temperature region to a high temperature region when all other gradients are zero. Similarly, since $V_i$ is positive, Eq. 25 indicates that the flux of each component is in the opposite direction of pressure gradient, i.e. migrating from high pressure to low pressure regions when all other gradients are zero. Since the values of $S_i$ and $V_i$ are different for each component, their fluxes are different, resulting in inhomogeneous distributions of components along the gradients. Further complications occur as the migration introduces both the concentration and entropy gradients with the latter generating a temperature gradient. The final steady state is reached with the balance of contributions of $\nabla c_j$, $\nabla V$ or $\nabla P$, and $\nabla T$ to flux so the net flux is zero under the external constraints of the mass conservation and the constant temperature in the system. More detailed discussions on individual cases will be given in the next several sections.

From the above discussions, it is clear that only $L_{\xi_j}$:s are the kinetic coefficients, and the cross-effect coefficients are thermodynamic properties represented by derivatives in Eq. 20 in general and Eq. 24 and Eq. 25 for mass transports. The derivatives between two potentials are relatively easy to measure as temperature, pressure, and electrical field are controlled experimentally, while the derivative between molar quantities is straightforward to predict computationally [1,10,11]. This represents an integration of complimentary experimental and computational views of the same phenomenon connected by the Maxwell relation as follows with various derivatives between potentials further discussed in Section 4.10

$$\frac{\partial Y^a}{\partial X^b} = \frac{\partial^2 \Phi}{\partial X^b \partial X^a} = \frac{\partial Y^b}{\partial X^a} \qquad \textit{Eq. 31}$$

$$\frac{\partial Y^a}{\partial Y^b} = \frac{\partial^2 \Phi}{\partial Y^b \partial X^a} = -\frac{\partial X^b}{\partial X^a} \qquad \textit{Eq. 32}$$

## 4.1 Kinetic coefficients, $L_{\xi_j}$

It is natural that theoretical calculations of $L_{\xi_j}$ are commonly approached from the kinetics of internal processes. Fundamentally, any internal processes introduce local short- or global long-range disturbances, resulting in a barrier against the internal process from its environment or the surroundings of the internal process. This is true even in Maxwell's thought experiment where the Maxwell's daemon must make some disturbances when opening or closing a small "massless" door between two chambers of gas. Consequently, $L_{X_j^a}$ may be generally represented by the following equation

$$L_{\xi_j} = L_{\xi_j}^0 e^{-\frac{Q_{\xi_j}}{k_B T}} \qquad \textit{Eq. 33}$$

where $L_{\xi_j}^0$ and $Q_{\xi_j}$ are the prefactor and the barrier for the internal process, respectively, both being functions of all independent variables, and $k_B$ is the Boltzmann constant. A schematical energy profile as a function of one internal process in Figure 2 depicts the transition from one state to another state through an energy barrier [1]. More complex energy landscape with multiple internal processes such as fractal free energy landscapes with simple basins, metabasins and fractal basins in structural glasses [30] can in principle be considered as being composed of many such individual energy profiles shown in Figure 2 at various time and spatial scales.

*Figure 2: Schematic diagram of energy landscape as a function of one internal process [1].*

Internal processes with $D_j > 0$ may be broadly categorized into following cases in terms of the values of $Q_{\xi_j}$ with respect to $k_B T$

I. $Q_{\xi_j} \leq 0$. The system is unstable locally with respect to the internal process, resulting in a local stochastic evolution that is affected by interactions with its environment such as spinodal decompositions where self-assembled structures form [31] or in general the dissipated structures [32].

II. $Q_{\xi_j} \approx 0$. The internal process is near barrierless. A small disturbance from the surroundings will activate the internal process for it to continue perpetually, such as the electrical current in a superconductor.

III. $0 < Q_{\xi_j} < k_B T$. The barrier is small so that intermediate states along the pathway of the internal process are not distinguishable due to the time and spatial resolutions of measurement techniques and can thus only be described by probability in space predicted in terms of wave functions such as migrations of electrons and phonons in electrical and thermal conductors.

IV. $Q_{\xi_j} > k_B T$. The internal process is activated thermally with low probability, and the intermediate states along the pathway can be discretely investigated such as atomic diffusion.

V. $Q_{\xi_j} \gg k_B T$. The internal process is a rare event such as migrations of electrons and phonons in electrical and thermal insulators and transition of diamond to graphite under ambient conditions.

The present discussion concerns mainly internal processes in Case III and IV, and they all experience the scenario described by Case I during overcoming the barriers. It may be mentioned that we have developed a multiscale entropy approach (termed *zentropy*) to predict instability [14,20], which may have some implications in understanding the Case I and II internal processes in terms of their occurrences and driving forces and will be discussed in Section 5, including postulations on superconductivity in Section 5.5.

### 4.2 Boltzmann transport equation for electrical and thermal conductions

For electrical and thermal conductions in Case III, the Boltzmann transport equation (BTE) is commonly used [33–36]. The BTE describes the behavior of a nonequilibrium system in terms of a balance between scattering in and out of each possible state, with scalar scattering rates. In the widely used BoltzTraP program [35,37], the BTE is linearized under the relaxation time approximation (RTA) to describe the transport distribution function which is then used to calculate the moments of the generalized transport coefficients and the charge and heat currents. The kinetic coefficients are evaluated under two extreme cases, i.e. with only the electrical gradient or only the temperature gradient applied externally, respectively. The former is slightly different from the case without a temperature gradient in the system as discussed below. In the latter, there is usually an additional constraint, i.e. without electrical current in the system.

Let us first consider the transport of electrons in conductors or n-type semiconductors with electrons added to the conduction band. When an electrical field is applied externally to a system initially under equilibrium with homogeneous temperature and electron distributions, the initial internal process is the transport of electrons, i.e. $d\xi = dc_e$, resulting in an electrical current with a flux of electrons as follows

$$J_e = -L_e \nabla\mu_e = L_e E \qquad \textit{Eq. 34}$$

where $L_e$ is the electrical conductivity, and $\mu_e$ the chemical potential of electrons. By evaluating $J_e$ under given $E$ applied externally to the system from the BTE simulations, $L_e$ can be obtained by Eq. 34. It should be emphasized that both $L_e$ and $\mu_e$ depend on all independent internal variables and externally applied electrical field (see Eq. 12).

At the same time, the migrating electrons carry heat with them, inducing a heat current, $J_Q$, in the system as follows

$$J_Q = TS_e J_e = -TS_e L_e \nabla\mu_e \qquad \textit{Eq. 35}$$

where $S_e$ is the partial entropy of electrons, i.e. the heat carried by electrons. It is important to point out that this heat conduction is induced by the heat carried by the electrons in the charge current, not by the external temperature gradient. This results in the Peltier effect due to the fact that an electrical current is accompanied by a heat current in a homogeneous conductor even at constant temperature. The Peltier coefficient is thus evaluated by the division of heat current to the electrical current as follows, noting that the electrical current is in the opposite direction of that of electron flux

$$\Pi = \frac{J_Q}{-J_e} = -TS_e \qquad \textit{Eq. 36}$$

In the second case, a temperature gradient is applied externally to the system and generates an internal process of heat current as follows

$$J_Q = -L_Q \nabla T \qquad \textit{Eq. 37}$$

where the thermal conductivity $L_Q$ can be evaluated by measuring $J_Q$ for given $\nabla T$. Before the temperature gradient is applied, the electrons are homogeneously distributed in the system. When the temperature gradient is applied, the heat current results in another internal process, i.e. the migration of electrons, thus an electrical current as follows

$$J_e = -L_e \nabla \mu_e = -L_e \left( \frac{\partial \mu_e}{\partial c_e} \nabla c_e + \frac{\partial \mu_e}{\partial T} \nabla T \right) = -L_e (\Phi_{ee} \nabla c_e - S_e \nabla T) \qquad \textit{Eq. 38}$$

where $\Phi_{ee}$ is the thermodynamic factor of electrons. As shown by Eq. 35, this electrical current further contributes to heat conduction in the system. Furthermore, there can be a variation of volume along the way as shown by Eq. 24, which is omitted here. With the initial condition of $\nabla c_e = 0, J_e = L_e S_e \nabla T$, i.e. electrons migrate from the hot end to the cold end in the system instantaneously.

Under the condition that there is no electrical current across the system in a steady state, the inhomogeneous distribution of electrons results in an induced internal electrical field, i.e. a voltage between the two ends of the system due to electrostatic interactions. This voltage can be measured by forming an electrical circuit with an external voltage in the opposite direction, resulting in zero current in the circuit. By equating Eq. 38 to zero, one obtains the following equation, noting the negative charge of electrons,

$$J_e = -L_e (-\nabla V_e - S_e \nabla T) = 0 \qquad \textit{Eq. 39}$$

The Seebeck coefficient is thus defined as follows

$$S_{T,e} = \frac{\nabla V_e}{\nabla T} = -S_e \qquad \textit{Eq. 40}$$

Combining Eq. 36 and Eq. 40, one obtains the Thomson relation, i.e.

$$\Pi = TS_{T,e} \qquad \textit{Eq. 41}$$

The above discussion demonstrates that both Peltier and Seebeck coefficients are thermodynamic quantities related to $\partial\mu_e/\partial T$ as defined in Eq. 24. This derivative between two potentials is related to the partial entropy of electron through the Maxwell relation from the combined law in terms of Gibbs energy used in Eq. 32 with $T$ and $c_e$ as its natural variables as follows

$$S_{T,e} = \frac{\partial\mu_e}{\partial T} = \frac{\partial^2 G}{\partial T \partial c_e} = -\frac{\partial S}{\partial c_e} = -S_e \qquad \textit{Eq. 42}$$

For p-type thermoelectric materials with positively charged holes added to the valence band, one has

$$J_h = -L_h(\nabla V_h - S_h \nabla T) = 0 \qquad \textit{Eq. 43}$$

$$S_{T,h} = \frac{\nabla V_h}{\nabla T} = S_h \qquad \textit{Eq. 44}$$

where $L_h$, $\nabla V_h$, $S_h$, and $S_{T,h}$ are the kinetic coefficient, voltage, partial entropy, and Seebeck coefficient of holes in p-type thermoelectric materials, respectively. One thus has negative and positive Seebeck coefficients for n- and p-type thermoelectric materials, respectively. It needs to emphasize that this voltage is induced by the gradient of electron or hole concentrations in the system due to the heat conductions from the temperature gradient. As there is no external electrical field applied, the effect of temperature gradient on the chemical potential of electrons or holes is balanced by the internal voltage due to the concentration gradient of electrons or holes, resulting in homogeneous chemical potential of electrons or holes in the system and zero electrical current as shown by Eq. 39 or Eq. 43. Since both p- and n-type materials can be described by the same formula, only electrons will be used in the rest of the paper unless their differentiation is necessary.

It may be mentioned that the Green-Kubo formalism [38,39] based on statistical mechanics relates the kinetic coefficients to the integrated time correlation functions which involve the microscopic particle fluxes and heat flux. Recently, Green et al. derived a formula to describe the evolution of density excitations of the electron system when subject to a periodic potential with static or dynamical perturbations[40]. It is based on Green-Kubo formalism with a Mori memory function approach to include dissipation effects at all orders in the relaxation interaction. In their approach, the excitations of a given wave vector are kept as an active subspace, and those of different wave vectors are projected into a bath described by a memory function, while the slower variables within the active subspace are defined by the conservation laws. Their approach facilitates the approximation of the dissipation process into a single memory function that includes relaxation, dissipation, and quantum coherence. It is shown that this approach offers a practical method to evaluate the conductivity with existing electronic-structure codes and avoids the complications and the limitations of the Green-Kubo formula in the thermodynamic limit and the neglect of quantum coherence in the BTE approach. The potential importance of interband coherence on the transport process in systems with more than one band close to the Fermi level is emphasized beyond weak scattering captured by a relaxation time.

### 4.3 Atomic mobility by kinetic simulations and DFT-based calculations

For atomic diffusion in Case IV in Section 4.1, classic or *ab-initio* molecular dynamic (MD) [41–43] and kinetic Monte Carlo (KMC) [44–47] simulations directly follow the migration of diffusion component as a function of time and extract the self or tracer diffusivity ($D_i^*$) using Einstein's relation in terms of the mean square displacement (MSD) [48]. The tracer-diffusivity can be further converted to atomic mobility ($M_i$), the kinetic coefficient discussed above ($L_i$), and intrinsic diffusivity (${}^iD_{ik} = L_i \frac{\partial \mu_i}{\partial c_k}$) in the lattice-fixed frame of reference, and chemical diffusivity ($D_{ik} = \sum_j L'_{ij} \frac{\partial \mu_j}{\partial c_k}$) in the volume-fixed frame of reference with a complex relation between $L'_{ij}$ and $L_i$. Andersson and Ågren [49] presented an elegant discussion among all the diffusion coefficients and their computational implementation. The relationships among these diffusion coefficients and kinetic coefficients are shown in Figure 3.

*Figure 3: Relationships among tracer diffusivity, atomic mobility, kinetic L parameters, and intrinsic and chemical diffusivities.*

As indicated by Figure 1 and discussed above, thermodynamics and kinetics are closely related to each other, and it is thus natural that atomic diffusion process can be viewed from either thermodynamics or kinetics of the internal process. This can be accomplished by evaluating the free energy of vacancy formation and the individual microstates along the diffusion pathways. The free energy of vacancy formation can be obtained straightforwardly by the DFT-based first-principles calculations under the quasiharmonic phonon approximations, which also results in the vacancy concentration as a function of temperature [50,51]. In the classic transition state theory (TST) [52,53], the jump frequency for the migration of atom over the barrier is written in terms of enthalpy of barrier and an effective frequency, $\nu^*$, represented by the vibrational frequencies at the equilibrium and transition states with the imaginary vibrational modes at the transition state removed as follow

$$\nu^* = \frac{\prod_{i=1}^{3N-3} \nu_i}{\prod_{j=1}^{3N-4} \nu'_j} \qquad \textit{Eq. 45}$$

where $\nu_i$ and $\nu'_j$ are the vibrational frequencies at the equilibrium and transition states, respectively. The enthalpy of migration and the vibrational frequencies can be obtained from the DFT-based first-principles calculations [54] in combination with the nudged elastic band (NEB) method [55] and the five frequency model [56] along with the predicted vacancy concentration. We were then able to predict the tracer diffusion coefficients for pure elements and dilute solutions for fcc [57–61], bcc [62], and hcp [62–65] structures, showing remarkable agreement with available experimental data. A similar approach was also developed for prediction of interstitial diffusion coefficients by Wimmer et al. [66]. It should be mentioned that this approach is predictive with all inputs from DFT-based first-principles calculations without empirical parameters. More recently, Allnatt et al.[67] developed a 13 frequency model to address diffusion kinetics for the full anisotropic three-dimensional hcp structure and verified the some of the results with Monte Carlo simulations.

In next several sections, some classic cross phenomena, i.e. thermoelectricity for electrical conduction and thermodiffusion for atomic diffusion, both induced by the temperature gradient, upfill diffusion for migration of fast diffusion component against its concentration gradient induced by inhomogeneity of slow diffusion components, electromigration for atomic diffusion and the electrocaloric effect for heat conduction, both induced by electrical field, and electromechanical effect for interactions between mechanical deformation and electrical field are discussed in detail by applying the theoretical framework presented above.

## 4.4 Thermoelectricity: A cross phenomenon between electrical conduction and temperature

The thermoelectricity case is discussed in Section 4.2 for a system under external electrical field or external temperature gradient, respectively. It is shown that the Seebeck coefficient is defined by the derivative of the chemical potential of electrons with respect to temperature as shown by Eq. 42 and Eq. 44. At the same time, the Peltier coefficient and the Thomson relation are also obtained.

Based on this fundamental understanding, we predicted values of Seebeck coefficients for several n- and p-type thermoelectric materials using DFT-based first-principles calculations without empirical parameters [68,69]. In those calculations, the temperature dependence of the chemical potential of electrons are evaluated from the electron density of states (e-DOS) using the Mermin statistics [70,71] extended to DFT-based calculations at finite temperatures [50,72,73] with the electrons explicitly treated [74,75]. We start from the conservation equation

$$\int n(\varepsilon, V) f d\varepsilon = N_e \qquad \textit{Eq. 46}$$

where $n(\varepsilon, V)$ is the e-DOS at the band energy $\varepsilon$ and volume $V$, $N_e$ the total number of electrons, and $f$ the Fermi-Dirac distribution expressed as follows

$$f = \frac{1}{exp\left[\frac{\varepsilon + e\varphi}{k_B T}\right] + 1} \qquad \textit{Eq. 47}$$

where $e$ is the charge of electron, and $e\varphi = -(\mu_e - \varepsilon_F)$ with $\varepsilon_F$ being the Fermi energy. After a few rearrangements, we obtain the formulation to calculate the Seebeck coefficient under the isobaric condition as follows, [68,69]

$$S_{T,e} = -\left(\frac{\partial\varphi}{\partial T}\right)_P$$
$$= -\frac{k_B T \alpha_V}{e c_e}\int \frac{\partial n(\varepsilon, V)}{\partial V} f d\varepsilon - \frac{1}{eTc_e V}\int n(\varepsilon, V) f(1-f)(\varepsilon + e\varphi) d\varepsilon \qquad \textit{Eq. 48}$$

$$c_e = \frac{1}{V}\int n(\varepsilon, V) f(1-f) d\varepsilon \qquad \textit{Eq. 49}$$

where $\alpha_V$ is the volume thermal expansion coefficient, and $c_e$ the mobile charge carrier concentration. Therefore, the Seebeck coefficient can be computed from the readily accessible e-DOS without involving the electrical conductivity as in BTE, which is a much more challenging quantity to compute and requires the calculation of electron group velocity and the estimation of the relaxation time [76].

The integral $\int n(\varepsilon, V) f(1-f) d\varepsilon$ in Eq. 49 can be considered as the total number of mobile carriers in a thermoelectric solid, i.e. the number of electrons participating in the conduction process at finite temperatures. This can be understood as the pair product of $f(1-f)$ represents the probability that the electrons occupied "$f$ number" of electronic states with energy $\varepsilon$, transmitted to (or the vice versa), the "$1-f$ number" of unoccupied electronic states with energy $\varepsilon$ at finite temperature. Consequently, $n(\varepsilon, V)f(1-f)$ represents the e-DOS of charge carriers at finite temperatures.

For DFT-based computations, the issue related to the commonly underestimated band gap is resolved by following the approach developed by Singh [77] with the Engel-Vosko generalized gradient approximation (GGA) [78] plus spin-orbit coupling as implemented in the WIEN2k package for obtaining accurate electronic structures [79]. However, this approach has low accuracy for the total energy, which is instead computed by means of the Perdew-Burke-Ernzerhof exchange-correlational functional revised for solids (PBEsol) [80] as implemented in the Vienna *ab initio* simulation package (VASP, version 5.3) [74,75]. Furthermore, our mixed space approach [11,81–83] is used to account for the dipole-dipole interactions for a phonon calculation of polar insulators in the real space using the supercell method. For simplicity, we assume the rigid band approximation, [34] i.e. the band structure is assumed to remain unchanged from doping or from the thermal electronic excitation at finite temperatures. Our approach demonstrated excellent agreement with experimental observations as shown in Figure 4 for PbTe [68], SnSe [68], and $La_{3-x}Te_4$ [69]. In Figure 4(ii), the Seebeck coefficient and electrical conductivity reported in the literature [84] using the previous version of the BoltzTraP program [85] are superimposed, showing worse agreement with experimental data in comparison with our calculations without empirical parameters.

*Figure 4: Calculated Seebeck coefficients for (i) PbTe for various p- and n-type doping levels* [68]*; (ii) p-type SnSe* [68]*; (iii)* $La_{2.75}Te_4$ [69].

It is important to emphasize that regardless of whether electrons are driven by external temperature gradient or external electrical field, it is the ***migration of electrons*** that governs the kinetic process of electrical conduction, i.e. the electrical conductivity and the chemical potential of electrons. Furthermore, both electrical conductivity and the chemical potential of electrons are functions of temperature and electron concentration, resulting in the cross-phenomenon of thermoelectricity. A commonly used parameter measuring the performance of thermoelectric materials is the figure of merit [86] defined as follows

$$z_T = \frac{L_e S_{T,e}^2}{L_Q} T \qquad \textit{Eq. 50}$$

which includes the two kinetic coefficients $L_e$ and $L_Q$ and the thermodynamic Seebeck coefficient $S_{T,e}$. The square of $S_{T,e}$ shows its importance in affecting the performance of thermoelectric materials and can be reliably predicted by our approach discussed here.

### 4.5 Thermodiffusion: A cross phenomenon between atomic diffusion and temperature

For thermodiffusion, the chemical potential of a diffusion component drives the transport of the component from a high chemical potential region to a low chemical potential region until its chemical potential becomes homogeneous in the system [19,87]. As in the case of thermoelectricity, when a temperature gradient is applied externally to a system initially at equilibrium with homogeneous compositions, the chemical potentials of component in the system become inhomogeneous, resulting in an internal process for the migration of diffusion component as shown by Eq. 24 or Eq. 25. It is important to emphasize again that the fundamental process of thermodiffusion is the ***migration of the diffusion component***, driven by the gradient of its conjugate chemical potential, and the chemical potential and kinetic coefficient are functions of temperature and compositions of all components plus all independent internal variables as shown by Eq. 12.

Let us start with a system with homogeneous concentration, i.e. $\nabla c_j = 0$. When a temperature gradient is applied initially, two internal processes take place immediately in the system, i.e. the volume change and heat conduction, which will drive the initial diffusion flux as follows by re-writing Eq. 24

$$J_i = -L_i(-P_i \nabla V - S_i \nabla T) \qquad \textit{Eq. 51}$$

For a system with positive thermal expansion, $\nabla V$ and $\nabla T$ point to the same direction, so does $J_i$ for all diffusion components, though it should be noted that there are both liquid and solid phases that exhibit negative thermal expansion, i.e. $\nabla V$ and $\nabla T$ can point to the opposite directions in some systems [1,20,88].

As soon as the diffusion starts, the concentration gradients will be created and contribute to the flux as shown by Eq. 24. As typical physical experiments are conducted for closed systems, the fluxes of all diffusion components must be zero at both ends of the system as the boundary condition. Therefore, when the system reaches a steady state with heat conduction as the only internal process in the system, $\nabla V$, $\nabla T$ and $\nabla c_j$ balance each other to result in zero diffusion flux for each diffusion component, i.e.

$$J_i = -L_i \nabla \mu_i = -L_i \left( \sum_j \Phi_{ij} \nabla c_j - P_i \nabla V - S_i \nabla T \right) = 0 \qquad \textit{Eq. 52}$$

For non-ideal, non-dilute solutions, $L_i$, $\Phi_{ij}$, $P_i$, and $S_i$ can strongly depend on temperature and compositions of the solutions, resulting in a diffusion component switching their segregation regions as a function of composition and temperature [89–96]. Eq. 52 further indicates that at the steady state, $\nabla\mu_i = 0$ for each diffusion component, i.e. homogeneous chemical potential.

As most experiments reported in the literature are for binary systems, let us write Eq. 52 for a binary A-B system under steady state as follows

$$J_A = -L_A\nabla\mu_A = -L_A(\Phi_{AA}\nabla c_A + \Phi_{AB}\nabla c_B - P_A\nabla V - S_A\nabla T) = 0 \qquad \textit{Eq. 53}$$

$$J_B = -L_B\nabla\mu_B = -L_B(\Phi_{BA}\nabla c_A + \Phi_{BB}\nabla c_B - P_B\nabla V - S_B\nabla T) = 0 \qquad \textit{Eq. 54}$$

Eliminating $\nabla c_A$ from the equations results in

$$(\Phi_{BB}\Phi_{AA} - \Phi_{AB}\Phi_{BA})\nabla c_B = (P_B\Phi_{AA} - P_A\Phi_{BA})\nabla V + (S_B\Phi_{AA} - S_A\Phi_{BA})\nabla T \qquad \textit{Eq. 55}$$

The Soret coefficient in thermodiffusion [97] is usually evaluated experimentally in the literature by the negative ratio of the $\nabla c_B$ with respect to $\nabla T$ as follows

$$S_{T,B} = -\frac{\nabla c_B}{c_B\nabla T} = -\frac{1}{c_B}\frac{S_B\Phi_{AA} - S_A\Phi_{BA}}{\Phi_{BB}\Phi_{AA} - \Phi_{AB}\Phi_{BA}}\left(1 + \frac{P_B\Phi_{AA} - P_A\Phi_{BA}}{S_B\Phi_{AA} - S_A\Phi_{BA}}\frac{\nabla V}{\nabla T}\right) \qquad \textit{Eq. 56}$$

Graphically, the Soret coefficient is thus the negative slope of the concentration in log scale plotted with respect to temperature (see e.g., the plot by Duhr and Braun [19]). Eq. 56 shows that $S_{T,B} = 0$ when the one of the following conditions is met

$$S_B\Phi_{AA} - S_A\Phi_{BA} = 0 \qquad \textit{Eq. 57}$$

$$S_B\Phi_{AA} - S_A\Phi_{BA} + (P_B\Phi_{AA} - P_A\Phi_{BA})\frac{\nabla V}{\nabla T} = 0 \qquad \textit{Eq. 58}$$

For systems with negative thermal expansion at certain composition and temperature ranges [1,20,88], it may be more likely that the Soret coefficient changes its sign with respect to composition and temperature.

As pointed out by Duhr and Braun [19] and commonly agreed in the literature, a generally applied assumption for describing Soret effect is that it should be treated as a transport phenomenon. It is certainly true that Soret coefficients can be evaluated from kinetic processes as discussed in Section 4.2, such as the Green-Kubo formalism [38,39] for transport properties in liquid of binary systems using MD simulations [98–102]. Nevertheless, there were important efforts in the literature in addition to the present work to show that the Soret coefficient is a thermodynamic property of the system, rather than a transport property. Duhr and Braun [19] divided the nonequilibrium system into a succession of regions with small differences of local temperature and concentration and utilized a linearized Boltzmann distribution under local equilibria as follows

$$S_{T,B}dT = -\frac{dc_B}{c_B} = -\frac{d\mu_B}{k_BT} = \frac{S_B}{k_BT}dT \qquad \textit{Eq. 59}$$

where $c_B$, $\mu_B$, and $S_B$ are the concentration, chemical potential, and partial entropy of the particle $B$. In the publications by Duhr and Braun [19,103] $G$ was used in the place of $\mu_B$ in Eq. 59, noting that $\mu_B = \mu_B^0 + k_BTlnc_B$ with $\mu_B^0$ being the reference state for $\mu_B$. They further correlated $\mu_B$ to the particle radius by considering that for solid particles, only the solvation energy at their surface can be temperature dependent. The Soret coefficient was thus obtained as follows

$$S_{T,B} = \frac{S_B}{k_BT} \qquad \textit{Eq. 60}$$

It is noted that in their paper, there was a minus sign on the right-hand side of Eq. 60, which is probably a typo.

It is evident that Eq. 56 and Eq. 60 are substantially different. The issue is due to the relation defined by Eq. 59. It is noted that $\mu_B$ is a function of $c_B$, $V$, and $T$ as shown in Eq. 8, Eq. 12, Eq.

24, or Eq. 53 and cannot be written as a function of either $c_B$ or $T$ as in Eq. 59. The general differential form of $\mu_B$ in a binary system is to be written as follows

$$d\mu_B = \frac{\partial \mu_B}{\partial c_B} dc_B + \frac{\partial \mu_B}{\partial c_A} dc_A + \frac{\partial \mu_B}{\partial V} dV + \frac{\partial \mu_B}{\partial T} dT \qquad \textit{Eq. 61}$$

Furthermore, Eq. 60 implies that the Soret coefficient is always positive as $S_B$ is positive, which is in agreement with the experimental data by Duhr and Braun [19,103], but not true in general as reported in the literature [89–96]. While $dc_B + dc_A = 0$ is commonly used in the literature, it is a good approximation for most liquid systems, but not true in general as the net flux of all diffusion components may not be zero, resulting in a net vacancy flow, i.e. the Kirkendall effect in solids to be discussed in Section 4.6.

Kocherginsky and Gruebele made an important contribution in a series of publications [104–106] through a systematic framework for chemical transport. Based on the continuity equation for diffusive transport and the Smoluchowski equation, they obtained the flux of a component $i$ as follows

$$J_i = -M_i c_i \nabla \mu_i \qquad \textit{Eq. 62}$$

where $M_i$ (denoted by $U_i$ in their work), $c_i$ and $\mu_i$ (denoted by $\mu_{gi}$ in their work) are Einstein's mobility, concentration, and physicochemical potential of component $i$, respectively. By comparing Eq. 24 and Eq. 62, one gets $L_i = M_i c_i$, which was shown by Andersson and Ågren [49] (see also Figure 3) for the lattice-fixed frame of reference. Even though the present author has some disagreements with Kocherginsky and Gruebele on how to extend the dependence of $\nabla \mu_i$ to other gradients of independent variables (see Eq. 20 and Eq. 24) as discussed recently [27,107], Eq. 62 derived by Kocherginsky and Gruebele [104–106] from kinetic considerations is significant as it demonstrates that all cross-effect coefficients for atomic diffusion are related to the derivatives of chemical potential with respect to other independent variables.

Köhler and Morozov [108] reviewed the experimental observations and theoretical work in the literature for the Soret effect in non-ionic molecular binary and ternary liquid mixtures, and Piazza and Parola [109] and Würger [110] reviewed the Soret effect in the ionic and surfacted colloids and micellar solutions. For binary systems, Hartmann et al. [95] defined the parameters for each pure substance and obtained the following equation for Soret coefficient defined by component 1 in a binary system as follows

$$S_{T,1} = \frac{f_1 - f_2 + C(\partial V^E/\partial x_1)}{1 + \partial ln\gamma_1/\partial ln x_1} \qquad \textit{Eq. 63}$$

where $\gamma_1$ and $x_1$ are the chemical activity coefficient and mole fraction of component 1, $V^E$ is the non-ideal excess volume of mixing, $C$ is a constant, and $f_1$ and $f_2$ are model parameters related to Soret coefficients of pure components 1 and 2, respectively. They qualitatively correlated the model parameters to characteristic diameters of the solvent and solute, their interaction parameters, the solvent volume fraction, and the partial excess volume of mixing, thus implying that the Soret coefficient is a thermodynamic quantity. It can be seen that Eq. 56 and Eq. 63 contain similar thermodynamic quantities such as thermodynamic factor and the effect of volume denoted by $\partial ln\gamma_1/\partial ln x_1$ and $V^E$ in Eq. 63, and partial entropies in Eq. 56 may be related to $f_1$ and $f_2$ in Eq. 63 as shown below

$$S_i = S_i^0 - R ln x_i + \Delta S_i^{ex} \qquad \textit{Eq. 64}$$

where $S_i^0$ is the partial entropy of the pure component $i$, and $\Delta S_i^{ex}$ the excess non-ideal partial entropy of the component $i$ in the solution.

More recently, Schraml et al. [96] presented a detailed investigation of water ($H_2O$), ethanol ($ETH$), and triethylene glycol ($TEG$) ternary and the three constitutive binary systems and observed a characteristic sign change of the Soret coefficient as a function of temperature and/or composition. They correlated the decay of the Soret coefficient with respect to concentration to negative excess volumes of mixing as shown by Eq. 63, i.e. the decrease of volume with the increase of concentration of the minor component. Since the minor component migrates towards the cold side, the higher concentration of the minor component at the low temperature region reduces its volume further and increases the ratio of $\nabla V/\nabla T$ (see Eq. 56 and Eq. 58). When the excess volume reaches the minimum, further increase of the concentration decreases the ratio $\nabla V/\nabla T$. This change along with the interplay of other properties in Eq. 56 and Eq. 58 results in the sign change of the Soret coefficients in the systems studied.

Furthermore, Schraml et al. [96] showed the connection between the sign change of the Soret coefficients of the binaries and the signs of the Soret coefficients of the corresponding ternary mixtures as depicted in Figure 5. In particular, they observed that at least one ternary composition exists, where all three Soret coefficients vanish simultaneously, and no steady-state separation is observable. They postulated that this composition is on the path connecting the two compositions of zero Soret coefficients in the two binary systems. This can be understood by writing the flux equations for a ternary system as follows

$$J_A = -L_A \nabla \mu_A = -L_A(\Phi_{AA} \nabla c_A + \Phi_{AB} \nabla c_B + \Phi_{AC} \nabla c_C - P_A \nabla V - S_A \nabla T) = 0 \qquad \textit{Eq. 65}$$

$$J_B = -L_B \nabla \mu_B = -L_B(\Phi_{BA} \nabla c_A + \Phi_{BB} \nabla c_B + \Phi_{BC} \nabla c_C - P_B \nabla V - S_B \nabla T) = 0 \qquad \textit{Eq. 66}$$

$$J_C = -L_C \nabla \mu_C = -L_C(\Phi_{CA} \nabla c_A + \Phi_{CB} \nabla c_B + \Phi_{CC} \nabla c_C - P_C \nabla V - S_C \nabla T) = 0 \qquad \textit{Eq. 67}$$

Using these three equations, one can calculate the Soret coefficients of each component for given compositions in a ternary system. As observed by Schraml et al. [96], there is one composition for $S_{T,H_2O} = S_{T,ETH} = 0$ in the $H_2O - ETH$ binary system and $S_{T,H_2O} = S_{T,TEG} = 0$ in the $H_2O - TEG$ binary system, while in the $ETH - TEG$ binary system this happens near pure $ETH$. When the third component is added, Schraml et al. [96] discussed that there should be three composition pathways from each binary into the ternary with one Soret coefficient kept zero. They further articulated that these three composition pathways meet at one point with all three Soret coefficients being zero at the same time, i.e. $S_{T,H_2O} = S_{T,ETH} = S_{T,TEG} = 0$. This can be understood that when two Soret coefficients are zero, the third one must be zero due to the mass conversation at the steady state with $\nabla c_A + \nabla c_B + \nabla c_C = 0$, though this approximation may bring some error if there is a net vacancy flow due to the Kirkendall effect as discussed in Section 4.6.

*Figure 5: Signs of the Soret coefficients in the water (H2O), ethanol (ETH), and triethylene glycol (TEG) ternary system at 25 C [96]. The colored regions denote regions of negative Soret coefficients of the respective components. Point Z marks the intersection of the boundaries of the three colored regions, where all three Soret coefficients vanish simultaneously. The steady state optical signal vanishes along the dashed line.*

Based on the discussion above, it is clear that to predict the Soret coefficients, one needs to accurately model the thermodynamic properties of the solutions as a function of temperature, volume, and compositions. While if one would like to simulate thermodiffusion processes, the mobilities of diffusion components also need to be modeled. A very interesting work was reported by Höglund and Ågren [87] who simulated the thermodiffusion in an Fe-32%Ni-0.14%C (weight percent, wt%) alloy using an experimental temperature profile reported by Okafor et al. [111] and the Dictra software package [112,113] with thermodynamic and mobility databases developed by the CALPHAD method [1,114–118]. In the lattice-fixed frame of reference, $L_{NiC} = L_{CNi} = 0$ were adopted in accordance with Eq. 24. In the number/volume-fixed frame of reference, they derived the cross-coefficient as follows (see Figure 3)

$$\frac{L'_{kT}}{T^2} = \sum_{i=1}^{n} (\delta_{ki} - u_k a_i) \frac{u_i}{V_S} y_{Va} M_{iVa} Q_i^* \frac{1}{T} \qquad \textit{Eq. 68}$$

where $k$ represents $Fe$, $Ni$, or $C$; the Kronecker delta $\delta_{kk} = 1$ and $\delta_{ki} = 0$ for $k \neq i$; $a_{Fe} = a_{Ni} = 1$ for substitutional elements and $a_C$=0 for interstitial elements; $V_S$ is the molar volume per mole of substitutional atoms; $u_i$ is the mole fraction per mole of substitutional atoms; $y_{Va}$ is the mole fraction of vacant lattice sites; $M_{iVa}$ is the mobility of component $i$ through vacancy mechanism (see Figure 3); and $Q_i^*$ is the heat of transport for component $i$ commonly used in the literature.

As the mobilities of the substitutional atoms are many orders of magnitude lower than those of the interstitials, $Q_{Fe}^* = Q_{Ni}^* = 0$ are assumed, and Eq. 68 is approximated as follows for $C$

$$\frac{L'_{CT}}{T^2} = \frac{u_C}{V_S} y_{Va} M_{CVa} Q_C^* \frac{1}{T} \qquad \textit{Eq. 69}$$

The flux of $C$ is obtained as follows

$$J'_C = -D_{CC}^{Fe\prime} \frac{1}{V_S}\left(\nabla u_C + Q_C^* \frac{1}{T d\mu_C / du_C} \nabla T\right) \qquad \textit{Eq. 70}$$

where $D_{CC}^{Fe'}$ is the interdiffusion coefficient of $C$ (see Figure 3), and the full derivative means that the effects of other components are included. Under the steady-state condition for a closed system with $J'_C = 0$, $Q_C^*$ can be obtained as

$$Q_C^* = -T\frac{d\mu_C}{du_C}\frac{\nabla u_C}{\nabla T} = -T\frac{d\mu_C}{du_C}\frac{du_C}{dT} = \frac{1}{T}\frac{d\mu_C}{du_C}\frac{du_C}{d(1/T)} \qquad \textit{Eq. 71}$$

where $\nabla u_C/\nabla T$=$du_C/dT$ is used though the physical meaning of the derivative is not very clear. Using Eq. 56, one can obtain the following equation for the Soret coefficient

$$S_{T,C} = -\frac{\nabla u_C}{\nabla T} = \frac{Q_C^*}{T}/\frac{d\mu_C}{du_C} \qquad \textit{Eq. 72}$$

On the other hand, $Q_C^*$ can be obtained from the definition given in Eq. 1 and 2 in the publication by Höglund and Ågren [87] in combination with Eq. 24 in the present paper with $\nabla V = 0$ as follows

$$Q_C^* = T\frac{\partial \mu_C}{\partial T} = -TS_C \qquad \textit{Eq. 73}$$

Höglund and Ågren [87] assumed a constant $Q_C^*$ and found that $Q_C^* = -44\ kJ/mol$ can best reproduce the $C$ concentration profile after 102 hours reported by Okafor et al. [111], who assumed that the steady state were reached after 102 hours and obtained $Q_C^* = -12.3\ kJ/mol$. The negative value of $Q_C^*$ is in agreement with Eq. 73 as $S_C$ is positive. The simulated concentration profiles are shown in Figure 6 for 102 and 13889 (marked infinite) hours superimposed with experimental data by Okafor et al. [111] after 102 hours. It is evident that the steady state was not reached after 102 hours. Furthermore, even with $Q_{Fe}^* = Q_{Ni}^* = 0$, both Fe and Ni should diffuse due to the $C$ concentration profile. More detailed simulation results will be presented in a future publication.

*Figure 6: Simulated $C$ concentration profiles for 102 and 13889 hours* [87] *with experimental data from ref* [111] *after 102 hours superimposed.*

### 4.6 Uphill diffusion: A cross phenomenon between diffusion components

Diffusion in multicomponent systems under the isothermal condition is important for materials processing and joining of dissimilar materials [119–121]. Following the discussion on thermodiffusion, let us consider an isolated single-phase multicomponent system with an initial inhomogeneous distribution of some diffusion components and a homogeneous temperature profile. The inhomogeneous composition results in variations of chemical potentials which drive the diffusion of individual components from high chemical potential regions to low chemical potential regions. One scenario worth further discussion is when one component diffuses much faster than other components and thus takes much short time to reach the chemical equilibrium than other slow diffusion components. In the case that the chemical potential of the fast diffusion component is significantly affected by a slow diffusion component, the concentration of the fast diffusion component can become more inhomogeneous than before, i.e. so-called uphill diffusion where the fast diffusion component migrates against its concentration gradient.

Uphill diffusion was postulated by Darken [122] who pointed out that "for a system of more than two components it is no longer necessarily true that a given element tends to diffuse toward a region of lower concentration even within a single phase region", due to non-ideal interactions in the solution phase that is so great that the concentration gradient and the chemical potential gradient may be of different signs [123]. Darken further performed experimental validation of his postulations through four weld-diffusion experiments with pairs of steel of virtually the same carbon content, but differing markedly in alloy contents [124]. In two diffusion couples with the single face-centered-cubic (fcc) austenite solid solution phase, i.e. (I) Fe-3.8Si-0.25Mn-0.49C/Fe-0.05Si-0.88Mn-0.45C and (II) Fe-3.8Si-0.25Mn-0.49C/Fe-0.14Si-6.45Mn-0.58C (all in wt%), Darken observed that C diffusion from the low concentration region to the high concentration region at 1050°C after about two weeks. The diffusion couple II has contributions from both Si and Mn, while the diffusion couple I is primarily due to the Si concentration difference.

For the sake of simplicity, let us consider the diffusion couple I in the Fe-Si-C ternary system and write the flux of C as follows by omitting the effects of partial volume and partial entropy

$$J_C = -L_C \nabla \mu_C = -L_C(\Phi_{CC} \nabla c_C + \Phi_{CSi} \nabla c_{Si} + \Phi_{CFe} \nabla c_{Fe}) \qquad \textit{Eq. 74}$$

In usual situations, the last two terms are less important when $\nabla c_{Si}$ and $\nabla c_{Fe}$ are small, and the flux is controlled by the first term, resulting in $J_C$ and $\nabla c_C$ being in opposite directions, i.e. the diffusion of C from high concentration to low concentration regions. However, in dissimilar metal welds, the gradients of other components can be rather steep [125,126]. In the Fe-Si-C diffusion couple I, the Si contents in two parts of the diffusion couple are very different. The chemical potential of C at 1050°C is significantly increased by the Si content as shown in Figure 7(i) calculated using a CALPHAD thermodynamic database and Thermo-Calc software [112,113]. Since the mobility of Si is orders of magnitude lower than that of C, the second term $\Phi_{CSi} \nabla c_{Si}$ in Eq. 74 will drive C from high Si content region to low Si content region in order to reduce the chemical potential gradient of C. With a CALPHAD mobility database and Dictra software [112,113], the diffusion process in the diffusion couple I can be simulated [112], and the C concentration profile after 13 days is plotted in Figure 7(ii) in comparison with the experimental data by Darken [124], showing excellent agreement. Furthermore, Darken drew a schematic diagram showing the change in composition of two points on opposite sides of the weld in ultimately approaching uniformity of composition (see figure 6 in ref. [124]). Such a diagram for the diffusion couple I after 13 days is plotted in Figure 7(ii) in qualitative agreement with Darken's schematic diagram though much steeper.

*Figure 7: (i) Chemical potential of C in Fe-Si-0.45C alloys plotted with respect to the Si content at 1050°C with the TCFE9 thermodynamic database* [113]*; (ii) C composition profile in diffusion couple I after 13 days; (iii) C and Si compositions in the diffusion couple I with the numbers used to calculate the distance from the high Si side with the formula shown in the diagram.*

On the other hand, Smoluchowski mentioned a study of a diffusion couple between Fe-0.80C and Fe-4.0Co-0.80C alloys [127] without change in the carbon distribution after several days at 1000°C. It is indeed that the chemical potential of C in the Fe-0.80C alloy is increased only slightly by the Co content at 1000°C as shown in Figure 8, in agreement with Darken's suggestion [124,128]. Darken also noted that the uphill diffusion in liquid was observed more than a decade earlier by Hartley [129] and continues to be a very important phenomenon in a wide range of applications [130,131], such as transport of ionic species in aqueous solutions and crossing of distillation boundaries that are normally forbidden in multicomponent mixtures with azeotropes.

*Figure 8: Chemical potential of C in the Fe-0.80C alloy is increased only slightly by the Co content at 1000°C, calculated using a CALPHAD thermodynamic database and Thermo-Calc software* [112,113].

Another commonly observed uphill diffusion is when the solution is unstable with respect to composition fluctuations, referred to as spinodal decomposition [13,130,132,133]. The limit of stability of a solution is located at $\Phi_{ii} = 0$, and inside a spinodal the solution is unstable with $\Phi_{ii} < 0$, and the flux and concentration gradient of the component can thus point to the same direction as shown in many equations above, e.g., Eq. 52. This uphill diffusion thus irreversibly transfers an initially homogenous unstable solution to an inhomogeneous stable solution and ultimately a miscibility gap of the same phase with two distinct composition sets. The instability-driven spinodal decompositions exist in both binary and multicomponent systems and play a central role in the formation of patterns [13] and dissipative structures [32].

An important phenomenon should be discussed before the end of this section, i.e. the Kirkendall effect [134–136]. During the time of Darken's work, it was commonly believed that atomic diffusion in solid took place through direct exchange of atoms or ring mechanism with equal diffusion coefficients of binary elements. During that time period, Huntington and Seitz [137] suggested the vacancy mechanism based on evaluation of the activation energy for self-diffusion in copper using the electron theory. Kirkendall measured the diffusion coefficients of Zn in the Cu-Zn fcc solution ($\alpha$-brass) using diffusion couples between $\beta$-brass with 56.41-58.63wt%Zn (an intermetallic phase) and pure Cu at three temperatures [134] and concluded that Zn diffuses faster than Cu in $\alpha$-brass [135]. Subsequently, using single-phase diffusion couples between Cu-30wt%Zn $\alpha$-brass and pure Cu with insoluble thin Mo wires inserted in the bonding interface, Smigelskas and Kirkendall [136] observed the shift of Mo wires towards the original $\alpha$-brass, i.e. a shrinkage of the original $\alpha$-brass, indicating that Zn diffuses out of the high Zn $\alpha$-brass faster than Cu diffuses into the region. The Kirkendall effect and vacancy diffusion mechanism were accepted by the community a couple of years later [138]. This effect can be written in general as follows

$$J_{Va} = -\sum_{i \in S} J_i \qquad \textit{Eq. 75}$$

where $J_{Va}$ is the flux of vacancy, and $i \in S$ denotes all substitutional components. It shows that the unbalanced fluxes of substitutional components on the right-hand side of Eq. 75 can result in a net vacancy flow and may induce the formation of voids.

One additional effect of isothermal diffusion processes is that the diffusion of components can in principle result in a heat conduction due to the partial entropy differences of components similar to the discussion in Section 4.2 represented by Eq. 35 and can be written as

$$J_Q = T \sum_i S_i J_i \qquad \textit{Eq. 76}$$

It is noted that the heat flux in diffusion of atoms is probably much smaller than that in electron conduction.

## 4.7 Electromigration: A cross phenomenon between atomic diffusion and electrical field

Electromigration concerns the cross phenomenon of atomic diffusion under an external electrical field and is the most serious and persistent reliability issue in interconnect metallization and flip chip solder joints in electronic devices [139–141]. Early research activities on electromigration was reviewed by Black [142] as a critical issue in electronic devices due to the formation of voids at interfaces between metals and semiconductors and the formation of metallic whiskers resulting in electrical short circuit. Systematical investigations on electromigration were reported for Al and Al-alloys [143–145] and Cu and Cu-alloys [146–148]. Particularly, Blech [144] investigated the electromigration of Al as a function of temperature, electrical current, and sample size and observed significant diffusion of Al at high temperature, high electrical current, and large sample size. Electromigration related to Pb-free solder joints has also been extensively investigated including effects of thermodiffusion, electrical currents, and stress on the formation of Kirkendall voids as discussed in Section 4.6 above [149–152].

For electromigration under an externally applied electrical field, the initial internal process is the electrical current, followed by heat conduction and internal stress induced by electrical field and electrical current, that ultimately induces atomic diffusion. The flux for electromigration can thus be written as follows based on Eq. 20

$$J_i = -L_i \nabla \mu_i = -L_i \left( \sum_j \Phi_{ij} \nabla c_j - S_i \nabla T - \varepsilon_i \nabla \sigma - \theta_i \nabla E \right) \quad \textit{Eq. 77}$$

where partial strain $\varepsilon_i = \frac{\partial \varepsilon}{\partial c_i}$ and partial electrical displacement $\theta_i = \frac{\partial \theta}{\partial c_i}$ are introduced with the Gibbs energy as the characteristic function [10,21,22]. The summation in Eq. 77 includes a term for electrons due to the electrical current. It is noted that $L_i$, $\Phi_{ij}$, $\varepsilon_i$, and $\theta_i$ are functions of all internal variables shown in Eq. 77 plus additionally microstructural features such as dislocations, grain boundaries, and surfaces, which become more pronounced with the size reduction of electronic devices [141].

Kirchheim [153] and Basaran et al. [154] considered the contributions to flux from the concentration gradient of the diffusion component, the stress gradient, and the electrical current plus a source term for the non-equilibrium vacancy concentration with a characteristic vacancy generation or annihilation time when applying Fick's second law. One may consider that the electrical current in their model includes both the contributions of electron density and externally imposed electrical field expressed in Eq. 77. It is noted that the diffusion of vacancy is related to the diffusion of atomic components, and the vacancy concentration depends on all variables in Eq. 77 and is included in $L_i$ unless it is controlled externally as an independent variable such as radiation. The net vacancy flux is denoted by Eq. 75. It is thus not clear whether the vacancy needs to be separately considered as shown by Kirchheim [153] and Basaran et al. [154] as the mechanisms for non-equilibrium vacancy and the parameter for vacancy generation or annihilation time are not clearly defined.

For pure metals such as Al and Cu, $\theta_i \approx 0$, and one may also assume homogeneous heating due to the electrical current with $\nabla c_e \approx 0$, $\nabla T \approx 0$ and $\nabla \sigma \approx 0$. Eq. 77 can thus be approximated as follows

$$J_A = -L_A \nabla \mu_A = -L_A \Phi_{Ae} \nabla c_e \qquad \textit{Eq. 78}$$

Therefore, Al or Cu diffuses in the same direction of the decrease of electron concentration, i.e. the direction of electron flow from cathode to anode, and the vacancy diffuses in the opposite direction ($J_{Va} = -J_A$, see Eq. 75) resulting in the formation void on the cathode side [139]. The heat generated by the electrical current can be calculated by Eq. 76 with the summation including both the element and electron. For multicomponent systems with a temperature gradient, thermal expansion or contraction, the equations derived in Sections 4.5 and 4.6 can be utilized along with a mechanical energy term [153,154] and the thermodynamic and atomic mobility databases by the CALPHAD method [1,114–118].

### 4.8 Electrocaloric effect: A cross phenomenon between heat conduction and electrical field

The electrocaloric effect (ECE) concerns the cross phenomenon of heat conduction under an external electrical field, promising for both heating and cooling devices [155–161], in the opposite direction of pyroelectricity with voltage generation due to heating or cooling. It may be mentioned that the caloric effect can be realized additionally by external magnetic and mechanical fields [160], which will be briefly discussed in Section 4.10. It is similar to electromigration with the electrons as the only diffusion component. When an external electrical field is applied to the system, an electrical current is generated with electrons carrying heat with them, resulting in a heat conduction as follows from the general form of Eq. 20

$$J_Q = -L_Q \nabla T = -L_Q \left( \frac{\partial T}{\partial S} \nabla S + \sum\nolimits_i \frac{\partial T}{\partial c_i} \nabla c_i + \frac{\partial T}{\partial \sigma} \nabla \sigma + \frac{\partial T}{\partial E} \nabla E \right) \qquad \textit{Eq. 79}$$

The derivatives in Eq. 79 are represented below

$$\frac{\partial T}{\partial S} = \frac{T}{C_P} \qquad \textit{Eq. 80}$$

$$\frac{\partial T}{\partial c_i} = \frac{\partial^2 \Phi}{\partial S \partial c_i} = \frac{\partial \mu_i}{\partial S} = \sum\nolimits_j \frac{\partial \mu_i / \partial c_j}{\partial S / \partial c_j} = \sum\nolimits_j \frac{\Phi_{ij}}{S_j} \qquad \textit{Eq. 81}$$

$$\frac{\partial T}{\partial \sigma} = \frac{\partial^2 \Phi}{\partial S \partial \sigma} = -\frac{\partial \varepsilon}{\partial S} = -\frac{1}{S_\varepsilon} \qquad \textit{Eq. 82}$$

$$\frac{\partial T}{\partial E} = \frac{\partial^2 \Phi}{\partial S \partial E} = -\frac{\partial \theta}{\partial S} = -\frac{1}{S_\theta} \qquad \textit{Eq. 83}$$

where $C_P$, $S_\varepsilon = \frac{\partial S}{\partial \varepsilon}$, and $S_\theta = \frac{\partial S}{\partial \theta}$ are isobaric heat capacity under constant stress, and partial entropy with respect to strain and electrical displacement, respectively. The discussions on electrocaloric effects in the literature usually assume $\nabla c_i = 0$ and $\nabla \sigma = 0$, and Eq. 79 becomes

$$J_Q = -L_Q \left( \frac{T}{C_P} \nabla S - \frac{1}{S_\theta} \nabla E \right) \quad \textit{Eq. 84}$$

More recently, Chen et al. [22] discussed the thermodynamics of the direct and indirect methods used in characterizing the ECE materials. In the direct method, the temperature increase under the *adiabatic condition* as a function of electrical field is measured, i.e. $(\Delta T/\Delta E)_{\Delta Q=0}$. In the literature, the adiabatic condition with $\Delta Q = 0$ is often treated as identical to the isentropic condition with $\Delta S = 0$, which is true for a closed, equilibrium system, but not true for closed, non-equilibrium systems with internal processes as shown by Eq. 1. The differential form of $(\Delta T/\Delta E)_S$ is shown by Eq. 83 with the temperature change obtained through integration

$$\Delta T_{direct} = \int_{E_1}^{E_2} \frac{\partial T}{\partial E} dE = -\int_{E_1}^{E_2} \frac{1}{S_\theta} dE \quad \textit{Eq. 85}$$

In the indirect method, the heat production under the *isothermal condition* as a function of electrical field is measured, i.e. $(\Delta Q/\Delta E)_T$, where $\Delta Q$ is again usually approximated to be the same as $\Delta S$. Its differential form can be obtained as follows

$$\frac{\partial S}{\partial E} = -\frac{\partial^2 G}{\partial E \partial T} = \frac{\partial \theta}{\partial T} \quad \textit{Eq. 86}$$

The entropy change in the indirect method can be calculated through integration of Eq. 86 and is then used to calculate the anticipated temperature increase using the heat capacity of the materials as follows

$$\Delta S = \int_{E_1}^{E_2} \frac{\partial S}{\partial E} dE = \int_{E_1}^{E_2} \frac{\partial \theta}{\partial T} dE = \int_{T_1}^{T_{2,indirect}} \frac{C_P}{T} dT \qquad \textit{Eq. 87}$$

Consequently, $\Delta T_{indirect} = T_{2,indirect} - T_1$ can be obtained. While $\Delta T_{direct}$ from Eq. 85 and $\Delta S$ and $\Delta T_{indirect}$ from Eq. 87 can all be used to characterize the performance of ECE materials, $\Delta T_{direct}$ and $\Delta T_{indirect}$ are likely different from each other since both methods start from the same state, i.e. $(T_1, S_1, E_1)$, but end at different states, i.e. $\left(T_{2,direct},\ E_2\right)_{S_1}$ vs $\left(S_{2,indirect},\ E_2\right)_{T_1}$. Therefore, it is in general that $\Delta T_{indirect} \neq \Delta T_{direct}$.

In both methods, the integration is important because the materials properties change nonlinearly with respect to the electrical field. This is particularly true near critical points or phase transition boundaries commonly referred as morphotropic phase boundaries (MPBs). Unfortunately, as shown by Chen et al. [22] (noting the sign difference of their Eq. 1), Eq. 87 is usually approximated in the literature by the following equation

$$T_1 \Delta S = C_P \Delta T_{indirect} \qquad \textit{Eq. 88}$$

This can thus introduce large errors due to the dramatic temperature dependence of $C_P$ near MPBs where ECE is mostly investigated. Chen et al. [22] also pointed out the importance of domains and domain interactions in microstructure as also discussed in the literature [158,162–165] and will be further examined in Section 5.

### 4.9 Electromechanical effect: A cross phenomenon between electrical current and stress

The electromechanical effect concerns the interactions between electrical field and elastic deformation usually under isothermal condition without atomic diffusion [166–170]. Giant electromechanical effects have been observed near MPBs [171–174]. In piezoelectricity, electrical charge is accumulated in response to externally applied elastic deformation. While in the converse piezoelectric effect, the externally applied electrical field creates elastic deformation in the system.

In piezoelectricity, the electrical current under an external applied strain or stress can be obtained from Eq. 34, respectively

$$J_e = -L_e \nabla\mu_e = -L_e\left(\frac{\partial\mu_e}{\partial c_e}\nabla c_e + \frac{\partial\mu_e}{\partial\varepsilon}\nabla\varepsilon\right) = -L_e(\Phi_{ee}\nabla c_e + \sigma_e\nabla\varepsilon) \qquad \textit{Eq. 89}$$

$$J_e = -L_e\nabla\mu_e = -L_e(\Phi_{ee}\nabla c_e + \varepsilon_e\nabla\sigma) \qquad \textit{Eq. 90}$$

where $\Phi_{ee}$ is the thermodynamic factor and can be calculated as shown in Section 4.4, and $\sigma_e = \frac{\partial\mu_e}{\partial\varepsilon} = \frac{\partial\sigma}{\partial c_e}$ and $\varepsilon_e = \frac{\partial\mu_e}{\partial\sigma} = -\frac{\partial\varepsilon}{\partial c_e}$ are the partial stress and strain with respect to electron concentration, respectively. Under the condition of zero electrical current, the voltage generated is as follows under an external applied strain or stress, similar to Eq. 39, respectively.

$$\frac{\nabla V_e}{\nabla\varepsilon} = \sigma_e \qquad \textit{Eq. 91}$$

$$\frac{\nabla V_e}{\nabla\sigma} = \varepsilon_e \qquad \textit{Eq. 92}$$

In the converse piezoelectric effect, the elastic deformation is induced by an externally applied electrical field. The mechanical equilibrium can be defined by vanishing stress gradient as follows

$$\nabla\sigma = \frac{\partial\sigma}{\partial\varepsilon}\nabla\varepsilon + \frac{\partial\sigma}{\partial c_e}\nabla c_e + \frac{\partial\sigma}{\partial E}\nabla E = c\nabla\varepsilon + \sigma_e\nabla c_e - \theta_\varepsilon\nabla E = 0 \qquad \textit{Eq. 93}$$

where $c$ represents the elastic coefficient tensor, and $\theta_\varepsilon = \frac{\partial\theta}{\partial\varepsilon} = -\frac{\partial\sigma}{\partial E}$ is the partial electrical displacement with respect to strain.

### 4.10 General discussion of cross phenomena

In the previous sections, the general flux equation denoted by Eq. 20 is applied to cross phenomena on transport of electrons and atomic components induced by externally controlled temperature, electrical, stress, and strain fields. As shown by Eq. 20 and the discussions above, the cross phenomena can be represented by the derivatives of one potential to its conjugate molar quantity, other potentials, and other non-conjugate molar quantities, which are the second derivatives of free energy to its natural variables. The choice of free energy depends on how the system is controlled from the surroundings with relevant terms subtracted from the right-hand side of the general free energy formula shown by Eq. 10, i.e. the ensemble of the system as mentioned in Section 2. The derivatives between potential and molar quantities represent well known physical properties in the literature and are shown in Table 1 which were discussed in the recent overview by the author [1], except the last row which is for derivatives related to chemical reactivity of components due to external stimuli, $\partial N_j/\partial Y^a$, or other components for the last cell in the row, $\partial N_j/\partial N_i$. As discussed by the author [1], the diagonal properties in Table 1 are the derivatives of potentials with respect to their respective conjugate molar quantities $\left(\frac{\partial Y^a}{\partial X^a}\right)$ and are all positive for stable systems. The limit of stability is reached when they become zero $\left(\frac{\partial Y^a}{\partial X^a} = +0\right)$ and their inverses, i.e. the derivatives of molar quantities with respect to their respective conjugate potentials, diverge and become positive infinite, i.e.

$$\frac{\partial X^a}{\partial Y^a} = +\infty \qquad \textit{Eq. 94}$$

On the other hand, the off-diagonal properties can be either positive or negative, so are their divergencies

$$\left(\frac{\partial X^a}{\partial Y^b}\right)_{a \neq b} = \pm\infty \qquad \textit{Eq. 95}$$

This was discussed recently in terms of the zentropy theory [20] and will be further presented in Section 5. It should be mentioned that every derivative between potential and molar quantity in Table 1 denotes an internal process that represents how the system responds to the external stimulus and must result in entropy production for it to happen irreversibly as dictated by $Td_{ip}S > 0$ in Eq. 1. This includes thermal expansion, $\partial V/\partial T$ for the volumetric flow with respect to the external temperature change [20] that will be discussed in Section 5.3.

On the other hand, the derivatives between any two potentials are presented in Table 2 for all the cross-phenomenon coefficients. The last row of Table 2 denotes the derivatives between one chemical potential and other potentials and can represent either transport cross phenomena or chemical reactions or their combinations. For transport cross phenomena due to externally applied potentials, they include the thermodiffusion and electromigration discussed in the present work and stressmigration [175–177] and magnetomigration [178] discussed in the literature. The same quantities can be applied to electrons to result in thermoelectricity and electrochemical and electrocaloric effects discussed in the present work and the magnetochemical effect [179–182] in the literature. While for chemical reactions, they represent whether the formation of a component is enhanced or lessened by the external potentials as the increase or decrease of chemical potential of a components indicates the increase of decrease of its moles in a stable system. One example of chemical reactions is a quantum reaction for the formation of quasiparticles from electrons in quantum materials that will be discussed in Section 5.5.

By combining Table 1 and Table 2, one can define many cross phenomena using both derivatives with respect to potential, molar quantity, or a mixture of them, such as the electromechanical effects with respect to externally applied stress, strain, or a mixture of strain and stress [21]. It is noted that even though the derivatives between potentials can be understood straightforwardly through the combined law of thermodynamics as demonstrated in the present paper, they are usually not considered useful and rarely discussed in thermodynamic textbooks. This is probably due to the deeply rooted belief that thermodynamics is for equilibrium only as engraved by the combined law derived by Gibbs [5,183,184]. Since each potential is homogeneous in an equilibrium system, the derivatives between potentials are indeed not very meaningful when only equilibrium is considered. However, as discussed above, they play a central role in understanding and predicting internal processes and cross phenomena as the bridge between thermodynamics and kinetics.

The general flux equation, Eq. 20, can be applied to all irreversible processes inside a system [185–187] driven by any or all five sets of potentials, i.e. temperature, mechanical, electrical, magnetic, and chemical with one or more of them controlled externally from the surroundings in experiments. Both Table 1 and Table 2 are symmetrical due to the Maxwell relations through the second derivatives of free energy to its natural variables. This is particularly valuable for quantities in Table 2 as the derivatives between potentials, i.e. the lower left portion of the table, are relatively easy to measure experimentally, while the derivatives between molar quantities, i.e. the upper right portion of the table, are ideal for theoretically investigations, which are the focus of Section 5.

# 5 Zentropy theory for prediction of entropy and cross phenomenon coefficients

## 5.1 General discussion on entropy, fluctuation, and critical phenomena

The importance of temperature in all kinetic processes is reflected by the temperature dependence of kinetic coefficients shown by Eq. 33. The derivative of a free energy to temperature gives one of the most important quantities in science, i.e. entropy, which is the core of the second law of thermodynamics shown by Eq. 1 in the present paper and can be fundamentally understood by the Boltzmann distribution as discussed in Section 4.2. Entropy is intrinsically multi-scale from its presence in black holes [188–190], societies [191], forests [192], and quantum systems [17,193,194].

Thermodynamic modeling for metallic and some oxide and molten salt systems has been progressed significantly in last half century in terms of the semi-empirical CALPHAD method [1,114–118] as a function of temperature, compositions, and spontaneous magnetization, while the thermodynamic modeling with respect to ferroelectric polarization and stress/strain has been largely based on the Landau-Ginsburg-Devonshire (LGD) formalism [195,196] led by Cross and co-workers [197–199]. All those models are very useful for practical applications, particularly the extrapolations of the CALPHAD databases from binary and ternary systems to multicomponent systems for computational design of engineering materials [200–202] and the LGD formalism for ferroelectric materials [21]. However, those models are phenomenological, and their model parameters are evaluated from experimental data though DFT-based first-principles calculations have played more and more important roles recently [118,203].

While the DFT-based first-principles calculations can provide reliable free energy of a given configuration under the quasiharmonic approximation [50,51,204,205], including the recent stochastic self-consistent harmonic approximation [206], the prediction of anharmonicity remains challenging [207,208]. The state-of-the-art approaches are the *ab initio* molecular dynamic simulations (AIMD) for high temperatures [209] and the quantum Monte Carlo (QMC) simulations for low temperatures [210]. In addition to their highly computational intensity, they are not capable to systematically predict the critical point where a system changes from being stable to unstable, and the anharmonicity is at its maximum due to the divergency of molar quantities, including the thermodiffusion near critical point [211] and quantum critical point (QCP) [212] with the divergence of the quasiparticle effective mass (QEM) [213–216] which will be further discussed in Section 5.3. In Section 4.10, the concept of instability was introduced in terms of the divergence of the derivative of a molar quantity to its conjugate potential. A system usually changes from stable to metastable before reaching the limit of stability and becoming unstable, except at the critical point where the metastability region vanishes and the system switches from being stable to unstable directly.

The critical phenomena can be understood by the renormalization group (RG) theory in terms of statistical continuum limit with the fluctuations of many length or energy scales cascaded in a Hamiltonian that enters the partition function of the system [217–219], and the challenge is shifted to evaluating the Hamiltonians for various cases. Another challenge in implementing the RG theory is the explicit incorporation of multiscale interactions in the Hamiltonian. Consequently, predictive approaches are still lacking for critical phenomena at low temperatures for quantum critical points [210] and at high temperatures where Lennard-Jones potentials are usually used [220,221] or by fitted equation of state [222]. The more recently developed fluctuation theorem (FT) [15–17,223–226] also concerns the fluctuations in nonequilibrium steady-state systems, which may be considered as an extension of the equilibrium fluctuation often conveniently assumed in the RG theory. The development of FT started with the discussion that the 2$^{nd}$ law of thermodynamics has a statistical probability to be "violated" in the individual microscopic trajectories of those fluctuations though not macroscopically with all the microscopic trajectories combined [15]. There have been a number of experimental and theoretical studies followed [225,227–230]. As discussed below, this is probably because an additional entropy contribution due to the probability of the microscopic trajectory is not accounted for [224]. This additional entropy contribution has been related to the information or internal statistical microstates of fluctuations [229,231,232] and will further be discussed in Section 5.4.

The lack of quantification of free energy of ferroelectric materials was also mentioned by Scott [158] who pointed out that there is no known analytic equation of state for most ferroelectric materials, and the theories based on the LGD formalism do not include ferroelectric domains even though much of the entropy change involves the creation and destruction of domains [162–165]. Scott further noted that in the LGD approximation, one typically assumes that all coefficients except one are independent of temperature, resulting in the entropy change at a first-order phase transition equal to the square of the polarization change timed by 2π. In other words, the Lorentz factor comes from a quadratic free energy expansion in polarization, which is rarely valid.

Even though the LGD formalism describes the ferroelectric (FE) and paraelectric (PE) behaviors well macroscopically, the modern microscopic theory of polarization (MTP) [233–236] is needed to understand thermodynamics of ferroelectric polarization. In the MTP framework, effective Hamiltonians have been constructed by representing the energy surface in terms of a Taylor expansion around the high-symmetry cubic structure along with several approximations for computation efficiency. Their model parameters were fitted to DFT-based first-principles calculations followed by Monte Carlo or molecular dynamics simulations [162,164,237–239]. Significant nonzero local polarization are predicted in the cubic PE state for $BaTiO_3$ [162,237] and $PbTiO_3$ [238] in agreement with the experimental observations in the literature [240–242] and the orthorhombic-like 90º domain structure in $PbTiO_3$ in agreements with experiments [164] and AIMD simulations [165]. These results further demonstrate the important contributions of fluctuations and domain walls to free energy of materials, which is discussed in the next section.

### 5.2 Zentropy theory for multiscale entropy

Entropy can be presented in terms of the integration of heat capacity (see Eq. 87), which is convenient in experiments, or Boltzmann distribution (see discussions in Sections 4.2 and 4.4), which is convenient in theory and computations. In the literature, one often differentiates the thermal electronic, phonon, and configurational entropies as they are at different scales, governed by different physics, and calculated by different methods [50,204,243]. Nevertheless, they are all configurational at their respective scales and distinctive physics. The distribution of electrons is represented by the Fermi-Dirac statistics as shown by Eq. 47, while the phonon distribution is characterized by the Bose–Einstein statistics for indistinguishable particles. For classic distinguishable particles, the Maxwell-Boltzmann statistics applies.

Hierarchically, a macroscopic system of investigation (termed as macrostate in the present paper) can be considered as a statistical mixture of various microscopic configurations of the system (termed as microstates in the present paper), while each microstate can further be considered as a statistical mixture of various sub-microscopic configurations of the system. The scale higher than the macrostate is treated as the surroundings of the system [10,11], which dictates the statistical ensemble to be used to study the system as mentioned in Section 2 with a proper free energy represented by Eq. 10. The configurational entropy of the macrostate can be represented by Gibbs-Boltzmann entropy as follows,

$$S^{conf} = -k_B \sum_k p_k ln p_k \qquad \textit{Eq. 96}$$

where $p_k$ is the probability of microstate $k$ in the macrostate.

It is important to realize that in this hierarchical multiscale framework, each microstate and sub-microstate must fill the complete space of the system and experience the same constraints of the statistical ensemble of the system. It is thus evident that the total entropy of the macrostate needs to include the entropy of each microstate in addition to the configurational entropy among them as follows [14],

$$S = \sum_k p_k S^k + S^{conf} = \sum_k p_k (S^k - k_B ln p_k) = S_{0K} + \int_0^T \frac{C_h}{T} dT \qquad \textit{Eq. 97}$$

where $S^k$ is the entropy of microstate $k$ which can be further decomposed into its sub-microstates with the same formula as Eq. 97, and $S_{0K}$ and $C_h$ are the entropy at 0 $K$ and the heat capacity of the macrostate. The superscript is used in $S^k$ as the subscript is already used for partial entropy in the present paper. The integration from the experimentally measured heat capacity in Eq. 97 is usually obtained under the convention of the third law of thermodynamics, i.e. $S_{0K} = 0$ at 0 $K$, which is also reflected by DFT stating that there is only one ground state at 0 K fully defined by its electron density distribution[8,9]. Nevertheless, contributions to $S_{0K}$ could be included if needed and available.

From Eq. 97, the free energy and partition function of the macrostate and the probability of each microstate can be obtained as follows

$$\Phi = \sum_k p_k \Phi^k + k_B T \sum_k p_k ln p_k \qquad \textit{Eq. 98}$$

$$Z = e^{-\frac{\Phi}{k_B T}} = \sum_k e^{-\frac{\Phi^k}{k_B T}} = \sum_k Z_k \qquad \textit{Eq. 99}$$

$$p_k = \frac{Z_k}{Z} = \frac{Z_k}{\sum_k Z_k} = e^{-\frac{\Phi^k - \Phi}{k_B T}} \qquad \textit{Eq. 100}$$

The above equations represent a nested formula for many length or energy scales, resembling the RG theory discussed above though without explicit considerations of interactions between microstates, thus considerably simplifying the calculations. This is possible only if the microstates are ergodic, i.e. the interactions between microstates in the macrostate are represented internally in one or more microstates, similar to the ergodic dynamical systems considered in FT [16]. When the ergodicity is not met, an additional mean-field term needs to be added as shown by two approaches in predicting the critical point in $Ce$ [244,245]. It is noted that the free energy is used in the partition function in Eq. 99 rather than the total energy used in the partition function shown by Landau and Lifshitz [246]. In their derivation, the quantum states are treated as the microstates with their entropy implicitly assumed to be zero, i.e. $S^k = 0$, resulting in the configurational entropy among the microstates (Eq. 96) representing the total entropy of the system with the quasi-classical approximation. The detailed discussion on this comparison will be presented in a separate publication.

We have recently termed the nested formula, i.e. Eq. 96 to Eq. 100, as zentropy theory in an overview of our previous studies [20]. It is shown that the zentropy theory is capable of predicting critical points in magnetic materials and associated anomaly including the positive and negative divergency of volume with respect to temperature, i.e. positive and negative thermal expansions with all inputs from DFT-based quasiharmonic calculations without empirical parameters (see Section 5.3). The significances of the zentropy theory, which includes Eq. 96 to Eq. 100, are as follows

(i) The nested form of Eq. 97 covers any scales of investigations through the combination of all lower scales and the scale of interest represented by $S^k$ and $S^{conf}$ and can thus in principle describe any systems of interest from the quantum scale to black holes.
(ii) In the MSE discipline, the macrostates of interest are microstructures consisting of individual phases and their temporal and spatial arrangements. To start with, one can take each individual phase as a macrostate of investigation and define its microstates in terms of atomic, magnetic, electrical, and defect configurations. The entropy of each microstate, $S^k$, includes the contributions from thermal electrons and phonons that can be predicted by DFT-based first-principles calculations using the Fermi-Dirac and the Bose–Einstein statistics, respectively [50,247].

(iii) $S^{conf}$ represents the statistical fluctuation of microstates and is the key to predict the critical phenomena and anharmonicity at high temperatures. It is partially represented by phenomenological models such as the LGD formalism [195,196] and the CALPHAD compound energy formalism [248].

(iv) The zentropy theory integrates the bottom-up approach based on *quantum mechanics* ($S^k$ in Eq. 97) and the top-down approach based on *statistical mechanics* ($S^{conf}$ in Eq. 97), i.e. (ii) and (iii) above, and connects the combination of the quantum mechanics and statistical mechanics to the macroscopic *thermodynamics* ($C_h$ and its integration in Eq. 97). Such characteristics of the zentropy theory preserve the quantum information from the DFT electron level to the macroscopic level of experimental investigations and enable the prediction of the nonlinear, emergent macroscopic behaviors of macroscopic functionalities at finite temperatures, including their ultimate extreme at critical points [20] and potential applicability to superconductivity to be discussed in Section 5.5.

## 5.3 Applications of zentropy theory to magnetic and ferroelectric transitions

It is mentioned above that thermal expansion is one type of cross phenomenon as it concerns the volumetric flow driven by temperature, though it is not commonly related to cross phenomena in the literature. The volumetric flow with respect to temperature is represented by the third term in Eq. 20, i.e. $\frac{\partial Y_j^a}{\partial X_j^{b\neq a}}\nabla X_j^{b\neq a}$ with $Y_j^a$ for pressure and $X_j^{b\neq a}$ for entropy. The inverse of the corresponding Maxwell relation is

$$\frac{\partial S}{\partial(-P)} = \frac{\partial V}{\partial T} \qquad \textit{Eq. 101}$$

We applied the zentropy theory to magnetic transitions in anti-invar $Ce$ and invar $Fe_3Pt$ with the microstates defined by magnetic spin configurations [20]. The thermal electronic and phonon entropic contributions for each magnetic microstate are predicted from the DFT-based first-principles calculations. Since the transition in $Ce$ is between nonmagnetic (NM) ground-state microstate and ferromagnetic (FM) high temperature non-ground-state microstate, we started with these two microstates, but had to add a mean-field magnetic flipping term in free energy as commonly done in the literature [244] in order to obtain the critical point.

It is known that the co-existence of the NM and FM microstates results in spin-flipping magnetic (SFM) microstates, generating domain walls. We subsequently added the antiferromagnetic (AFM) microstate and predicted the critical point and associated anomalies from the decrease of probability of the NM microstate and the increase of probabilities of the AFM and FM microstates without the mean-field term [1,245,249]. The predicted temperature-pressure potential phase diagram using the zentropy theory is show in Figure 9(i) with a critical point and the two-phase region represented by the line in agreement with the Gibbs phase rule [10,11]. Various experimental data are well reproduced with details discussed in the early publication [20,245].

To show the divergency of volume at the critical point, the pressure axis is replaced by its conjugate molar quantity, i.e. volume, resulting in a temperature-volume potential-molar phase diagram plotted in Figure 9(ii). It is noted that the two-phase region changes from a line in the temperature-pressure potential phase diagram to an area in the temperature-volume potential-molar phase diagram. This does not violate the Gibbs phase rule as the Gibbs phase rule concerns only the number of independent potentials without changing the number of phases in equilibrium and needs to be modified if applied to phase diagrams including molar quantities [10]. Several isobaric volume curves are plotted in Figure 9(ii) for $Ce$, depicting the thermal expansion anomaly marked by the pink diamond symbols. It can be seen that at the critical point marked by the green circle $\frac{\partial T}{\partial V} = 0$ and $\frac{\partial V}{\partial T} = +\infty$ in accordance with Eq. 95. There is a temperature range on each isobaric volume curve that the thermal expansion shows abnormally larger than usual. The further is it away from the critical point, the less is the anomaly. In both Figure 9(i) and (ii), the agreement between experiments and predictions is remarkable with all inputs from DFT-based first-principles calculations and no empirical parameters.

*Figure 9: Predicted phase diagrams of $Ce$ (i) temperature-pressure* [245] *with symbols for experimental data and (ii) temperature-volume* [249] *with purple diamond squares for thermal expansion anomaly and other symbols for experimental data*.

While for $Fe_3Pt$, the magnetic transition is between FM and paramagnetic (PM) macrostates, and many more SFM microstates are needed. We found that the 12 atom supercell with 9 $Fe$ atoms is large enough to predict experimentally observed anomalies [1,249,250]. This supercell results in $2^9 = 512$ magnetic microstates with 37 being unique due to symmetry. With their free energies predicted from DFT-based first-principles calculations including both thermal electronic and phonon contributions, the predicted temperature-pressure [250] is calculated using the zentropy theory and shown in Figure 10(i) with a critical point marked by the red circle in good agreement with experimental data. As in the case of $Ce$, the temperature-volume phase diagram of $Fe_3Pt$ is plotted in Figure 10(ii) with several isobaric volume curves. The filled black circles on the ambient pressure (0GPa) volume curve are experimental data with details discussed in the early publication [249], showing remarkable agreement with predicted values, including the negative thermal expansion temperature range. The purple diamonds mark the anomalous regions of negative thermal expansions, including the negative divergences at the critical point marked by the green circle, i.e. $\frac{\partial T}{\partial V} = 0$ and $\frac{\partial V}{\partial T} = -\infty$ in accordance with Eq. 95. It was concluded that the negative thermal expansion originates from the decreasing probability of the ground-state FM microstate and the increasing probabilities of those non-ground-state SFM microstates with their volumes smaller than that of the FM microstate.

*Figure 10: Predicted phases diagrams of $Fe_3Pt$ (i) temperature-pressure* [250] *with experimental data superimposed, noting pressure decrease from left to right, and (ii) temperature-volume* [249] *with experimental data (black circles) on the ambient pressure (0GPa) volume curve and the purple diamonds for the anomalous regions of negative thermal expansions.*

We have applied the zentropy theory to the AFM-PM transitions in several rare-earth nicklates [251] and the FE-PE transitions in ferroelectric materials [165,252] and observed remarkable agreement between predicted and measured transition temperatures. Those results are being prepared for publications.

Another interesting experimental phenomenon related to the limit of stability is the divergence of QEM at a QCP [213–216] as mentioned in Section 5.1 with the examples of $YbRh_2Si_2$ [213], $YbRh_2(Si_{0.95}Ge_{0.05})_2$ [214], Al-doped CrAs [253], and $FeSe_{1-x}S_x$ [254,255] shown in Figure 11. Monthoux et al. [256] pointed out that quasiparticles reduce to bare electrons in the hypothetical absence of the electron-electron and electron-ion interactions. In conductors, a quasiparticle may be considered as an electron plus a co-moving screening cloud, while in superconductors, the quasiparticles can have effective masses two or more orders of magnitude greater than that of electrons and hence move relatively slowly [256]. According to itinerant spin fluctuation theory [212], the QEM divergence at a QCP is due to the abundance of low-lying and long-range spin fluctuations, mediating the interactions between the heavy quasiparticles, giving rise to pronounced deviations from Landau Fermi liquid behavior [213–216]. This is exactly how the zentropy theory works.

*Figure 11: QCP phase diagrams: (i) temperature with respect to magnetic field of* $YbRh_2Si_2$ *with antiferromagnetic (AF), non-Fermi liquid (NFL), and Landau Fermi liquid (LFL) phase regions* [213], *(ii) temperature with respect to magnetic field of* $YbRh_2Si_2$ *and* $YbRh_2(Si_{0.95}Ge_{0.05})_2$ *with AF (blue left), NFL (yellow), and LFL (blue, right) phase regions* [214], *(iii) temperature with respect to pressure with contour map of electrical resistivity of Al-*

*doped CrAs and pure CrAs* [253], *and (iv) temperature with respect to composition of* $FeSe_{1-x}S_x$ *with contour maps for electrical resistivity (ρ) and a parameter (A*) related to quasiparticle effective mass* [255].

Let us discuss further the divergency of QEM at a QCP in terms of the limit of stability in Section 4.10 and the present section for $Ce$ and $Fe_3Pt$ with more details presented in the literature [1,10,11]. One should keep in mind that even though a QCP is theoretically defined as a point at $0\ K$ where a microstate becomes unstable to new forms of microstates, experimental measurements are always performed at finite temperatures with thermal fluctuations. Let us consider a simple internal process by adding one electron to an existing quasiparticle represented by the following reaction

$$q_{n-1} + e = q_n \qquad \textit{Eq. 102}$$

The combined law of thermodynamics can be re-written from Eq. 3 as follows with $d\xi = dn$ and its driving force per quasiparticle as $D_n = -(\mu_{q_n} - \mu_{q_{n-1}} - \mu_e) = -\Delta\mu_{q_n}$

$$dU = Y^a dX^a + \Delta\mu_{q_n}\mathrm{d}n \qquad \textit{Eq. 103}$$

where $\mu_e$, $\mu_{q_n}$, $\mu_{q_{n-1}}$ are the chemical potentials of electron and quasiparticle with $n$ and $n-1$ electrons, respectively, and $Y^a$ and $X^a$ are the conjugate variables representing the external controlling variables, i.e. pressure, magnetic field, or composition (chemical potential) of dopants as shown in Figure 11. The limit of stability of the system in terms of Eq. 95 can be written as

$$\frac{\partial n}{\partial Y^a} = \pm\infty \qquad \textit{Eq. 104}$$

where the positive and negative signs represent the change of $Y^a$ or $X^a$ towards QCP or away from QCP as the driving force $D_n$ changes sign accordingly. To make the model more realistic, one needs to consider a distribution of quasiparticle sizes determined by thermodynamic equilibrium with a distribution of electrons as shown in Section 4.4. When $n$ diverges, its effective mass, $m^*_{q_n} \approx n\, m^0_e$ for relatively slow-moving quasiparticles [256] with $m^0_e$ being the resting mass of electron, also diverges as shown in Figure 11.

### 5.4 On microscopic "violation of second law of thermodynamics"

As mentioned above, the development of fluctuation theorem started with the discussion by Evans et al. [15] on probability of "violations of second law of thermodynamics" in a nonequilibrium steady state far from equilibrium. They derived an expression for the ratio of the probabilities to find a fluid with an induced shear stress in the direction of or opposite to the externally imposed shear rate on a phase space trajectory segment of a certain duration in a dynamical state. Through MD simulations, they observed statistically significant probability that the induced shear stress is in the opposite direction of the externally imposed shear rate, resulting in negative entropy production evaluated from the induced shear stress, thus an apparent "violation of the second law of thermodynamics" for those events or internal processes.

Independently, Jarzynski [224] derived an expression for the equilibrium free energy difference between two states of a system, $\Delta F$, in terms of an ensemble of nonequilibrium (finite-time) measurements with the work, $W$, as follows

$$\Delta F = -\frac{1}{k_B T} ln\left(\overline{e^{-\frac{W}{k_B T}}}\right) \qquad \textit{Eq. 105}$$

The work is computed for each trajectory in the ensemble with the quasiequilibrium statistical mechanics invoked. This is remarkable as $\Delta F$ represents an equilibrium property independent of path, while $W$ is realized by trajectories with finite rate and

$$\bar{W} \geq \Delta F \qquad \textit{Eq. 106}$$

The equality in Eq. 106 holds for infinitely slow internal processes between two states, i.e. the conventionally defined reversible processes. Jarzynski [224] also showed the applicability of the above equations to the cases where the system of interest and its surroundings together constitute a larger, isolated Hamiltonian system.

The validity of Eq. 105 and Eq. 106 does not prevent some trajectories with $W < \Delta F$, and many theoretical, computational, and experimental investigations have demonstrated the existence of such trajectories [15,225,227–230]. Particularly, using a single-electron transistor two-level system with $\Delta F = 0$ and a single thermodynamic trajectory level coupled to a single heat bath, Maillet et al. [229] experimentally demonstrated that the amount of work up to large fractions of $k_B T$ can be extracted from the two carefully designed out-of-equilibrium driving cycles with probabilities substantially greater than 1/2. These observations result in an apparent "violation of the second law of thermodynamics" for those individual trajectories with the external intervention and energy consumption to drive and select those trajectories. By including the degrees of freedom of external intervention such as a feedback controller, Sagawa and Ueda [232] concluded the consistency with the conventional second law of thermodynamics because of the energy cost of the controller. However, this does not resolve the issue related to those trajectories with $W < \Delta F$ as internal processes because second law of thermodynamics is not about the total entropy change, but the entropy production of each independent internal process.

Therefore, the interpretations of results from the theory, computations, and experiments in the framework of fluctuation theorem are worth another look, particularly with each individual trajectory as an internal process. We will approach this in terms of the relative magnitude of $Q_{\xi_j}$ and $k_B T$ discussed in Section 4.1. It is noted that Jarzynski [224] emphasized that $W$ must not be much greater than $k_B T$ in order to verify Eq. 105 experimentally. If the fluctuations in $W$ between trajectories are much larger than $k_B T$ (Cases II and III discussed in Section 4.1), the ensemble average in Eq. 105 will be dominated by trajectories with much lower $W$ (Case IV discussed in Section 4.1). The situation here is similar to the reversible Brownian motion in a macroscopically equilibrium system, which is used for the prediction of diffusivity in terms of DFT-based first-principles calculations discussed in Section 4.3 [57–64], with $\Delta F = 0$.

Let us follow each individual atom in a reversible Brownian motion and consider its every motion as an internal process or trajectory. When the atom sits still at its equilibrium position, such as at 0 K, both the work and entropy are zero. Next, when the atom starts to fluctuate from its equilibrium position at low temperatures, the entropy production of the internal process is represented by the phonon dispersion and the associated vibrational entropy calculated from the probability of each vibrational frequency that the atom can possess through integration over all frequencies and probabilities through the Bose-Einstein statistics. For conductors, thermal electronic entropy can be included through Fermi-Dirac statistics. When the fluctuation is large enough, the probability for the atom to move to the next vacant site becomes significant. This internal process now generates a new configurational entropy due to the probability for the atom to jump over to the previous vacant site or fall back to the new vacant site that it leaves behind. This new entropy contribution can be represented by Eq. 96 as

$$S^{conf} = -k_B(p_+ lnp_+ + p_- lnp_-) \qquad \textit{Eq. 107}$$

with $p_+$ and $p_-$ being the probability jumping forward and backward, respectively. This results in an additional positive entropy production with its maximum value as $k_B ln2$ with $p_+ = p_- = 0.5$, assuming that the two microstates are identical with the same entropy. The entropy production due to the Brownian motion can thus be written from Eq. 97 as

$$\Delta_{ip} S = (S^+ - S^-) + k_B ln2 = k_B ln2 \qquad \textit{Eq. 108}$$

with $S^+$ and $S^-$ being the entropy of two microstates, respectively. Consequently, the second law of thermodynamics is maintained as the internal process of a single trajectory generates a positive entropy production.

The work along the trajectory for every successful jump of a Brownian motion is performed by the thermal energy. Part of the work could in principles be extracted by an external intervention, being a fraction of $TS^{conf} = k_B T ln2$. While in the experiments of the single-electron transistor two-level system by Maillet et al. [229], the electron is manipulated by external force, and they did point out the Shannon entropy difference between the equilibrium microstates before and after the quench, which is related to $S^{conf}$. The fluctuations can be considered in similar fashion for nonequilibrium steady state systems such as various cross phenomena discussed in Section 4 by dividing the nonequilibrium system into a succession of regions with small differences of local temperature and concentration [15]. Furthermore, the probability of a successful jump can be used to evaluate the diffusivity in the DFT-based first-principles calculations discussed in Section 4.3 [57–64]. It seems that in most discussions related to the "violation of second law of thermodynamics", $S^{conf}$ is not included, resulting in incomplete accounting of entropy production in a single trajectory.

Let us discuss further the entropy production of internal processes in general. In our previous work [14], it was shown that the entropy production of an internal process can be divided into four parts: (1) heat generation $\left(d_{ip}Q\right)$, (2) consumption of some nutrients $(dN_i^n)$, (3) production of some wastes $\left(dN_j^w\right)$, and (4) reorganization of its configurations $\left(d_{ip}S^{config}\right)$, as follows,

$$d_{ip}S = \frac{d_{ip}Q}{T} - \sum\nolimits_i S_i^n dN_i^n + \sum\nolimits_j S_j^w dN_j^w + d_{ip}S^{config} \qquad \textit{Eq. 109}$$

where $S_i^n$ and $S_j^w$ are the partial entropies of nutrient $i$ and waste $j$ of the internal process, respectively. For individual trajectories considered here, $dN_i^n = dN_j^w = 0$ with no mass exchange, $d_{ip}Q = 0$ without heat exchange between the internal process and its surroundings, and $d_{ip}S^{config}$ is obtained from Eq. 97 as follows

$$d_{ip}S = d_{ip}S^{config} = \sum\nolimits_k d(p_k S^k) - k_B \sum\nolimits_k d(p_k ln p_k) \qquad \textit{Eq. 110}$$

The first term in Eq. 110 is due to the thermal electronic and vibrational contributions of each microstate. The second term makes a positive contribution to the entropy production of each trajectory due to introduction of new microstates and is the key to demonstrate the validity of the second law of thermodynamics in both microscopical and macroscopical scales. The fluctuations and their probabilities of individual trajectories are related to the relative magnitude of barrier and $k_B T$ and nested in our zentropy theory in terms of Eq. 96 to Eq. 100. It is further noted that the scales are relatives as mentioned at the beginning of Section 5.1, and a macrostate in one investigation can be a microstate in a larger scale of investigation [14].

### 5.5 Postulations on superconductivity as a cross phenomenon with criticality

Superconductivity is a phenomenon discovered in 1911 where electrical resistance of a material vanishes at temperatures below a critical temperature $T_C$, and correspondingly, its electrical conductivity diverges. It is a critical phenomenon with Schottky anomaly and a critical point between the 1st and 2nd order transitions [257]. The transition temperatures can be altered through composition, pressure, and magnetic field (see Figure 11). The Bardeen–Cooper–Schrieffer (BCS) theory [258,259] was developed in 1957 as the microscopic theory of superconductivity in terms of the condensation of Cooper pairs. This paired state of electrons is originated from the electron–phonon interaction and has a lower energy than the Fermi energy. Due to the weak pairing interaction ($\sim 10^{-3} eV$), thermal energy ($k_B T$ with $k_B$=8.617 $10^{-5} eV/K$) can easily break the pairs, resulting in conventional superconductors only at low temperatures. The BCS theory can explain the conventional low $T_C$ superconductors, but fails for high $T_C$ superconductors [260], and the early predictions based on the phonon-induced attraction indicated a ceiling to $T_C$ below the boiling point of liquid nitrogen [256,261].

The current approaches used in the community to predict $T_C$ for conventional BCS superconductors is based on Eliashberg's model for interactions between electrons and lattice vibrations [262]. McMillan [261] and Allen and Dynes [263] made further developments, with the latter pointing out that Eliashberg theory appears to place no bounds on achievable values of $T_C$ determined by two key parameters with one for the attractive effects of the electron-phonon coupling and another for the repulsive effects of direct Coulomb interactions between electrons. Ashcroft [264,265] predicted the superconductivity of metallic hydrogen either as a monatomic or paired metal using the following mean field formula in terms of the BCS theory

$$T_C = 0.85\Theta_D e^{-1/n(\varepsilon_F)\phi_{el-ph}} \qquad \textit{Eq. 111}$$

where $\Theta_D$ is the Debye temperature derived from the highest-frequency vibrational mode in the system, $n(\varepsilon_F)$ the e-DOS at the Fermi level (see Eq. 46), and $\phi_{el-ph}$ an effective electron-phonon attractive interaction [266,267]. This mean field formula is also used in predicting of $T_C$ of various hydrides [268–272] with strong anharmonicity included [273].

Stewart [274] compared the fundamental properties of 9 unconventional superconducting classes of materials including 4f-electron heavy fermions, cuprates, Fe-based superconductors, organic superconductors, and new emerging topological superconductors with the hope that common features would emerge to help theory explain and predict these phenomena. It was concluded that "a common understanding - in the manner of a BCS theory - is not possible" with the chief question being "what interaction is coupling the electrons?" Monthoux et al.[256] also pointed out the importance of understanding of effective charge-charge interaction between quasiparticles and its dependences on phonons or spins. Typical superconducting phase diagrams are presented in Figure 12 for (i) cooperates [275], (ii) $CeM_2X_2$ [256], (iii) $Ba(Fe_{1-x}M_x)_2As_2$ [276], and (iv) carbon- and hydrogen-doped $H_3S$ [277], showing that the transitions between insulator, semiconductor, conductor, and superconductor of a material can be controlled through pressure and composition (doping).

*Figure 12: Superconducting phase diagrams, (i) cuprates* [275], *(ii)* $CeM_2X_2$ [256], *(iii)* $Ba(Fe_{1-x}M_x)_2As_2$ [276], *and (iv) carbon- and hydrogen-doped* $H_3S$ [277].

The author's interest on superconductivity started when working on processing of the $MgB_2$ superconductor [278,279]. When the original concept of the zentropy theory was developed in 2008 [244], the author's group explored the applicability of the zentropy theory to challenges facing the scientific community. After some literature search, five topics were identified: (1) thermoelectric materials, (2) multiferroic materials, (3) coexistence of conductivity and transparency, (4) superconductivity, and (5) metal-insulator transition. Five people collected relevant information in the literature, gave short presentations, and led the lively discussions during a retreat held on April 12, 2008. As discussed in Section 4.4, we have made good progress in predicting Seebeck coefficient [68,69] for topic 1. For topic 2, we are able to predict magnetic transitions [1,245,249,250] and are working on manuscripts applying the zentropy theory for more complex magnetic materials and ferroelectric materials [252]. We have not made any progresses on topics 3. We have been actively working on topics 4 and 5 as they are closely related as mentioned above.

Right after the retreat, the author wrote five postulations on superconductivity, which are re-organized as follows

I. Postulation One: Superconductivity at 0 K. A conductor could be made to be superconducting at 0 K through proper external constrains such as deformation or doping.

II. Postulation Two: Low $T_C$ superconductors. For crystals with high entropy due to phonons, the electron-phonon interactions will interfere with the migration of electron pairs at finite temperatures. When temperature is high enough, the above interactions will break the Cooper pairs, and superconductivity is lost. This group of materials exhibits low $T_C$ and follows the BCS theory, including $MgB_2$.

III. Postulation Three: High $T_C$ superconductors. The layered structures in complex phases with or without magnetism significantly reduce the entropy due to phonons, thus preserving the Cooper pairs within the layers. There can be two scenarios here: (i) the entropy due to phonons becomes dominant with the increase of temperature, and the superconductivity is lost before magnetic disordering; (ii) as the magnetic disordering lifts the constraints on the entropy due to phonons, the superconductivity is lost at the same time of the magnetic disordering temperature.

IV. Postulation Four: Doping Effects. Some possible effects of doping are: (a) increase (decrease) the number of electron pairs for higher (lower) $T_C$; (b) stiff (soften) the lattice and reduce (increase) the entropy due to phonons for higher (lower) $T_C$; (c) enhance (diminish) the magnetic strength for higher (lower) $T_C$.

V. Postulation Five: Transition between superconductor and non-superconductor. Similar to SFM microstates proposed for Invar materials such as $Fe_3Pt$ [249,250] (which was submitted for publication in March 2008), the transition between superconductor and non-superconductor can be proposed as follows: (a) there are two extreme states in terms of superconductivity: superconducting state or non-superconducting state; (b) there can be many states between these two states; (c) the thermal populations of various states vary as a function of temperature, pressure, or magnetic field; (d) various mixed states can be represented through the Boltzmann distributions expressed in terms of partition functions (now called zentropy theory); (e) the transition between the superconductor and non-superconductor can be predicted using the zentropy theory in terms of Eq. 98 to Eq. 100; (f) the critical transition point with respect to magnetic field should also be reflected at 0 K. Therefore, one should be able to see a transition between the superconductor and non-superconductor on the temperature-pressure diagram at 0K and its extension to higher temperatures with respect to pressure (similar to Figure 9 and Figure 10). Similarly, one should be able to see a transition in the temperature-magnetic field diagram at 0 K, which extends to higher temperatures with the decrease of the magnetic field. Therefore, 0 K stability should be examined first in detail; (g) The key challenge is to define the superconducting and various non-superconducting microstates.

Along these intuitive postulations, the author's team applied the zentropy theory to predict the properties of $BaFe_2As_2$ [280–283]. $BaFe_2As_2$ is a parent compound of iron arsenide superconductors [284] and is orthorhombic and tetragonal at low and high temperatures, respectively. It was recently made into a superconductor through epitaxial superlattice heterostructures [285]. The ground state of $BaFe_2As_2$ is a spin density wave (SDW) AFM microstate with a stripe-like Fe spin ordering pattern within the *ab* plane and antiparallel nearest-neighbor Fe spins along the *c* axis [280–283]. Figure 13(i) plots the predicted SDW ordering temperature ($T_{SDW}$) and the characteristic temperature ($T_*$) as a function of pressure using the zentropy theory with experimental measurements superimposed, showing remarkable agreement [282]. $T_{SDW}$ and $T_*$ are defined when the probability of the SDW-AFM microstate equals to 0.5 and 0.9999, respectively. The Fermi surfaces of the SDW and Stripe microstates are plotted in Figure 13(ii). The Stripe microstate is similar to the SDW microstate except that the nearest-neighbor Fe spins along the *c* axis are parallel [281,282]. The blue and red colors in Figure 13(ii) mark the single sheet hole-pocket and single sheet electron-pocket, respectively, with G, Z, Y, and T denoting the high symmetry points in the reciprocal space. The shape of the predicted Fermi surface near the Z points of the SDW microstate agree with the measurement reported in the literature [286].

*Figure 13: Orthorhombic-$BaFe_2As_2$ (i) SDW ordering temperature ($T_{SDW}$, red curve) and characteristic temperature ($T_*$, blue curve) plotted with respect to pressure with experimental data (symbols)* [282], *and (ii) Fermi surface, Isosurfaces: (a) Stripe and (b) SDW; Cuts: (c) Stripe and (d) SDW; Top views of isosurfaces: (e) Stripe and (f) SDW* [281].

As demonstrated above, the essential component of the zentropy theory is to assemble the microstates at 0 K that are ergodic for the investigation and predict their free energies by DFT-based calculations. While we should continue to use our scientific intuitions to guide the design of important microstates, the author believes that the integration of domain knowledge and deep neural network machine learning approaches [287–289] has the potential to develop robust approaches to assemble the ergodic microstates for complex systems including superconductors aided by high throughput DFT-based calculations [290,291] and high throughput thermodynamic modeling [292–295]. The power of such integrated approach is demonstrated in our recent work in determining the crystal structure of $NdBi_2$ [296]. First, all the $AB_2$-type configurations (26,055) were identified within a large dataset of DFT-relaxed or experimental structures (~1.3 millions at the time of extraction, now over 4 millions) [289]. Following the substitution with $NdBi_2$ into the microstates, all of the candidates were examined by our deep neural network machine learning models (SIPFENN: structure-informed prediction of formation energy using neural networks) [287,288] to identify configurations with low formation enthalpy, resulting in 500 $AB_2$-type configurations with low values up to 12 kJ/mol-atom variation. With initial DFT-based verifications, 20 configurations with the lowest values were selected for more detailed DFT-based calculations to obtain the energy versus volume equation of state and its stable crystal structure in accordance with the experimental structure data by X-ray diffractions [296].

The insulating and conducting microstates can be defined by their band gaps, which can be predicted by the DFT-based calculations with continuously improved accuracy and efficiency [68,297,298], while the superconducting microstate is often discussed in connection with the Fermi surface[299] (see Figure 13 (ii)), but a succinct differentiation between superconducting and normal conducting microstates is lacking. Considering that the electronic properties in the DFT-based calculations are represented by electronic band structure, density of states, charge density, and Fermi surface [281,300–302], their combination may uniquely define the superconducting microstate of a superconductor at 0 K.

Among various superconductors studied, two are highlighted here, i.e. an optimally doped $HgBa_2CuO_{4+d}$ (Hg-1201) [303] and $FeSe_{1-x}S_x$ [304] as shown in Figure 14. For Hg-1201 at lowest pressures, the behavior is metallic with a sharp superconducting transition and a well-defined tetragonal structure with a fluctuating 1D linear oxygen ordering along the two tetragonal axes. As pressure increases, $T_C$ increases slightly up to 5 GPa and then decreases with a widened superconducting transition, reaching 0 K at 11.5 GPa (see blue open circles Figure 14(i)). At higher pressures, it becomes an insulator. At 8 GPa, the superstructure diffraction spots start to appear and become more intense and diffuse at higher pressures (see green filled squares in Figure 14(i)). Since no extra diffuse lines develop in the *a–b* plane, the ordering is proposed to be along the *c* axis, resulting in a 3D ordering of the oxygen. Various phase regions are depicted in Figure 14(i) as a function of hole concentration in Hg-1201, including the superconducting phase (SC in the figure). These ordered structures can be considered as the microstates in Hg-1201 and will be used to study the transitions between superconductor, conductor and insulator in terms of the zentropy theory.

A large number of experimental [305,306,315,307–314] and computational [316–318] investigations has been reported for $FeSe$ due to its unique properties as an exceptional member of the family of Fe-based superconductors [308] with complex phase diagrams involving magnetic, superconducting, and structural transitions. For example, the effects of $S$ [305] are shown in Figure 14(iii) with similar ones for $Te$ [315]. It is hoped that the zentropy theory will be able to help to develop better understanding of the superconducting transition through the quantum fluctuations of superconducting and non-superconducting microstates and aid the discovery of new superconductors.

*Figure 14: Phase diagrams of (i) $T_C$ and normalized intensity of the superstructure spots as a function of pressure in Hg-1201* [303], *(ii) $T$ plotted against hole concentration in Hg-1201 with hole concentration for (i) marked by the vertical rectangle and various phase regions* [303], *(iii) $T$ plotted against pressure in $FeSe_{1-x}S_x$* [304] *with various phase regions: SDW (see Figure 13(i)), Tetra (tetragonal), Ortho (orthorhombic), M (magnetic), Nematic (no magnetic long-range), and SC (superconducting).*

### 5.6 Cross phenomena involving phase transformations with multiple internal processes

When commenting on Darken's work [124], Bhadeshia [319] generalized Darken's statement to "for a system of more than two components, it is no longer necessarily true that a given element tends to diffuse toward a region of lower chemical potential", which happens when some elements are trapped by a moving interface between two phases [320]. When a local equilibrium is assumed at the phase interface, the chemical potential of each element is the same in both phases at the interface, and the interface migration is controlled by diffusion in each phase from high chemical potential regions to low chemical potential regions in that phase. When the phase interfaces deviate from local equilibrium, the chemical potentials of some elements in the new phase increase, and those of other elements decrease with respect to their chemical potentials in the parent phase. At the extreme, some or all elements are completely trapped from the parent phase into the new phase and have the same compositions in both phases, resulting in their conjugate potentials being different in the two phases. The phase transformation is thus termed paraequilibrium for complete trapping of substitutional elements or partitionless for complete trapping of all elements, respectively [321,322]. It is important to point out that when independent variables are changed from potentials to conjugate molar quantities, the Gibbs phase rule needs to be modified as shown by Hillert [10]. If there are additional constraints to a system such as the coherency of phase interface, the Gibbs phase rule needs to be further revised [323].

The intermediate scenarios between local equilibrium, paraequilibrium, and partitionless transitions are determined by the interplay of thermodynamic and kinetic properties of the systems, resulting in a wide range of dissipation of energy and dissipative structures [13,32,322,324,325]. As depicted in Figure 15 for Gibbs energy diagram of an alloy system A-B, Ågren [326] showed that the chemical potential of element B in the growing α phase, $\mu_B^{\alpha/\beta}$, is higher than that in the parent β phase, $\mu_B^{\beta/\alpha}$, due to the diffusion over the interface between α and β that requires a driving force, and the chemical potential for local equilibrium is between these two values (not drawn). The opposite is true for the chemical potential of element A. This is called solute drag, a highly important topic for phase transformations and properties of multicomponent materials [322,326–333]. Therefore, it seems that the element B migrates from high chemical potential region to low chemical potential region across the interface, due to the overall reduction of free energy as pointed out by Bhadeshia [319].

*Figure 15: Gibbs energy diagram of an alloy system A-B with the effect of a driving force for diffusion acting over an interface between $\alpha$ and $\beta$, and the tangents describing the chemical potentials on both sides of the interface rotated relative to each other* [326].

This can be further related to the coupling of atomic diffusion when crossing the phase interface. When multiple internal processes occur *simultaneously* at the same location, i.e. the migration of interface and the diffusion across the interface in the present case, they are coupled, and a new driving force needs to be defined to describe the coupled outcome as shown by Eq. 5. For example, in the case of paraequilibrium, since the diffusion of all substitutional elements is prevented due to the much faster migration of interface than the atomic diffusion rates, the substitutional elements are no longer independent, but act collectively. Their chemical potentials are combined into one chemical potential, $\mu_{\in S}$, as follows, resulting in a revised local equilibrium condition[321],

$$\mu_{\in S} = \sum_{i \in S} c_i \mu_i^{\alpha} = \sum_{i \in S} c_i \mu_i^{\beta} \qquad \textit{Eq. 112}$$

In a study on dissolution of cementite ($MC_3$ with $M$ representing the mixture of Fe and Cr) at 910°C in an Fe-2.06Cr-3.91C (at. pct) alloy, the author found that the newly formed austenite inherited the Cr content of cementite due to the low diffusivity of Cr, while the high diffusivity of C resulted in a nearly instantaneous dissolution of cementite with its volume fraction decreased from 14.1 to 12.4 % [120] in a fraction of second, most likely by paraequilibrium transformation. After the C activity reached homogenous in the system, the further dissolution of cementite was controlled by the slower diffusion of Cr, resulting in the local equilibrium at the interface. This local equilibrium condition increased the Cr concentration in $MC_3$ at the interface to a very high value as shown by the isothermal section in Figure 16(i) [120]. Consequently, Cr diffuses from the interface towards the center of $MC_3$ particles as shown by the concentration profile in Figure 16(ii) after 1000 seconds holding at 910°C [120]. Computer simulations under the local equilibrium conditions showed good agreement with experimental measurements in Figure 16(ii).

*Figure 16: Dissolution of cementite at 910°C in an Fe-2.06Cr-3.91C (at. pct) alloy, (i) Isothermal section [120], (ii) Cr concentration profile after 1000 s [120], (iii) isothermal section with C activity plotted on the x-axis* [334], *(iv) transmission electron microscope (TEM) bright field image of cementite exhibiting Widmanstatten plates of α-bcc phase (γ-fcc at dissolution temperature)* [334], *and (v) TEM bright field image of extraction replica showing lamellar $M_7C_3$ structure ($u_{Cr} = 0.54$ at point A) and untransformed $MC_3$ (darker area)* [334].

Furthermore, Figure 16(i) shows that the compositions of $MC_3$ near the interface fall initially into the $\gamma + M_7C_3 + M_{23}C_6$ three-phase region and move into the $\gamma + M_7C_3$ two-phare region when the activity of C increases with more $MC_3$ dissolved. This can be seen more clearly with the C composition replaced by its chemical activity in the isothermal section shown in Figure 16(iii) [334]. It is evident that both $MC_3$ compositions at the center and near the interface, denoted by the horizontal dashed line and the top dotted line, respectively, are within the $\gamma + M_7C_3$ stable two-phase regions. Indeed, the $MC_3(low\ Cr) \rightarrow \gamma + MC_3(high\ Cr)$ transformation was observed inside $MC_3$ particles (Figure 16(iv)) [334], along with the decomposition of $MC_3$ into the lamellar structure of $\gamma + M_7C_3$ (Figure 16(v)) [334].

This brings up an interesting topic on classification of phase transitions [335,336]. The cross phenomena discussed in the present article becomes more complex due to multiple internal processes with moving interfaces as mentioned above. The simplest cases are grain boundaries in polycrystals and domain walls in ferromagnetic and ferroelectric materials. In polycrystals, particularly nanograins, both the grain growth and redistributions of elements between the grain interior and grain boundary need to be considered as two independent internal processes simultaneously [12,337]. In ferroelectrics, the polarization domains need to be included explicitly in thermodynamic models in order to predict the FE-PE transitions as pointed out by Scott [158] and discussed in Section 5.1.

It is noted that phase transformations are usually classified by three schemes in terms of thermodynamics, mechanism, and morphology [335,336]. They are related to the three inner circles in Figure 1(ii), replotted in Figure 17 with the thermodynamic scheme being at the interface between the thermodynamics and kinetics circles, the mechanistic scheme within the circle of kinetics, and the morphological scheme between the crystallography and kinetics circles.  Each scheme thus focuses on difference aspects of a phase transformation [336], and when combined together they give a complete understanding of phase transformations.

*Figure 17:Classification schemes of kinetic processes in relation to fundamental components of materials science and engineering.*

In the thermodynamic scheme, phase transformations are classified by the derivatives of free energy with respect to potentials, which are their conjugate molar quantities.  The order of a phase transition is assigned when the highest order of derivatives between two phases is discontinuous.  For example, in a first-order transformation, the molar quantities of the two phases are different, while for the second-order transformation, all quantities in Table 1 are different between the two phases.  The first- and second-order transformations meet each other at a critical point as shown in Figure 9 and Figure 10.  Higher-order transformations are less discussed in the literature and can be important for smaller scale phase transitions that may include the superconducting transition mentioned above [338–341].  Furthermore, as pointed out by Ågren [336], one should use the molar quantity of the individual phases in defining the order of a transformation, not those of the system.

The microstructural scheme is divided into two major groups: homogenous and heterogeneous with the former taking place everywhere simultaneously and the latter starting from some regions and propagating through the system. While a heterogenous transformation is certainly first-order in the thermodynamic scheme, a homogeneous transformation can be either first- or second-order. Though simple to use, the microstructural scheme has less value as it depends on length- and time-scale of investigation and is qualitative in nature with limited fundamental understanding and practical usefulness.

The mechanistic scheme reflects the core activities in the MSE discipline, i.e. to understand the fundamental mechanisms of phase transformations so they can be controlled through chemistry and external stimuli to generate desired microstructures and properties. At the same time, it is a very complex one due to multiple, simultaneous internal processes with cross effects discussed in the present article. Furthermore, since phase transformations take place in three-dimensional (3D) space, while investigations are usually performed in two dimensions (2D), it is important to keep in mind that observations may be incomplete. For example, a pearlite colony in steels looks like that it consists of alternating lamellae of ferrite and cementite in 2D which were thought to be formed by repeated nucleation and growth, while 3D sectioning showed that a pearlite consists of only one ferrite crystal and one cementite crystal, and it is their cooperative growth with shortened diffusion distance that results in the 3D interwoven lamellar structure [342,343]. This demonstrates the importance to have a 3D perspective when investigating complex transformation mechanisms, which can be facilitated by 3D phase-field simulations [13,344].

## 6 Summary and outlooks

In the present article, the combined law of thermodynamics is presented without removing the contributions from irreversible internal processes and is thus applicable to both equilibrium and non-equilibrium systems. It is pointed out that the driving force for the flux of a molar quantity is the gradient of its conjugate potential, which enables the development of physics-based kinetic equations rather than the phenomenological ones proposed by Onsager. The kinetic coefficient matrix is thus diagonal and naturally fulfills the Onsager reciprocal relations for a symmetrical kinetic coefficient matrix. The dependence of the conjugate potential on other independent variables of the internal processes is manifested by the cross phenomena with the coupling coefficients being the derivatives of the potential to the other independent variables. The commonly studied cross phenomena are analyzed in terms of both fundamental understanding and DFT-based computational methodology through the Maxwell relations and our zentropy theory, including thermoelectricity, thermodiffusion, uphill diffusion, electromigration, electrocaloric and electromechanical effects, thermal expansion, and quantum criticality. The successful predictions of Seebeck coefficients in thermoelectric materials, critical phenomena with magnetic transitions and property divergence, and negative thermal expansion in Invar $Fe_3Pt$ without empirical parameters are reviewed. It is anticipated that this new theory of cross phenomena and the associated predictive computational approach will have the potential to enable more efficient discovery of materials with emergent behaviors such as superconductivity along with better understanding of their fundamentals.

## 7 Acknowledgements

The author feels privileged to work with all his current and former students and numerous collaborators over the years at Penn State and around the world as reflected in the references cited in this paper and previous overview paper [1]. He would like to particularly thank Yi Wang, Shun-Li Shang, and Jorge Sofo for simulating discussions on cross phenomena and superconductivity.  The author is grateful for financial supports from many funding agencies in the United States as listed in the cited references and previous overview paper [1], plus some more recent ones including the National Science Foundation (NSF CMMI-2050069), the Department of Energy (DE-AR0001435, DE-NE0008945), and Office of Naval Research (N00014-21-1-2608). The high-performance computing facilities include the Roar supercomputer at the Pennsylvania State University's Institute for Computational and Data Sciences (ICDS), the National Energy Research Scientific Computing Center (NERSC) supported by the U.S. Department of Energy Office of Science User Facility operated under Contract No. DE-AC02-05CH11231, and the Extreme Science and Engineering Discovery Environment (XSEDE) supported by NSF with Grant No. ACI-1548562.  The author would like to thank Johnn Ågren, Graeme Murch, and Irina Belova for many stimulating discussions on cross phenomena, John Mauro, Digby MacDonald, and Susan Sinnott for critical reading of the manuscript, and inspiring communications with Graeme Ackland, Alan Allnatt, Ping Ao, H.K.D.H. Bhadeshia, Dieter Braun, Nigel Goldenfeld, Martin Gruebele, Christopher Jarzynski, Werner Köhler, Manfred Martin, Piergiulio Tempesta, Anton Van Der Ven, George Verros, and Han-Ill Yoo.

## 8 References

[1] Liu Z-K. Computational thermodynamics and its applications. Acta Mater [Internet]. 2020;200:745–792. Available from: https://linkinghub.elsevier.com/retrieve/pii/S1359645420306054.

[2] Onsager L. Reciprocal Relations in Irreversible Processes. I. Phys Rev [Internet]. 1931;38:405–426. Available from: https://link.aps.org/doi/10.1103/PhysRev.37.405.

[3] Liu ZK, Chen L-Q, Spear KE, et al. An Integrated Education Program on Computational Thermodynamics, Kinetics, and Materials Design [Internet]. https://www.tms.org/pubs/journals/JOM/0312/LiuII/LiuII-0312.html. 2003. Available from: https://www.tms.org/pubs/journals/JOM/0312/LiuII/LiuII-0312.html.

[4] Liu ZK. Perspective on Materials Genome®. Chinese Sci Bull [Internet]. 2014;59:1619–1623. Available from: http://link.springer.com/10.1007/s11434-013-0072-x.

[5] Gibbs JW. The collected works of J. Willard Gibbs: Vol. I Thermodynamics. New Haven: Yale University Press; 1948.

[6] McNaught AD, Wilkinson A, editors. Gibbs energy (function), G. IUPAC Compend Chem Terminol [Internet]. Research Triangle Park, NC: International Union of Pure and Applied Chemistry (IUPAC); 2008. Available from: https://goldbook.iupac.org/terms/view/G02629.

[7] McNaught AD, Wilkinson A, editors. Helmholtz energy (function), A. IUPAC Compend Chem Terminol [Internet]. 2nd ed. Research Triangle Park, NC: International Union of Pure and Applied Chemistry (IUPAC); 2014. Available from: https://goldbook.iupac.org/terms/view/H02772.

[8] Burke K. Perspective on density functional theory. J Chem Phys [Internet]. 2012 [cited 2022 Jan 14];136:150901. Available from: https://aip-scitation-org.ezaccess.libraries.psu.edu/doi/abs/10.1063/1.4704546.

[9] Medvedev MG, Bushmarinov IS, Sun J, et al. Density functional theory is straying from the path toward the exact functional. Science [Internet]. 2017 [cited 2018 May 16];355:49–52. Available from: http://www.ncbi.nlm.nih.gov/pubmed/28059761.

[10] Hillert M. Phase Equilibria, Phase Diagrams and Phase Transformations. Second. Cambridge: Cambridge University Press; 2007.

[11] Liu Z-K, Wang Y. Computational Thermodynamics of Materials [Internet]. Cambridge: Cambridge University Press; 2016. Available from: http://ebooks.cambridge.org/ref/id/CBO9781139018265.

[12] Darling KA, Tschopp MA, VanLeeuwen BK, et al. Mitigating grain growth in binary nanocrystalline alloys through solute selection based on thermodynamic stability maps. Comput Mater Sci [Internet]. 2014;84:255–266. Available from: https://linkinghub.elsevier.com/retrieve/pii/S0927025613006356.

[13] Chen L-Q. Phase-field models for microstructure evolution. Annu Rev Mater Res [Internet]. 2002 [cited 2017 Jan 31];32:113–140. Available from: http://www.annualreviews.org/doi/10.1146/annurev.matsci.32.112001.132041.

[14] Liu ZK, Li B, Lin H. Multiscale Entropy and Its Implications to Critical Phenomena, Emergent Behaviors, and Information. J Phase Equilibria Diffus [Internet]. 2019 [cited 2019 Sep 18];40:508–521. Available from: http://link.springer.com/10.1007/s11669-019-00736-w.

[15] Evans DJ, Cohen EGD, Morriss GP. Probability of second law violations in shearing steady states. Phys Rev Lett [Internet]. 1993 [cited 2022 Jan 7];71:2401–2404. Available from: https://link.aps.org/doi/10.1103/PhysRevLett.71.2401.
[16] Evans DJ, Searles DJ. The Fluctuation Theorem. Adv Phys [Internet]. 2002 [cited 2021 Jul 13];51:1529–1585. Available from: http://www.tandfonline.com/doi/abs/10.1080/00018730210155133.
[17] Seifert U. From Stochastic Thermodynamics to Thermodynamic Inference. Annu Rev Condens Matter Phys [Internet]. 2019 [cited 2021 Oct 7];10:171–192. Available from: https://www.annualreviews.org/doi/10.1146/annurev-conmatphys-031218-013554.
[18] Ross D, Strychalski EA, Jarzynski C, et al. Equilibrium free energies from non-equilibrium trajectories with relaxation fluctuation spectroscopy. Nat Phys [Internet]. 2018 [cited 2020 Dec 14];14:842–847. Available from: https://doi.org/10.1038/s41567-018-0153-5.
[19] Duhr S, Braun D. Thermophoretic Depletion Follows Boltzmann Distribution. Phys Rev Lett [Internet]. 2006 [cited 2021 Apr 22];96:168301. Available from: https://link.aps.org/doi/10.1103/PhysRevLett.96.168301.
[20] Liu Z-K, Wang Y, Shang S-L. Zentropy Theory for Positive and Negative Thermal Expansion. J Phase Equilibria Diffus [Internet]. 2022; Available from: https://link.springer.com/10.1007/s11669-022-00942-z.
[21] Chen L-Q. Phase-Field Method of Phase Transitions/Domain Structures in Ferroelectric Thin Films: A Review. J Am Ceram Soc [Internet]. 2008;91:1835–1844. Available from: https://onlinelibrary.wiley.com/doi/10.1111/j.1551-2916.2008.02413.x.
[22] Chen X, Li S, Jian X, et al. Maxwell relation, giant (negative) electrocaloric effect, and polarization hysteresis. Appl Phys Lett [Internet]. 2021 [cited 2021 Dec 23];118:122904. Available from: https://aip.scitation.org/doi/abs/10.1063/5.0042333.
[23] Nye JF. Physical Properties of Crystals: Their Representation by Tensors and Matrices. Oxford: Clarendon Press; 1985.
[24] Onsager L. The Motion of Ions: Principles and Concept. Science. 1969;166:1359–1364.
[25] Balluffi RW, Allen SM, Carter WC. Kinetics of Materials [Internet]. Kinet. Mater. John Wiley and Sons; 2005 [cited 2021 Nov 19]. Available from: https://onlinelibrary.wiley.com/doi/book/10.1002/0471749311.
[26] Prigogine I, de Groot SR. Thermodynamics of Irreversible Processes. New York: Interscience; 1961.
[27] Liu Z-K. Comment on “Thermodiffusion: The physico-chemical mechanics view” [J. Chem. Phys. 154, 024112 (2021)]. J Chem Phys [Internet]. 2021;155:087101. Available from: https://aip.scitation.org/doi/10.1063/5.0055842.
[28] Coleman BD, Truesdell C. On the reciprocal relations of Onsager. J Chem Phys [Internet]. 1960 [cited 2020 Jul 4];33:28–31. Available from: http://aip.scitation.org/doi/10.1063/1.1731098.
[29] Agren J. The Onsager reciprocity relations revisited. J Phase Equilibria Diffus. 2022;accepted.
[30] Charbonneau P, Kurchan J, Parisi G, et al. Fractal free energy landscapes in structural glasses. Nat Commun [Internet]. 2014;5:3725. Available from: http://www.nature.com/articles/ncomms4725.
[31] Chen L-Q. Computer simulation of spinodal decomposition in ternary systems. Acta

Metall Mater [Internet]. 1994 [cited 2020 Feb 9];42:3503–3513. Available from: https://linkinghub.elsevier.com/retrieve/pii/0956715194904820.
[32] Kondepudi D, Prigogine I. Modern Thermodynamics: From Heat Engines to Dissipative Structures. Hoboken, New Jersey: John Wiley & Sons Ltd.; 1998.
[33] Ziman JM. Electrons and Phonons [Internet]. Oxford: Oxford University Press; 2001. Available from: https://oxford.universitypressscholarship.com/view/10.1093/acprof:oso/9780198507796.001.0001/acprof-9780198507796.
[34] Scheidemantel TJ, Ambrosch-Draxl C, Thonhauser T, et al. Transport coefficients from first-principles calculations. Phys Rev B [Internet]. 2003 [cited 2021 Nov 20];68:125210. Available from: https://link.aps.org/doi/10.1103/PhysRevB.68.125210.
[35] Madsen GKH, Carrete J, Verstraete MJ. BoltzTraP2, a program for interpolating band structures and calculating semi-classical transport coefficients. Comput Phys Commun [Internet]. 2018;231:140–145. Available from: https://linkinghub.elsevier.com/retrieve/pii/S0010465518301632.
[36] Poncé S, Li W, Reichardt S, et al. First-principles calculations of charge carrier mobility and conductivity in bulk semiconductors and two-dimensional materials. Reports Prog Phys [Internet]. 2020 [cited 2021 Nov 20];83:036501. Available from: https://iopscience.iop.org/article/10.1088/1361-6633/ab6a43/meta.
[37] BoltzTraP2 [Internet]. Available from: https://gitlab.com/sousaw/BoltzTraP2.
[38] Green MS. Markoff Random Processes and the Statistical Mechanics of Time-Dependent Phenomena. II. Irreversible Processes in Fluids. J Chem Phys [Internet]. 1954 [cited 2021 Dec 10];22:398–413. Available from: http://aip.scitation.org/doi/10.1063/1.1740082.
[39] Kubo R. Statistical-Mechanical Theory of Irreversible Processes. I. General Theory and Simple Applications to Magnetic and Conduction Problems. J Phys Soc Japan [Internet]. 1957 [cited 2021 Nov 20];12:570–586. Available from: https://journals.jps.jp/doi/10.1143/JPSJ.12.570.
[40] Green BR, Troppenz M, Rigamonti S, et al. Memory Function Representation for the Electrical Conductivity of Solids. 2021 [cited 2021 Nov 20]; Available from: https://arxiv.org/abs/2110.02859.
[41] Sandberg N, Magyari-Köpe B, Mattsson TR. Self-Diffusion Rates in Al from Combined First-Principles and Model-Potential Calculations. Phys Rev Lett [Internet]. 2002 [cited 2016 Oct 25];89:065901. Available from: http://link.aps.org/doi/10.1103/PhysRevLett.89.065901.
[42] Fang HZ, Wang WY, Jablonski PD, et al. Effects of reactive elements on the structure and diffusivity of liquid chromia: An ab initio molecular dynamics study. Phys Rev B. 2012;85:014207.
[43] Zou CY, Shin YK, van Duin ACT, et al. Molecular dynamics simulations of the effects of vacancies on nickel self-diffusion, oxygen diffusion and oxidation initiation in nickel, using the ReaxFF reactive force field. Acta Mater. 2015;83:102–112.
[44] Van der Ven A, Ceder G, Asta M, et al. First-principles theory of ionic diffusion with nondilute carriers. Phys Rev B [Internet]. 2001 [cited 2016 Oct 25];64:184307. Available from: http://link.aps.org/doi/10.1103/PhysRevB.64.184307.
[45] Van der Ven A, Ceder G. First Principles Calculation of the Interdiffusion Coefficient in Binary Alloys. Phys Rev Lett [Internet]. 2005 [cited 2021 Jun 14];94:045901. Available

from: https://journals-aps-org.ezaccess.libraries.psu.edu/prl/abstract/10.1103/PhysRevLett.94.045901.
[46] Gabriel Goiri J, Krishna Kolli S, Van der Ven A. Role of short- and long-range ordering on diffusion in Ni-Al alloys. Phys Rev Mater. 2019;3:93402.
[47] Kolli SK, Van Der Ven A. Elucidating the Factors That Cause Cation Diffusion Shutdown in Spinel-Based Electrodes. Chem Mater [Internet]. 2021 [cited 2021 Dec 11];33:6421–6432. Available from: https://pubs-acs-org.ezaccess.libraries.psu.edu/doi/full/10.1021/acs.chemmater.1c01668.
[48] Einstein A. Über die von der molekularkinetischen Theorie der Wärme geforderte Bewegung von in ruhenden Flüssigkeiten suspendierten Teilchen. Ann Phys [Internet]. 1905 [cited 2021 Nov 5];322:549–560. Available from: https://onlinelibrary.wiley.com/doi/full/10.1002/andp.19053220806.
[49] Andersson J, Ågren J. Models for numerical treatment of multicomponent diffusion in simple phases. J Appl Phys [Internet]. 1992;72:1350–1355. Available from: http://aip.scitation.org/doi/10.1063/1.351745.
[50] Wang Y, Liu Z-K, Chen L-Q. Thermodynamic properties of Al, Ni, NiAl, and Ni3Al from first-principles calculations. Acta Mater [Internet]. 2004;52:2665–2671. Available from: https://linkinghub.elsevier.com/retrieve/pii/S1359645404000965.
[51] Shang S-L, Zhou B-C, Wang WY, et al. A comprehensive first-principles study of pure elements: Vacancy formation and migration energies and self-diffusion coefficients. ACTA Mater. 2016;109:128–141.
[52] Eyring H. The Activated Complex in Chemical Reactions. J Chem Phys [Internet]. 1935 [cited 2020 Feb 17];3:107–115. Available from: http://aip.scitation.org/doi/10.1063/1.1749604.
[53] Vineyard GH. Frequency factors and isotope effects in solid state rate processes. J Phys Chem Solids. 1957;3:121–127.
[54] Kohn W, Sham LJ. Self-consisten equations including exchange and correlation effects. Phys Rev. 1965;140:A1133–A1138.
[55] Henkelman G, Uberuaga BP, Jonsson H. A climbing image nudged elastic band method for finding saddle points and minimum energy paths. J Chem Phys. 2000;113:9901–9904.
[56] Le Claire AD. Solute diffusion in dilute alloys. J Nucl Mater. 1978;69–70:70.
[57] Mantina M, Wang Y, Arroyave R, et al. First-principles calculation of self-diffusion coefficients. Phys Rev Lett. 2008;100:215901.
[58] Mantina M, Shang SL, Wang Y, et al. 3d transition metal impurities in aluminum: A first-principles study. Phys Rev B. 2009;80:184111.
[59] Mantina M, Wang Y, Chen LQ, et al. First principles impurity diffusion coefficients. Acta Mater. 2009;57:4102–4108.
[60] Hargather CZ, Shang SL, Liu ZK, et al. A first-principles study of self-diffusion coefficients of fcc Ni. Comput Mater Sci. 2014;86:17–23.
[61] Hargather CZ, Shang SL, Liu ZK. A comprehensive first-principles study of solute elements in dilute Ni alloys: Diffusion coefficients and their implications to tailor creep rate. Acta Mater [Internet]. 2018;157:126–141. Available from: https://doi.org/10.1016/j.actamat.2018.07.020.
[62] Mantina M, Chen LQ, Liu ZK. Predicting Diffusion Coefficients from First-principles via Eyring's Reaction Rate Theory. Defect Diffus Forum [Internet]. 2009;294:1–13. Available

from: http://dx.doi.org/10.4028/www.scientific.net/DDF.294.1.
[63] Ganeshan S, Shang SL, Zhang H, et al. Elastic constants of binary Mg compounds from first-principles calculations. Intermetallics. 2009;17:313–318.
[64] Zhou B-C, Shang S-L, Wang Y, et al. Diffusion coefficients of alloying elements in dilute Mg alloys: A comprehensive first-principles study. Acta Mater [Internet]. 2016 [cited 2015 Nov 24];103:573–586. Available from: http://www.sciencedirect.com/science/article/pii/S1359645415300112.
[65] Shang SL, Hector LG, Wang Y, et al. Anomalous energy pathway of vacancy migration and self-diffusion in hcp Ti. Phys Rev B. 2011;83:224104.
[66] Wimmer E, Wolf W, Sticht J, et al. Temperature-dependent diffusion coefficients from ab initio computations: Hydrogen, deuterium, and tritium in nickel. Phys Rev B [Internet]. 2008 [cited 2016 Feb 28];77:134305. Available from: http://journals.aps.org.ezaccess.libraries.psu.edu/prb/abstract/10.1103/PhysRevB.77.134305.
[67] Allnatt AR, Belova IV, Murch GE. Diffusion kinetics in dilute binary alloys with the h.c.p. crystal structure. Philos Mag [Internet]. 2014 [cited 2016 Oct 26];94:2487–2504. Available from: http://www.tandfonline.com/doi/abs/10.1080/14786435.2014.916426.
[68] Wang Y, Hu Y-J, Bocklund B, et al. First-principles thermodynamic theory of Seebeck coefficients. Phys Rev B [Internet]. 2018 [cited 2018 May 7];98:224101. Available from: https://link.aps.org/doi/10.1103/PhysRevB.98.224101.
[69] Wang Y, Chong X, Hu YJ, et al. An alternative approach to predict Seebeck coefficients: Application to La 3−x Te 4. Scr Mater [Internet]. 2019;169:87–91. Available from: https://doi.org/10.1016/j.scriptamat.2019.05.014.
[70] Mermin ND. Thermal Properties of the Inhomogeneous Electron Gas. Phys Rev [Internet]. 1965;137:A1441–A1443. Available from: https://link.aps.org/doi/10.1103/PhysRev.137.A1441.
[71] McMahan AK, Ross M. High-temperature electron-band calculations. Phys Rev B. 1977;15:718.
[72] Wang Y, Chen DQ, Zhang XW. Calculated equation of state of Al, Cu, Ta, Mo, and W to 1000 GPa. Phys Rev Lett. 2000;84:3220–3223.
[73] Wang Y, Wang JJ, Zhang H, et al. A first-principles approach to finite temperature elastic constants. J Physics-Condensed Matter. 2010;22:225404.
[74] Kresse G, Furthmüller J. Efficiency of ab-initio total energy calculations for metals and semiconductors using a plane-wave basis set. Comput Mater Sci. 1996;6:15–50.
[75] Kresse G, Joubert D, Kresse D, et al. From ultrasoft pseudopotentials to the projector augmented-wave method. Phys Rev B Condens Matter. 1999;59:1758–1775.
[76] Hicks LD, Dresselhaus MS. Effect of quantum-well structures on the thermoelectric figure of merit. Phys Rev B. 1993;47:12727.
[77] Singh DJ. Doping-dependent thermopower of PbTe from Boltzmann transport calculations. Phys Rev B. 2010;81:195217.
[78] Engel E, Vosko SH. Exact exchange-only potentials and the virial relation as microscopic criteria for generalized gradient approximations. Phys Rev B. 1993;47:13164.
[79] Blaha P, Schwarz K, Tran F, et al. WIEN2k: An APW+lo program for calculating the properties of solids. J Chem Phys [Internet]. 2020;152:074101. Available from: http://aip.scitation.org/doi/10.1063/1.5143061.

[80] Perdew JP, Ruzsinszky A, Csonka GI, et al. Restoring the Density-Gradient Expansion for Exchange in Solids and Surfaces. Phys Rev Lett [Internet]. 2008 [cited 2022 Jan 14];100:136406. Available from: https://link.aps.org/doi/10.1103/PhysRevLett.100.136406.
[81] Wang Y, Wang JJ, Wang WY, et al. A mixed-space approach to first-principles calculations of phonon frequencies for polar materials. J Phys Condens Matter [Internet]. 2010;22:202201. Available from: https://iopscience.iop.org/article/10.1088/0953-8984/22/20/202201.
[82] Wang Y, Shang S, Liu Z-K, et al. Mixed-space approach for calculation of vibration-induced dipole-dipole interactions. Phys Rev B [Internet]. 2012;85:224303. Available from: https://link.aps.org/doi/10.1103/PhysRevB.85.224303.
[83] Wang Y, Shang S-L, Fang H, et al. First-principles calculations of lattice dynamics and thermal properties of polar solids. npj Comput Mater [Internet]. 2016 [cited 2017 Jan 16];2:16006. Available from: http://www.nature.com/articles/npjcompumats20166.
[84] Zhao L-D, Tan G, Hao S, et al. Ultrahigh power factor and thermoelectric performance in hole-doped single-crystal SnSe. Science [Internet]. 2016;351:141–144. Available from: https://www.science.org/doi/10.1126/science.aad3749.
[85] Madsen GKH, Singh DJ. BoltzTraP. A code for calculating band-structure dependent quantities. Comput Phys Commun [Internet]. 2006;175:67–71. Available from: https://linkinghub.elsevier.com/retrieve/pii/S0010465506001305.
[86] Goldsmid HJ. Introduction to Thermoelectricity [Internet]. Berlin, Heidelberg: Springer Berlin Heidelberg; [cited 2021 Dec 12]. Available from: http://link.springer.com/10.1007/978-3-662-49256-7.
[87] Höglund L, Ågren J. Simulation of Carbon Diffusion in Steel Driven by a Temperature Gradient. J Phase Equilibria Diffus [Internet]. 2010 [cited 2017 Sep 11];31:212–215. Available from: http://link.springer.com/10.1007/s11669-010-9673-0.
[88] Liu Z-K, Wang Y, Shang S-L. Origin of negative thermal expansion phenomenon in solids. Scr Mater [Internet]. 2011 [cited 2016 Jan 24];65:664–667. Available from: http://www.sciencedirect.com/science/article/pii/S1359646211003903.
[89] Iacopini S, Piazza R. Thermophoresis in protein solutions. Eur Lett. 2003;63:247–253.
[90] Kita R, Polyakov P, Wiegand S. Ludwig−Soret Effect of Poly( N -isopropylacrylamide): Temperature Dependence Study in Monohydric Alcohols. Macromolecules [Internet]. 2007 [cited 2021 Jul 17];40:1638–1642. Available from: https://pubs-acs-org.ezaccess.libraries.psu.edu/doi/full/10.1021/ma0621831.
[91] Kishikawa Y, Wiegand S, Kita R. Temperature Dependence of Soret Coefficient in Aqueous and Nonaqueous Solutions of Pullulan. Biomacromolecules [Internet]. 2010 [cited 2021 Dec 15];11:740–747. Available from: https://pubs.acs.org/sharingguidelines.
[92] Iacopini S, Rusconi R, Piazza R. The “macromolecular tourist": Universal temperature dependence of thermal diffusion in aqueous colloidal suspensions. Eur Phys J E [Internet]. 2006 [cited 2021 Dec 15];19:59–67. Available from: http://link.springer.com/10.1140/epje/e2006-00012-9.
[93] de Gans B-J, Kita R, Wiegand S, et al. Unusual Thermal Diffusion in Polymer Solutions. Phys Rev Lett [Internet]. 2003 [cited 2021 Jul 17];91:245501. Available from: https://link.aps.org/doi/10.1103/PhysRevLett.91.245501.
[94] Costesèque P, Loubet J-C. Measuring the Soret coefficient of binary hydrocarbon

mixtures in packed thermogravitational columns (contribution of Toulouse University to the benchmark test). Philos Mag [Internet]. 2003 [cited 2021 Dec 15];83:2017–2022. Available from: https://www.tandfonline.com/doi/full/10.1080/0141861031000108187.
[95] Hartmann S, Wittko G, Köhler W, et al. Thermophobicity of Liquids: Heats of Transport in Mixtures as Pure Component Properties. Phys Rev Lett [Internet]. 2012 [cited 2021 Jul 27];109:065901. Available from: https://link.aps.org/doi/10.1103/PhysRevLett.109.065901.
[96] Schraml M, Bataller H, Bauer C, et al. The Soret coefficients of the ternary system water/ethanol/triethylene glycol and its corresponding binary mixtures. Eur Phys J E [Internet]. 2021 [cited 2021 Dec 15];44:128. Available from: https://link.springer.com/10.1140/epje/s10189-021-00134-6.
[97] Rahman MA, Saghir MZ. Thermodiffusion or Soret effect: Historical review. Int J Heat Mass Transf [Internet]. 2014 [cited 2020 Oct 24];73:693–705. Available from: https://linkinghub.elsevier.com/retrieve/pii/S0017931014001859.
[98] Evteev A V, Levchenko E V, Belova I V, et al. Thermotransport in binary system: case study on Ni50Al50 melt. Philos Mag. 2014;94:3574–3602.
[99] Ahmed T, Wang WY, Kozubski R, et al. Interdiffusion and thermotransport in Ni–Al liquid alloys. Philos Mag [Internet]. 2018 [cited 2018 Jul 19];98:2221–2246. Available from: https://www.tandfonline.com/doi/full/10.1080/14786435.2018.1479077.
[100] Sarder U, Ahmed T, Wang WY, et al. Mass and thermal transport in liquid Cu-Ag alloys. Philos Mag [Internet]. 2019 [cited 2019 Mar 17];99:468–491. Available from: https://www.tandfonline.com/doi/full/10.1080/14786435.2018.1546958.
[101] Tang J, Xue X, Yi Wang W, et al. Activation volume dominated diffusivity of Ni50Al50 melt under extreme conditions. Comput Mater Sci [Internet]. 2020 [cited 2019 Sep 13];171:109263. Available from: https://www.sciencedirect.com/science/article/pii/S0927025619305622?dgcid=coauthor.
[102] Momenzadeh L, Belova I V., Murch GE. Simulation of the ionic conductivity, thermal conductivity and thermotransport of doped zirconia using molecular dynamics. Comput Condens Matter [Internet]. 2021 [cited 2022 Jan 12];28:e00583. Available from: https://linkinghub.elsevier.com/retrieve/pii/S2352214321000484.
[103] Duhr S, Braun D. Why molecules move along a temperature gradient. Proc Natl Acad Sci U S A [Internet]. 2006 [cited 2021 Apr 4];103:19678–19682. Available from: www.pnas.orgcgidoi10.1073pnas.0603873103.
[104] Kocherginsky N, Gruebele M. A thermodynamic derivation of the reciprocal relations. J Chem Phys [Internet]. 2013 [cited 2020 Jul 4];138:124502. Available from: http://aip.scitation.org/doi/10.1063/1.4793258.
[105] Kocherginsky N, Gruebele M. Mechanical approach to chemical transport. Proc Natl Acad Sci [Internet]. 2016 [cited 2021 Feb 14];113:11116–11121. Available from: https://www.pnas.org/content/113/40/11116.
[106] Kocherginsky N, Gruebele M. Thermodiffusion: The physico-chemical mechanics view. J Chem Phys [Internet]. 2021 [cited 2021 Apr 4];154:024112. Available from: http://aip.scitation.org/doi/10.1063/5.0028674.
[107] Kocherginsky N, Gruebele M. Response to “Comment on ‘Thermodiffusion: The physico-chemical mechanics view’” [J. Chem. Phys. 155, 087101 (2021)]. J Chem Phys [Internet]. 2021 [cited 2021 Sep 16];155:087102. Available from: https://aip-scitation-

org.ezaccess.libraries.psu.edu/doi/abs/10.1063/5.0060107.
[108] Köhler W, Morozov KI. The Soret Effect in Liquid Mixtures – A Review. J Non-Equilibrium Thermodyn [Internet]. 2016;41:151–197. Available from: https://www.degruyter.com/view/j/jnet.2016.41.issue-3/jnet-2016-0024/jnet-2016-0024.xml.
[109] Piazza R, Parola A. Thermophoresis in colloidal suspensions. J Phys Condens Matter [Internet]. 2008 [cited 2021 Dec 15];20:153102. Available from: https://iopscience.iop.org/article/10.1088/0953-8984/20/15/153102.
[110] Würger A. Thermal non-equilibrium transport in colloids. Reports Prog Phys [Internet]. 2010 [cited 2021 Jul 17];73:126601. Available from: https://iopscience-iop-org.ezaccess.libraries.psu.edu/article/10.1088/0034-4885/73/12/126601.
[111] Okafor ICI, Carlson ON, Martin DM. Mass transport of carbon in one and two phase iron-nickel alloys in a temperature gradient. Metall Trans A [Internet]. 1982;13:1713–1719. Available from: http://link.springer.com/10.1007/BF02647826.
[112] Andersson JO, Helander T, Höglund L, et al. THERMO-CALC & DICTRA, computational tools for materials science. CALPHAD. 2002;26:273–312.
[113] Thermo-Calc Databases [Internet]. Available from: https://www.thermocalc.com/products-services/databases/.
[114] Kaufman L, Bernstein H. Computer Calculation of Phase Diagrams [Internet]. New York: Academic Press Inc.; 1970. Available from: http://www.calphad.org/.
[115] Saunders N, Miodownik AP. CALPHAD (Calculation of Phase Diagrams): A Comprehensive Guide. Oxford; New York: Pergamon; 1998.
[116] Lukas HL, Fries SG, Sundman B. Computational Thermodynammics: The Calphad method. Cambridge: Cambridge University Press; 2007.
[117] Spencer PJ. A brief history of CALPHAD. CALPHAD. 2008;32:1–8.
[118] Liu ZK. First-Principles calculations and CALPHAD modeling of thermodynamics. J Phase Equilibria Diffus [Internet]. 2009;30:517–534. Available from: http://dx.doi.org/10.1007/s11669-009-9570-6.
[119] Kirkaldy JS, Young DJ. Diffusion in the condensed state. Institute of Metals; 1987.
[120] Liu Z-K, Höglund L, Jönsson B, et al. An experimental and theoretical study of cementite dissolution in an Fe-Cr-C alloy. Metall Trans A [Internet]. 1991;22:1745–1752. Available from: http://link.springer.com/10.1007/BF02646498.
[121] Helander T, Agren J. Computer Simulation of Multicomponent Diffusion in Joints of Dissimilar Steels. Metall Mater Trans A. 1997;28A:303–308.
[122] Darken LS. Diffusion in metal accompanied by phase change. Trans Am Inst Min Metall Eng. 1942;150:157–169.
[123] Darken LS. Diffusion, mobility and their interrelation through free energy in binary metallic systems. Trans Am Inst Min Metall Eng. 1948;175:184–201.
[124] Darken LS. Diffusion of carbon in austenite with a discontinuity in composition. Trans Am Inst Min Metall Eng. 1949;180:430–438.
[125] Lundin CD. Dissimilar Metal Welds—Transition Joints Literature Review. Weld J. 1982;61:S58–S63.
[126] Maurya AK, Pandey C, Chhibber R. Dissimilar welding of duplex stainless steel with Ni alloys: A review. Int J Press Vessel Pip [Internet]. 2021 [cited 2021 Dec 22];192:104439. Available from: https://linkinghub.elsevier.com/retrieve/pii/S0308016121001356.

[127] Smoluchowski R. Diffusion Rate of Carbon in Iron-Cobalt Alloys. Phys Rev [Internet]. 1942 [cited 2021 Dec 22];62:539–544. Available from: https://link.aps.org/doi/10.1103/PhysRev.62.539.
[128] Smoluchowski R, Darken LS, Paranjpe VG, et al. Diffusion of carbon in austenite with a discontinuity in composition - discussion. Trans Am Inst Min Metall Eng. 1949;185:304–304.
[129] Hartley GS. Diffusion and distribution in a solvent of graded composition. Trans Faraday Soc [Internet]. 1931 [cited 2021 Dec 22];27:10–29. Available from: https://pubs-rsc-org.ezaccess.libraries.psu.edu/en/content/articlehtml/1931/tf/tf9312700010.
[130] Krishna R. Uphill diffusion in multicomponent mixtures. Chem Soc Rev [Internet]. 2015 [cited 2021 Dec 22];44:2812–2836. Available from: http://xlink.rsc.org/?DOI=C4CS00440J.
[131] Deshmukh A, Boo C, Karanikola V, et al. Membrane distillation at the water-energy nexus: limits, opportunities, and challenges. Energy Environ Sci [Internet]. 2018 [cited 2021 Dec 22];11:1177–1196. Available from: http://xlink.rsc.org/?DOI=C8EE00291F.
[132] Hillert M. A solid-solution model for inhomogeneous systems. Acta Metall [Internet]. 1961 [cited 2016 Sep 7];9:525–535. Available from: http://linkinghub.elsevier.com/retrieve/pii/0001616061901559.
[133] Cahn JW. On spinodal decomposition in cubic crystals. Acta Metall [Internet]. 1962;10:179–183. Available from: https://linkinghub.elsevier.com/retrieve/pii/0001616062901141.
[134] Kirkendall E, Thomassen L, Upthegrove C. Rates of diffusion of copper and zinc in alpha brass. Trans Am Inst Min Metall Eng [Internet]. 1939;133:186–203. Available from: file://catalog.hathitrust.org/Record/001479728.
[135] Kirkendall EO. Diffusion of zinc in alpha brass. Trans Am Inst Min Metall Eng. 1942;147:104–109.
[136] Smigelskas AD, Kirkendall EO. Zinc diffusion in alpha-brass. Trans Am Inst Min Metall Eng. 1947;171:130–142.
[137] Huntington HB, Seitz F. Mechanism for Self-Diffusion in Metallic Copper. Phys Rev [Internet]. 1942 [cited 2021 Dec 22];61:315–325. Available from: https://journals-aps-org.ezaccess.libraries.psu.edu/pr/abstract/10.1103/PhysRev.61.315.
[138] Nakajima H. The Discovery and Acceptance of the Kirkendall Effect. JOM [Internet]. 1997 [cited 2021 Jul 11];49(6):15–19. Available from: https://www.tms.org/pubs/journals/JOM/9706/Nakajima-9706.html.
[139] Tu KN. Recent advances on electromigration in very-large-scale-integration of interconnects. J Appl Phys [Internet]. 2003 [cited 2021 Sep 16];94:5451. Available from: https://aip-scitation-org.ezaccess.libraries.psu.edu/doi/abs/10.1063/1.1611263.
[140] Chen C, Tong HM, Tu KN. Electromigration and Thermomigration in Pb-Free Flip-Chip Solder Joints. Annu Rev Mater Res [Internet]. 2010 [cited 2021 Sep 16];40:531–555. Available from: https://www.annualreviews.org/doi/abs/10.1146/annurev.matsci.38.060407.130253.
[141] Tu KN, Liu Y, Li M. Effect of Joule heating and current crowding on electromigration in mobile technology. Appl Phys Rev [Internet]. 2017 [cited 2021 Dec 24];4:011101. Available from: http://aip.scitation.org/doi/10.1063/1.4974168.
[142] Black JR. Electromigration—A brief survey and some recent results. IEEE Trans Electron

Devices [Internet]. 1969 [cited 2021 Dec 23];16:338–347. Available from: http://ieeexplore.ieee.org/document/1475796/.
[143] Ames I, d'Heurle FM, Horstmann RE. Reduction of Electromigration in Aluminum Films by Copper Doping. IBM J Res Dev [Internet]. 1970 [cited 2021 Dec 24];14:461–463. Available from: http://ieeexplore.ieee.org/document/5391634/.
[144] Blech IA. Electromigration in thin aluminum films on titanium nitride. J Appl Phys [Internet]. 1976 [cited 2021 Dec 23];47:1203–1208. Available from: http://aip.scitation.org/doi/10.1063/1.322842.
[145] Hu C-K, Rodbell KP, Sullivan TD, et al. Electromigration and stress-induced voiding in fine Al and Al-alloy thin-film lines. IBM J Res Dev [Internet]. 1995 [cited 2021 Dec 24];39:465–497. Available from: http://ieeexplore.ieee.org/document/5389486/.
[146] Lee KL, Hu CK, Tu KN. In situ scanning electron microscope comparison studies on electromigration of Cu and Cu(Sn) alloys for advanced chip interconnects. J Appl Phys [Internet]. 1995 [cited 2021 Dec 24];78:4428–4437. Available from: http://aip.scitation.org/doi/10.1063/1.359851.
[147] Hu C-K, Gignac L, Rosenberg R. Electromigration of Cu/low dielectric constant interconnects. Microelectron Reliab [Internet]. 2006 [cited 2021 Dec 24];46:213–231. Available from: https://linkinghub.elsevier.com/retrieve/pii/S0026271405001113.
[148] Barmak K, Cabral C, Rodbell KP, et al. On the use of alloying elements for Cu interconnect applications. J Vac Sci Technol B Microelectron Nanom Struct [Internet]. 2006 [cited 2021 Dec 24];24:2485. Available from: http://scitation.aip.org/content/avs/journal/jvstb/24/6/10.1116/1.2357744.
[149] Zeng K, Tu KN. Six cases of reliability study of Pb-free solder joints in electronic packaging technology. Mater Sci Eng R Reports [Internet]. 2002 [cited 2021 Dec 23];38:55–105. Available from: https://linkinghub.elsevier.com/retrieve/pii/S0927796X02000074.
[150] Kim D, Chang J, Park J, et al. Formation and behavior of Kirkendall voids within intermetallic layers of solder joints. J Mater Sci Mater Electron [Internet]. 2011 [cited 2021 Dec 25];22:703–716. Available from: http://link.springer.com/10.1007/s10854-011-0357-2.
[151] Chen C, Hsiao H-Y, Chang Y-W, et al. Thermomigration in solder joints. Mater Sci Eng R Reports [Internet]. 2012 [cited 2021 Dec 25];73:85–100. Available from: https://linkinghub.elsevier.com/retrieve/pii/S0927796X12000551.
[152] Liu Y, Lin S. A Critical Review on the Electromigration Effect, the Electroplastic Effect, and Perspectives on the Effects of Electric Current Upon Alloy Phase Stability. JOM [Internet]. 2019 [cited 2021 Dec 25];71:3094–3106. Available from: http://link.springer.com/10.1007/s11837-019-03661-y.
[153] Kirchheim R. Stress and electromigration in Al-lines of integrated circuits. Acta Metall Mater [Internet]. 1992;40:309–323. Available from: https://linkinghub.elsevier.com/retrieve/pii/095671519290305X.
[154] Basaran C, Lin M, Ye H. A thermodynamic model for electrical current induced damage. Int J Solids Struct [Internet]. 2003 [cited 2021 Dec 25];40:7315–7327. Available from: https://linkinghub.elsevier.com/retrieve/pii/S0020768303004761.
[155] Wiseman GG, Kuebler JK. Electrocaloric Effect in Ferroelectric Rochelle Salt. Phys Rev [Internet]. 1963 [cited 2021 Dec 23];131:2023–2027. Available from:

https://link.aps.org/doi/10.1103/PhysRev.131.2023.
[156] Lombardo G, Pohl RO. Electrocaloric Effect and a New Type of Impurity Mode. Phys Rev Lett [Internet]. 1965 [cited 2021 Dec 23];15:291–293. Available from: https://link.aps.org/doi/10.1103/PhysRevLett.15.291.
[157] Lu S-G, Zhang Q. Electrocaloric Materials for Solid-State Refrigeration. Adv Mater [Internet]. 2009 [cited 2021 Dec 23];21:1983–1987. Available from: https://onlinelibrary.wiley.com/doi/10.1002/adma.200802902.
[158] Scott JF. Electrocaloric Materials. Annu Rev Mater Res [Internet]. 2011 [cited 2021 Dec 23];41:229–240. Available from: https://www.annualreviews.org/doi/10.1146/annurev-matsci-062910-100341.
[159] Moya X, Kar-Narayan S, Mathur ND. Caloric materials near ferroic phase transitions. Nat Mater [Internet]. 2014 [cited 2021 Dec 23];13:439–450. Available from: http://www.nature.com/articles/nmat3951.
[160] Moya X, Mathur ND. Caloric materials for cooling and heating. Science [Internet]. 2020 [cited 2021 Dec 23];370:797–803. Available from: https://www.science.org/doi/10.1126/science.abb0973.
[161] Qian X, Han D, Zheng L, et al. High-entropy polymer produces a giant electrocaloric effect at low fields. Nature [Internet]. 2021 [cited 2021 Dec 23];600:664–669. Available from: https://www.nature.com/articles/s41586-021-04189-5.
[162] Zhong W, Vanderbilt D, Rabe KM. First-Principles Theory of Ferroelectric Phase-Transitions for Perovskites - The Case of BaTiO3. Phys Rev B. 1995;52:6301–6312.
[163] Porokhonskyy V, Damjanovic D. Domain wall contributions in Pb(Zr,Ti)O3 ceramics at morphotropic phase boundary: A study of dielectric dispersion. Appl Phys Lett [Internet]. 2010 [cited 2021 Dec 26];96:242902. Available from: http://aip.scitation.org/doi/10.1063/1.3455328.
[164] Kumar A, Rabe KM, Waghmare U V. Domain formation and dielectric response in PbTiO3: A first-principles free-energy landscape analysis. Phys Rev B [Internet]. 2013 [cited 2021 Jun 26];87:024107. Available from: https://journals-aps-org.ezaccess.libraries.psu.edu/prb/abstract/10.1103/PhysRevB.87.024107.
[165] Fang HZ, Wang Y, Shang SL, et al. Nature of ferroelectric-paraelectric phase transition and origin of negative thermal expansion in PbTiO3. Phys Rev B. 2015;91:24104.
[166] Caspari ME, Merz WJ. The Electromechanical Behavior of BaTi <math display="inline"> <mrow> <msub> <mrow> <mi mathvariant="normal">O</mi> </mrow> <mrow> <mn>3</mn> </mrow> </msub> </mrow> </math> Single-Domain Crystals. Phys Rev [Internet]. 1950 [cited 2021 Dec 28];80:1082–1089. Available from: https://link.aps.org/doi/10.1103/PhysRev.80.1082.
[167] Kulcsar F. Electromechanical Properties of Lead Titanate Zirconate Ceramics Modified with Certain Three-or Five-Valent Additions. J Am Ceram Soc [Internet]. 1959 [cited 2021 Dec 28];42:343–349. Available from: https://onlinelibrary.wiley.com/doi/10.1111/j.1151-2916.1959.tb14321.x.
[168] Somlyo A V, Somlyo AP. Electromechanical and pharmacomechanical coupling in vascular smooth muscle. J Pharmacol Exp Ther [Internet]. 1968 [cited 2021 Dec 28];159:129–145. Available from: http://www.ncbi.nlm.nih.gov/pubmed/4296170.
[169] Zhao J, Zhang QM, Kim N, et al. ELECTROMECHANICAL PROPERTIES OF RELAXOR FERROELECTRIC LEAD MAGNESIUM NIOBATE-LEAD TITANATE

CERAMICS. Japanese J Appl Phys Part 1-Regular Pap Short Notes Rev Pap. 1995;34:5658–5663.
[170] Park S-E, Shrout TR. Ultrahigh strain and piezoelectric behavior in relaxor based ferroelectric single crystals. J Appl Phys [Internet]. 1997;82:1804–1811. Available from: http://aip.scitation.org/doi/10.1063/1.365983.
[171] Fu H, Cohen RE. Polarization rotation mechanism for ultrahigh electromechanical response in single-crystal piezoelectrics. Nature [Internet]. 2000;403:281–283. Available from: http://www.nature.com/articles/35002022.
[172] Kutnjak Z, Petzelt J, Blinc R. The giant electromechanical response in ferroelectric relaxors as a critical phenomenon. Nature. 2006;441:956–959.
[173] Ahart M, Somayazulu M, Cohen RE, et al. Origin of morphotropic phase boundaries in ferroelectrics. Nature [Internet]. 2008 [cited 2020 Nov 1];451:545–548. Available from: http://www.nature.com/articles/nature06459.
[174] Li F, Cabral MJ, Xu B, et al. Giant piezoelectricity of Sm-doped Pb(Mg 1/3 Nb 2/3 )O 3 -PbTiO 3 single crystals. Science [Internet]. 2019 [cited 2020 Feb 1];364:264–268. Available from: https://www.science.org/doi/10.1126/science.aaw2781.
[175] Nakagawa T, Miyatake H, Maeda T, et al. Evaluation on Electromigration and Stressmigration of Metal Interconnections by Hardness Measurements. MRS Proc [Internet]. 1991 [cited 2022 Jan 1];225:167. Available from: http://link.springer.com/10.1557/PROC-225-167.
[176] Dalleau D, Weide-Zaage K. Three-Dimensional Voids Simulation in chip Metallization Structures: a Contribution to Reliability Evaluation. Microelectron Reliab [Internet]. 2001 [cited 2022 Jan 1];41:1625–1630. Available from: https://linkinghub.elsevier.com/retrieve/pii/S0026271401001512.
[177] Lee C-C, Chuang O, Hsieh C-P, et al. Simulation and Experimental Validations of EM/TM/SM Physical Reliability for Interconnects Utilized in Stretchable and Foldable Electronics. 2019 IEEE 69th Electron Components Technol Conf [Internet]. IEEE; 2019 [cited 2022 Jan 1]. p. 2009–2015. Available from: https://ieeexplore.ieee.org/document/8811204/.
[178] Rodrigues IR, Lukina L, Dehaeck S, et al. Effect of Magnetic Susceptibility Gradient on the Magnetomigration of Rare-Earth Ions. J Phys Chem C [Internet]. 2019 [cited 2022 Jan 1];123:23131–23139. Available from: https://pubs.acs.org/doi/10.1021/acs.jpcc.9b06706.
[179] Owen M. Magnetochemische Untersuchungen. Die thermomagnetischen Eigenschaften der Elemente. II. Ann Phys [Internet]. 1912 [cited 2022 Jan 1];342:657–699. Available from: https://onlinelibrary.wiley.com/doi/10.1002/andp.19123420404.
[180] Sessoli R, Gatteschi D, Tsai HL, et al. High-Spin Molecules: [Mn12O12(O2CR)16(H2O)4]. J Am Chem Soc [Internet]. 1993 [cited 2022 Jan 1];115:1804–1816. Available from: https://pubs-acs-org.ezaccess.libraries.psu.edu/doi/abs/10.1021/ja00058a027.
[181] Schneider A, Fu C-C, Waseda O, et al. Ab initio based models for temperature-dependent magnetochemical interplay in bcc Fe-Mn alloys. Phys Rev B [Internet]. 2021 [cited 2022 Jan 1];103:024421. Available from: https://link.aps.org/doi/10.1103/PhysRevB.103.024421.
[182] Raj D, Kumar Padhi S. The sporadic μ-pyridine bridge in transition metal complexes: A real bond or an interaction? Coord Chem Rev [Internet]. 2022 [cited 2022 Jan

1];450:214238. Available from: https://linkinghub.elsevier.com/retrieve/pii/S0010854521005129.
[183] Gibbs JW. Graphical methods in the thermodynamics of fluids. Trans Connect Acad II. 1873;April-May:309–342.
[184] Gibbs JW. On the equilibrium of heterogeneous substances. Am J Sci [Internet]. 1878 [cited 2016 Feb 25];s3-16:441–458. Available from: http://www.ajsonline.org/content/s3-16/96/441.citation.
[185] Bauer GEW, Saitoh E, Van Wees BJ. Spin caloritronics. Nat Mater [Internet]. 2012 [cited 2021 Feb 18];11:391–399. Available from: www.nature.com/naturematerials.
[186] Oster G, Perelson A, Katchalsky A. Network Thermodynamics. Nature. 1971;234:393–399.
[187] Jarzynski C. Diverse phenomena, common themes. Nat Phys [Internet]. 2015 [cited 2020 Dec 14];11:105–107. Available from: http://www.nature.com/articles/nphys3229.
[188] Bekenstein JD. Black Holes and Entropy. Phys Rev D [Internet]. 1973 [cited 2022 Mar 11];7:2333–2346. Available from: https://link.aps.org/doi/10.1103/PhysRevD.7.2333.
[189] Hawking SW. Black holes and thermodynamics. Phys Rev D. 1976;13:191–197.
[190] Hawking SW, Page DN. Communications in Mathematical Physics Thermodynamics of Black Holes in Anti-de Sitter Space. Commun Math Phys. 1983;87:577–588.
[191] Rovelli C. The order of time. Penguin Books; 2017.
[192] Quijano J, Lin H. Entropy in the Critical Zone: A Comprehensive Review. Entropy [Internet]. 2014 [cited 2018 Apr 30];16:3482–3536. Available from: http://www.mdpi.com/1099-4300/16/6/3482.
[193] Bhattacharya J, Nozaki M, Takayanagi T, et al. Thermodynamical property of entanglement entropy for excited states. Phys Rev Lett [Internet]. 2013 [cited 2020 Oct 16];110:091602. Available from: https://journals-aps-org.ezaccess.libraries.psu.edu/prl/abstract/10.1103/PhysRevLett.110.091602.
[194] Tempesta P. Beyond the Shannon–Khinchin formulation: The composability axiom and the universal-group entropy. Ann Phys (N Y) [Internet]. 2016;365:180–197. Available from: https://linkinghub.elsevier.com/retrieve/pii/S0003491615003176.
[195] Devonshire AF. Theory of barium titanate 1. Philos Mag. 1949;40:1040–1063.
[196] Chandra P, Littlewood PB. A Landau Primer for Ferroelectrics. In: Rabe KM, Ahn CH, Triscone J-M, editors. Phys Ferroelectr a Mod Perspect. Verlag Berlin Heidelberg: Springer; 2007. p. 69–116.
[197] Haun MJ, Furman E, Jang SJ, et al. Thermodynamic theory of PbTiO3. J Appl Phys. 1987;62:3331–3338.
[198] Haun MJ, Furman E, Jang SJ, et al. Thermodynamic theory of the lead zirconate-titanatesolid solution system, part I: Phenomenology. Ferroelectrics. 1989;99:13–25.
[199] Haun MJ, Furman E, Jang SJ, et al. Thermodynamic theory of the lead zirconate-titanate solid solution system, part v: Theoretical calculations. Ferroelectrics. 1989;99:63–86.
[200] Olson GB. Computational Design of Hierarchically Structured Materials. Science. 1997;277:1237–1242.
[201] Liu ZK. A Materials Research Paradigm Driven by Computation. JOM - J Miner Met Mater Soc. 2009;61 (10):18–20.
[202] Olson GB, Kuehmann CJ. Materials genomics: From CALPHAD to flight. Scr Mater [Internet]. 2014 [cited 2018 Jan 27];70:25–30. Available from:

https://linkinghub.elsevier.com/retrieve/pii/S1359646213004375.
[203] Wang B, Gu Y, Zhang S, et al. Flexoelectricity in solids: Progress, challenges, and perspectives. Prog Mater Sci [Internet]. 2019 [cited 2021 Dec 28];106:100570. Available from: https://linkinghub.elsevier.com/retrieve/pii/S0079642519300465.
[204] van de Walle A, Ceder G, G.Ceder. First-principles computation of the vibrational entropy of ordered and disordered Pd3V. Phys Rev B. 2000;61:5972–5978.
[205] van de Walle A. Multicomponent multisublattice alloys, nonconfigurational entropy and other additions to the Alloy Theoretic Automated Toolkit. CALPHAD [Internet]. 2009;33:266–278. Available from: https://linkinghub.elsevier.com/retrieve/pii/S0364591608001314.
[206] Errea I, Calandra M, Mauri F. Anharmonic free energies and phonon dispersions from the stochastic self-consistent harmonic approximation: Application to platinum and palladium hydrides. Phys Rev B [Internet]. 2014 [cited 2022 Jan 2];89:064302. Available from: https://link.aps.org/doi/10.1103/PhysRevB.89.064302.
[207] Fultz B. Vibrational thermodynamics of materials. Prog Mater Sci [Internet]. 2010;55:247–352. Available from: https://linkinghub.elsevier.com/retrieve/pii/S0079642509000577.
[208] Kapil V, Engel E, Rossi M, et al. Assessment of Approximate Methods for Anharmonic Free Energies. J Chem Theory Comput [Internet]. 2019;15:5845–5857. Available from: https://pubs.acs.org/doi/10.1021/acs.jctc.9b00596.
[209] Glensk A, Grabowski B, Hickel T, et al. Phonon Lifetimes throughout the Brillouin Zone at Elevated Temperatures from Experiment and Ab Initio. Phys Rev Lett [Internet]. 2019 [cited 2022 Jan 3];123:235501. Available from: https://journals-aps-org.ezaccess.libraries.psu.edu/prl/abstract/10.1103/PhysRevLett.123.235501.
[210] Berg E, Lederer S, Schattner Y, et al. Monte Carlo Studies of Quantum Critical Metals. Annu Rev Condens Matter Phys [Internet]. 2019 [cited 2022 Jan 3];10:63–84. Available from: https://www.annualreviews.org/doi/abs/10.1146/annurev-conmatphys-031218-013339.
[211] Thomaes G. Thermal Diffusion near the Critical Solution Point. J Chem Phys [Internet]. 1956;25:32–33. Available from: http://aip.scitation.org/doi/10.1063/1.1742842.
[212] Hertz JA. Quantum critical phenomena. Phys Rev B [Internet]. 1976 [cited 2022 Jan 28];14:1165–1184. Available from: https://link.aps.org/doi/10.1103/PhysRevB.14.1165.
[213] Gegenwart P, Custers J, Geibel C, et al. Magnetic-Field Induced Quantum Critical Point in YbRh2Si2. Phys Rev Lett [Internet]. 2002 [cited 2022 Feb 12];89:056402. Available from: https://link.aps.org/doi/10.1103/PhysRevLett.89.056402.
[214] Custers J, Gegenwart P, Wilhelm H, et al. The break-up of heavy electrons at a quantum critical point. Nature [Internet]. 2003 [cited 2022 Jan 28];424:524–527. Available from: http://www.nature.com/articles/nature01774.
[215] Sebastian SE, Harrison N, Batista CD, et al. Heavy holes as a precursor to superconductivity in antiferromagnetic CeIn3. Proc Natl Acad Sci [Internet]. 2009 [cited 2022 Feb 12];106:7741–7744. Available from: http://www.pnas.org/cgi/doi/10.1073/pnas.0811859106.
[216] Wölfle P, Schmalian J, Abrahams E. Strong coupling theory of heavy fermion criticality II. Reports Prog Phys [Internet]. 2017 [cited 2022 Feb 12];80:044501. Available from: https://iopscience.iop.org/article/10.1088/1361-6633/aa5751.

[217] Wilson KG. The renormalization group: Critical phenomena and the Kondo problem. Rev Mod Phys [Internet]. 1975 [cited 2017 Dec 7];47:773–840. Available from: https://link.aps.org/doi/10.1103/RevModPhys.47.773.
[218] Pelissetto A, Vicari E. Critical phenomena and renormalization-group theory. Phys REPORTS-REVIEW Sect Phys Lett. 2002;368:549–727.
[219] Dupuis N, Canet L, Eichhorn A, et al. The nonperturbative functional renormalization group and its applications. Phys Rep [Internet]. 2021;910:1–114. Available from: https://linkinghub.elsevier.com/retrieve/pii/S0370157321000156.
[220] Wilding NB. Simulation studies of fluid critical behaviour. J Phys Condens Matter [Internet]. 1997 [cited 2022 Jan 5];9:585–612. Available from: https://iopscience.iop.org/article/10.1088/0953-8984/9/3/002.
[221] Errington JR. Direct calculation of liquid–vapor phase equilibria from transition matrix Monte Carlo simulation. J Chem Phys [Internet]. 2003 [cited 2022 Jan 5];118:9915–9925. Available from: http://aip.scitation.org/doi/10.1063/1.1572463.
[222] Banuti DT, Raju M, Ihme M. Similarity law for Widom lines and coexistence lines. Phys Rev E [Internet]. 2017 [cited 2022 Jan 3];95:052120. Available from: http://link.aps.org/doi/10.1103/PhysRevE.95.052120.
[223] Crooks GE. Entropy production fluctuation theorem and the nonequilibrium work relation for free energy differences. Phys Rev E [Internet]. 1999 [cited 2022 Jan 7];60:2721–2726. Available from: https://link.aps.org/doi/10.1103/PhysRevE.60.2721.
[224] Jarzynski C. Nonequilibrium Equality for Free Energy Differences. Phys Rev Lett [Internet]. 1997 [cited 2020 Dec 14];78:2690–2693. Available from: https://link.aps.org/doi/10.1103/PhysRevLett.78.2690.
[225] Jarzynski C. Equalities and Inequalities: Irreversibility and the Second Law of Thermodynamics at the Nanoscale. Annu Rev Condens Matter Phys [Internet]. 2011 [cited 2020 Dec 14];2:329–351. Available from: http://www.annualreviews.org/doi/10.1146/annurev-conmatphys-062910-140506.
[226] Seifert U. Stochastic thermodynamics, fluctuation theorems and molecular machines. Reports Prog Phys [Internet]. 2012 [cited 2019 Feb 3];75:126001. Available from: http://stacks.iop.org/0034-4885/75/i=12/a=126001?key=crossref.a937f55c8fce3473876363a3edb3c282.
[227] Wang GM, Sevick EM, Mittag E, et al. Experimental Demonstration of Violations of the Second Law of Thermodynamics for Small Systems and Short Time Scales. Phys Rev Lett [Internet]. 2002 [cited 2019 Feb 2];89:050601. Available from: https://link.aps.org/doi/10.1103/PhysRevLett.89.050601.
[228] Sagawa T, Ueda M. Fluctuation Theorem with Information Exchange: Role of Correlations in Stochastic Thermodynamics. Phys Rev Lett [Internet]. 2012 [cited 2019 Feb 3];109:180602. Available from: https://link.aps.org/doi/10.1103/PhysRevLett.109.180602.
[229] Maillet O, Erdman PA, Cavina V, et al. Optimal Probabilistic Work Extraction beyond the Free Energy Difference with a Single-Electron Device. Phys Rev Lett [Internet]. 2019;122:150604. Available from: https://doi.org/10.1103/PhysRevLett.122.150604.
[230] Seifert U. Entropy and the second law for driven, or quenched, thermally isolated systems. Phys A Stat Mech its Appl [Internet]. 2020 [cited 2021 Oct 7];552:121822. Available from: https://linkinghub.elsevier.com/retrieve/pii/S0378437119310696.

[231] Toyabe S, Sagawa T, Ueda M, et al. Experimental demonstration of information-to-energy conversion and validation of the generalized Jarzynski equality. Nat Phys [Internet]. 2010 [cited 2019 Feb 2];6:988–992. Available from: www.nature.com/naturephysics.
[232] Sagawa T, Ueda M. Generalized Jarzynski Equality under Nonequilibrium Feedback Control. Phys Rev Lett [Internet]. 2010 [cited 2022 Jan 7];104:090602. Available from: https://link.aps.org/doi/10.1103/PhysRevLett.104.090602.
[233] Resta R. Theory of the electric polarization in crystals. Ferroelectrics [Internet]. 1992 [cited 2020 Nov 5];136:51–55. Available from: https://www-tandfonline-com.ezaccess.libraries.psu.edu/doi/abs/10.1080/00150199208016065.
[234] King-Smith RD, Vanderbilt D. Theory of polarization of crystalline solids. Phys Rev B [Internet]. 1993 [cited 2020 Nov 5];47:1651–1654. Available from: https://journals-aps-org.ezaccess.libraries.psu.edu/prb/abstract/10.1103/PhysRevB.47.1651.
[235] Rabe KM, Ahn CH, Triscone J. Physics of Ferroelectrics: A Modern Perspective. Berlin, Heidelberg: Springer; 2007.
[236] Resta R, Vanderbilt D. Theory of Polarization: A Modern Approach. In: Rabe KM, Ahn CH, Triscone J-M, editors. Phys Ferroelectr a Mod Perspect. Verlag Berlin Heidelberg: Springer; 2007. p. 31–68.
[237] Zhong W, Vanderbilt D, Rabe KM. Phase-Transitions in BaTiO3 From First Principles. Phys Rev Lett. 1994;73:1861.
[238] Waghmare U V, Rabe KM. Ab initio Statistical Mechanics of Ferroelectric Phase Transition in PbTiO3. Phys Rev B. 1997;55:6161.
[239] Nishimatsu T, Waghmare U V., Kawazoe Y, et al. Fast molecular-dynamics simulation for ferroelectric thin-film capacitors using a first-principles effective Hamiltonian. Phys Rev B - Condens Matter Mater Phys [Internet]. 2008 [cited 2021 Jun 26];78:104104. Available from: https://journals-aps-org.ezaccess.libraries.psu.edu/prb/abstract/10.1103/PhysRevB.78.104104.
[240] Sicron N, Ravel B, Yacoby Y, et al. Nature of the ferroelectric phase-transition in PbTiO3. Phys Rev B. 1994;50:13168–13180.
[241] Sicron N, Ravel B, Yacoby Y, et al. The ferroelectric phase transition in PbTiO3 from a local perspective. Phys B Condens Matter [Internet]. 1995 [cited 2019 Apr 15];208–209:319–320. Available from: https://www.sciencedirect.com/science/article/pii/092145269400687Q.
[242] Ravel B, Sicron N, Yacoby Y, et al. Order-disorder behavior in the phase transition of PbTiO3. Ferroelectrics [Internet]. 1995 [cited 2019 Apr 15];164:265–277. Available from: http://www.tandfonline.com/doi/abs/10.1080/00150199508221849.
[243] Asta M, McCormack R, de Fontaine D. Theoretical study of alloy phase stability in the Cd-Mg system. Phys Rev B [Internet]. 1993;48:748–766. Available from: https://link.aps.org/doi/10.1103/PhysRevB.48.748.
[244] Krisch M, Farber DL, Xu R, et al. Phonons of the anomalous element cerium. Proc Natl Acad Sci [Internet]. 2011 [cited 2022 Jan 26];108:9342–9345. Available from: https://www.pnas.org/content/108/23/9342.
[245] Wang Y, Hector Jr LG, Zhang H, et al. A thermodynamic framework for a system with itinerant-electron magnetism. J Phys Condens Matter [Internet]. 2009;21:326003. Available from: https://iopscience.iop.org/article/10.1088/0953-8984/21/32/326003.
[246] Landau LD, Lifshitz EM. Statistical Physics. Second Rev. Oxford, New York: Pergamon

Press Ltd.; 1970.
[247] Shang S-L, Wang Y, Kim D, et al. First-principles thermodynamics from phonon and Debye model: Application to Ni and Ni3Al. Comput Mater Sci. 2010;47:1040–1048.
[248] Hillert M. The compound energy formalism. J Alloys Compd [Internet]. 2001;320:161–176. Available from: https://linkinghub.elsevier.com/retrieve/pii/S092583880001481X.
[249] Liu ZK, Wang Y, Shang S. Thermal Expansion Anomaly Regulated by Entropy. Sci Rep [Internet]. 2014;4:7043. Available from: http://www.nature.com/articles/srep07043.
[250] Wang Y, Shang SL, Zhang H, et al. Thermodynamic fluctuations in magnetic states: Fe3Pt as a prototype. Philos Mag Lett [Internet]. 2010;90:851–859. Available from: http://www.tandfonline.com/doi/abs/10.1080/09500839.2010.508446.
[251] Catalano S, Gibert M, Fowlie J, et al. Rare-earth nickelates RNiO3: thin films and heterostructures. Reports Prog Phys [Internet]. 2018 [cited 2019 Jul 31];81:046501. Available from: http://stacks.iop.org/0034-4885/81/i=4/a=046501?key=crossref.c95ab197e6b880dc2da6fd71f5758d81.
[252] Liu ZK, Shang SL, Wang Y, et al. Zentropy Theory for Ferroelectrics: A Case Study of PbTiO3. Prep. 2022;
[253] Park S, Shin S, Kim S-I, et al. Tunable quantum critical point and detached superconductivity in Al-doped CrAs. npj Quantum Mater [Internet]. 2019 [cited 2022 Feb 13];4:49. Available from: http://www.nature.com/articles/s41535-019-0188-6.
[254] Licciardello S, Buhot J, Lu J, et al. Electrical resistivity across a nematic quantum critical point. Nature [Internet]. 2019 [cited 2022 Feb 13];567:213–217. Available from: http://www.nature.com/articles/s41586-019-0923-y.
[255] Licciardello S, Maksimovic N, Ayres J, et al. Coexistence of orbital and quantum critical magnetoresistance in FeSe1-xSx. Phys Rev Res [Internet]. 2019 [cited 2022 Mar 10];1:023011. Available from: https://link.aps.org/doi/10.1103/PhysRevResearch.1.023011.
[256] Monthoux P, Pines D, Lonzarich GG. Superconductivity without phonons. Nature [Internet]. 2007 [cited 2021 Mar 26];450:1177–1183. Available from: http://www.nature.com/articles/nature06480.
[257] Wang B, Luo X, Ishigaki K, et al. Two distinct superconducting phases and pressure-induced crossover from type-II to type-I superconductivity in the spin-orbit-coupled superconductors BaBi3 and ScBi3. Phys Rev B [Internet]. 2018 [cited 2021 Oct 27];98:220506. Available from: https://link.aps.org/doi/10.1103/PhysRevB.98.220506.
[258] Bardeen J, Cooper LN, Schrieffer JR. Microscopic Theory of Superconductivity. Phys Rev [Internet]. 1957 [cited 2021 Mar 25];106:162–164. Available from: https://journals.aps.org/pr/abstract/10.1103/PhysRev.106.162.
[259] Bardeen J, Cooper LN, Schrieffer JR. Theory of Superconductivity. Phys Rev [Internet]. 1957;108:1175–1204. Available from: https://link.aps.org/doi/10.1103/PhysRev.108.1175.
[260] Bednorz JG, Muller KA. Possible High T c Superconductivity in the Ba-La-Cu-0 System. Zeitschrift Fur Phys B-Condensed Matterr Phys B Condens Matter [Internet]. 1986 [cited 2022 Jan 8];64:189–193. Available from: http://link.springer.com/10.1007/BF01303701.
[261] McMillan WL. Transition Temperature of Strong-Coupled Superconductors. Phys Rev [Internet]. 1968 [cited 2022 Jan 29];167:331–344. Available from: https://link.aps.org/doi/10.1103/PhysRev.167.331.
[262] Eliashberg GM. Interactions between electrons and lattice vibrations in a superconductor.

Sov Phys JETP-USSR Phys JETP. 1960;11:696–702.
[263] Allen PB, Dynes RC. Transition temperature of strong-coupled superconductors reanalyzed. Phys Rev B [Internet]. 1975 [cited 2022 Jan 29];12:905–922. Available from: https://link.aps.org/doi/10.1103/PhysRevB.12.905.
[264] Ashcroft NW. Metallic Hydrogen: A High-Temperature Superconductor? Phys Rev Lett [Internet]. 1968 [cited 2022 Jan 29];21:1748–1749. Available from: https://link.aps.org/doi/10.1103/PhysRevLett.21.1748.
[265] Ashcroft NW. Hydrogen Dominant Metallic Alloys: High Temperature Superconductors? Phys Rev Lett [Internet]. 2004 [cited 2022 Jan 29];92:187002. Available from: https://link.aps.org/doi/10.1103/PhysRevLett.92.187002.
[266] Giustino F. Electron-phonon interactions from first principles. Rev Mod Phys [Internet]. 2017 [cited 2022 Jan 29];89:015003. Available from: https://link.aps.org/doi/10.1103/RevModPhys.89.015003.
[267] Karsai F, Engel M, Flage-Larsen E, et al. Electron–phonon coupling in semiconductors within the GW approximation. New J Phys [Internet]. 2018 [cited 2022 Apr 1];20:123008. Available from: https://iopscience.iop.org/article/10.1088/1367-2630/aaf53f.
[268] Eremets MI, Trojan IA, Medvedev SA, et al. Superconductivity in Hydrogen Dominant Materials: Silane. Science [Internet]. 2008 [cited 2022 Jan 30];319:1506–1509. Available from: https://www.science.org/doi/10.1126/science.1153282.
[269] Wang H, Tse JS, Tanaka K, et al. Superconductive sodalite-like clathrate calcium hydride at high pressures. Proc Natl Acad Sci [Internet]. 2012 [cited 2022 Jan 29];109:6463–6466. Available from: https://www.pnas.org/content/109/17/6463.
[270] Wang Y, Ma Y. Perspective: Crystal structure prediction at high pressures. J Chem Phys [Internet]. 2014 [cited 2022 Jan 30];140:040901. Available from: http://aip.scitation.org/doi/10.1063/1.4861966.
[271] Pickard CJ, Errea I, Eremets MI. Superconducting Hydrides Under Pressure. Annu Rev Condens Matter Phys [Internet]. 2020;11:57–76. Available from: https://www.annualreviews.org/doi/10.1146/annurev-conmatphys-031218-013413.
[272] Li Y, Hao J, Liu H, et al. The metallization and superconductivity of dense hydrogen sulfide. J Chem Phys [Internet]. 2014 [cited 2021 Oct 26];140:174712. Available from: http://aip.scitation.org/doi/10.1063/1.4874158.
[273] Errea I, Calandra M, Pickard CJ, et al. High-Pressure Hydrogen Sulfide from First Principles: A Strongly Anharmonic Phonon-Mediated Superconductor. Phys Rev Lett [Internet]. 2015 [cited 2022 Jan 30];114:157004. Available from: https://link.aps.org/doi/10.1103/PhysRevLett.114.157004.
[274] Stewart GR. Unconventional superconductivity. Adv Phys [Internet]. 2017 [cited 2022 Feb 7];66:75–196. Available from: https://www.tandfonline.com/doi/full/10.1080/00018732.2017.1331615.
[275] Vittoria Mazziotti M, Jarlborg T, Bianconi A, et al. Room temperature superconductivity dome at a Fano resonance in superlattices of wires. Europhys Lett [Internet]. 2021 [cited 2022 Feb 7];134:17001. Available from: https://iopscience.iop.org/article/10.1209/0295-5075/134/17001.
[276] Hosono H, Kuroki K. Iron-based superconductors: Current status of materials and pairing mechanism. Phys C Supercond its Appl [Internet]. 2015 [cited 2022 Feb 5];514:399–422. Available from: https://linkinghub.elsevier.com/retrieve/pii/S0921453415000477.

[277] Snider E, Dasenbrock-Gammon N, McBride R, et al. Room-temperature superconductivity in a carbonaceous sulfur hydride. Nature [Internet]. 2020 [cited 2022 Feb 9];586:373–377. Available from: https://www.nature.com/articles/s41586-020-2801-z.
[278] Liu ZK, Schlom DG, Li Q, et al. Thermodynamics of the Mg–B system: Implications for the deposition of MgB2 thin films. Appl Phys Lett. 2001;78:3678.
[279] Zeng XH, Pogrebnyakov A V, Kotcharov A, et al. In situ epitaxial MgB2 thin films for superconducting electronics. Nat Mater. 2002;1:35–38.
[280] Wang Y, Shang SL, Hui XD, et al. Effects of spin structures on phonons in BaFe2As2. Appl Phys Lett [Internet]. 2010;97:022504. Available from: http://aip.scitation.org/doi/10.1063/1.3464166.
[281] Wang Y, Saal JE, Shang SL, et al. Effects of spin structures on Fermi surface topologies in BaFe2As2. Solid State Commun [Internet]. 2011;151:272–275. Available from: https://linkinghub.elsevier.com/retrieve/pii/S0038109810007337.
[282] Wang Y, Shang SL, Chen LQ, et al. Magnetic excitation and thermodynamics of BaFe2As2. Int J Quantum Chem [Internet]. 2011;111:3565–3570. Available from: https://onlinelibrary.wiley.com/doi/10.1002/qua.22865.
[283] Kang J-H, Xie L, Wang Y, et al. Control of Epitaxial BaFe2As2 Atomic Configurations with Substrate Surface Terminations. Nano Lett [Internet]. 2018 [cited 2018 Nov 12];18:6347–6352. Available from: http://pubs.acs.org/doi/10.1021/acs.nanolett.8b02704.
[284] Kamihara Y, Watanabe T, Hirano M, et al. Iron-Based Layered Superconductor La[O 1- x F x ]FeAs ( x = 0.05−0.12) with T c = 26 K. J Am Chem Soc [Internet]. 2008 [cited 2022 Feb 9];130:3296–3297. Available from: https://pubs.acs.org/doi/10.1021/ja800073m.
[285] Kang J-H, Kim J-W, Ryan PJ, et al. Superconductivity in undoped BaFe2As2 by tetrahedral geometry design. Proc Natl Acad Sci [Internet]. 2020 [cited 2022 Feb 9];117:21170–21174. Available from: http://www.pnas.org/lookup/doi/10.1073/pnas.2001123117.
[286] Shimojima T, Ishizaka K, Ishida Y, et al. Orbital-Dependent Modifications of Electronic Structure across the Magnetostructural Transition in BaFe2As2. Phys Rev Lett [Internet]. 2010 [cited 2022 Feb 9];104:057002. Available from: https://link.aps.org/doi/10.1103/PhysRevLett.104.057002.
[287] Krajewski AM, Siegel JW, Xu J, et al. Extensible Structure-Informed Prediction of Formation Energy with improved accuracy and usability employing neural networks. Comput Mater Sci [Internet]. 2022 [cited 2020 Aug 31];208:111254. Available from: https://linkinghub.elsevier.com/retrieve/pii/S0927025622000593.
[288] SIPFENN: Structure-Informed Prediction of Formation Energy using Neural Networks. https://phaseslab.com/sipfenn.
[289] MPDD: Materials Property Descriptor Database. https://phaseslab.com/mpdd/.
[290] Wang Y, Liao M, Bocklund BJ, et al. DFTTK: Density Functional Theory ToolKit for high-throughput lattice dynamics calculations. CALPHAD [Internet]. 2021 [cited 2021 Oct 7];75:102355. Available from: https://linkinghub.elsevier.com/retrieve/pii/S0364591621001024.
[291] DFTTK: Density Functional Theory Tool Kits. https://www.dfttk.org/.
[292] Otis R, Liu Z-K. pycalphad: CALPHAD-based Computational Thermodynamics in Python. J Open Res Softw [Internet]. 2017 [cited 2017 Jan 11];5:1–11. Available from:

https://pycalphad.org.
[293] PyCalphad: Python library for computational thermodynamics using the CALPHAD method. https://pycalphad.org.
[294] Bocklund B, Otis R, Egorov A, et al. ESPEI for efficient thermodynamic database development, modification, and uncertainty quantification: application to Cu–Mg. MRS Commun [Internet]. 2019 [cited 2019 Jul 9];9:618–627. Available from: https://www.cambridge.org/core/product/identifier/S2159685919000594/type/journal_article.
[295] ESPEI: Extensible Self-optimizing Phase Equilibria Infrastructure. https://espei.org.
[296] Im S, Shang S-L, Smith ND, et al. Thermodynamic properties of the Nd-Bi system via emf measurements, DFT calculations, machine learning, and CALPHAD modeling. Acta Mater [Internet]. 2022 [cited 2021 Nov 26];223:117448. Available from: https://linkinghub.elsevier.com/retrieve/pii/S1359645421008272.
[297] Tao SX, Cao X, Bobbert PA. Accurate and efficient band gap predictions of metal halide perovskites using the DFT-1/2 method: GW accuracy with DFT expense. Sci Rep [Internet]. 2017 [cited 2022 Feb 10];7:14386. Available from: http://www.nature.com/articles/s41598-017-14435-4.
[298] Wing D, Ohad G, Haber JB, et al. Band gaps of crystalline solids from Wannier-localization–based optimal tuning of a screened range-separated hybrid functional. Proc Natl Acad Sci [Internet]. 2021 [cited 2022 Feb 10];118:e2104556118. Available from: http://www.pnas.org/lookup/doi/10.1073/pnas.2104556118.
[299] Horio M, Hauser K, Sassa Y, et al. Three-Dimensional Fermi Surface of Overdoped La-Based Cuprates. Phys Rev Lett [Internet]. 2018 [cited 2022 Feb 10];121:077004. Available from: https://link.aps.org/doi/10.1103/PhysRevLett.121.077004.
[300] Wang Y, Wang WY, Chen L-Q, et al. Bonding Charge Density from Atomic Perturbations. J Comput Chem. 2015;36:1008–1014.
[301] Wang Y, Lee SH, Zhang LA, et al. Quantifying charge ordering by density functional theory: Fe3O4 and CaFeO3. Chem Phys Lett. 2014;607:81–84.
[302] Wang WY, Tang B, Shang S-L, et al. Local lattice distortion mediated formation of stacking faults in Mg alloys. Acta Mater [Internet]. 2019 [cited 2022 Feb 11];170:231–239. Available from: https://linkinghub.elsevier.com/retrieve/pii/S1359645419301739.
[303] Izquierdo M, Freitas DC, Colson D, et al. Charge Order and Suppression of Superconductivity in HgBa2CuO4+d at High Pressures. Condens Matter [Internet]. 2021 [cited 2022 Feb 7];6:25. Available from: https://www.mdpi.com/2410-3896/6/3/25.
[304] Kreisel A, Hirschfeld P, Andersen B. On the Remarkable Superconductivity of FeSe and Its Close Cousins. Symmetry (Basel) [Internet]. 2020 [cited 2022 Jan 16];12:1402. Available from: https://www.mdpi.com/2073-8994/12/9/1402.
[305] Kothapalli K, Böhmer AE, Jayasekara WT, et al. Strong cooperative coupling of pressure-induced magnetic order and nematicity in FeSe. Nat Commun [Internet]. 2016 [cited 2022 Feb 12];7:12728. Available from: https://www.nature.com/articles/ncomms12728.
[306] Coldea AI, Watson MD. The Key Ingredients of the Electronic Structure of FeSe. Annu Rev Condens Matter Phys [Internet]. 2018 [cited 2022 Jan 16];9:125–146. Available from: https://www.annualreviews.org/doi/abs/10.1146/annurev-conmatphys-033117-054137.
[307] Böhmer AE, Kreisel A. Nematicity, magnetism and superconductivity in FeSe. J Phys

Condens Matter [Internet]. 2018 [cited 2022 Jan 16];30:023001. Available from: https://iopscience.iop.org/article/10.1088/1361-648X/aa9caa.
[308] Gati E, Böhmer AE, Bud'ko SL, et al. Bulk Superconductivity and Role of Fluctuations in the Iron-Based Superconductor FeSe at High Pressures. Phys Rev Lett. 2019;123:167002.
[309] Hsu F-C, Luo J-Y, Yeh K-W, et al. Superconductivity in the PbO-type structure a-FeSe. Proc Natl Acad Sci [Internet]. 2008 [cited 2022 Jan 30];105:14262–14264. Available from: http://www.pnas.org/cgi/doi/10.1073/pnas.0807325105.
[310] Imai T, Ahilan K, Ning FL, et al. Why Does Undoped FeSe Become a High-Tc Superconductor under Pressure? Phys Rev Lett [Internet]. 2009 [cited 2022 Jan 16];102:177005. Available from: http://dx.doi.org/10.1103/PhysRevLett.102.177005.
[311] Song C-L, Wang Y-L, Cheng P, et al. Direct Observation of Nodes and Twofold Symmetry in FeSe Superconductor. Science [Internet]. 2011 [cited 2022 Jan 16];332:1410–1413. Available from: www.sciencemag.org/cgi/content/full/332/6036/1407/DC1.
[312] He S, He J, Zhang W, et al. Phase diagram and electronic indication of high-temperature superconductivity at 65 K in single-layer FeSe films. Nat Mater [Internet]. 2013 [cited 2022 Jan 16];12:605–610. Available from: http://www.nature.com/articles/nmat3648.
[313] Baek S-H, Efremov D V., Ok JM, et al. Orbital-driven nematicity in FeSe. Nat Mater [Internet]. 2015 [cited 2022 Jan 30];14:210–214. Available from: http://www.nature.com/articles/nmat4138.
[314] Sprau PO, Kostin A, Kreisel A, et al. Discovery of orbital-selective Cooper pairing in FeSe. Science [Internet]. 2017 [cited 2022 Jan 16];357:75–80. Available from: https://www.science.org/doi/10.1126/science.aal1575.
[315] Mukasa K, Matsuura K, Qiu M, et al. High-pressure phase diagrams of FeSe1−xTex: correlation between suppressed nematicity and enhanced superconductivity. Nat Commun [Internet]. 2021 [cited 2022 Feb 7];12:381. Available from: http://www.nature.com/articles/s41467-020-20621-2.
[316] Subedi A, Zhang L, Singh DJ, et al. Density functional study of FeS, FeSe, and FeTe: Electronic structure, magnetism, phonons, and superconductivity. Phys Rev B [Internet]. 2008 [cited 2022 Jan 16];78:134514. Available from: https://journals-aps-org.ezaccess.libraries.psu.edu/prb/abstract/10.1103/PhysRevB.78.134514.
[317] Pokharel S, Fu H. Understanding the importance of local magnetic moment in monolayer FeSe. Phys Rev B [Internet]. 2021 [cited 2022 Jan 16];104:195110. Available from: https://link.aps.org/doi/10.1103/PhysRevB.104.195110.
[318] Yamada T, Tohyama T. Multipolar nematic state of nonmagnetic FeSe based on DFT+U. Phys Rev B [Internet]. 2021 [cited 2022 Jan 16];104:L161110. Available from: https://journals-aps-org.ezaccess.libraries.psu.edu/prb/abstract/10.1103/PhysRevB.104.L161110.
[319] Bhadeshia HKDH. A Commentary on: "Diffusion of Carbon in Austenite with a Discontinuity in Composition." Metall Mater Trans A [Internet]. 2010 [cited 2021 Dec 22];41:1605–1615. Available from: https://link.springer.com/10.1007/s11661-010-0276-5.
[320] Bhadeshia HKDH. Diffusional formation of ferrite in iron and its alloys. Prog Mater Sci [Internet]. 1985 [cited 2021 Dec 22];29:321–386. Available from: https://linkinghub.elsevier.com/retrieve/pii/0079642585900040.
[321] Hillert M. Paraequilibrium and other Restricted Equilibria. Bennett LH, Massalski TB,

Giessen BC, editors. MRS Proc [Internet]. 1982;19:295. Available from: http://link.springer.com/10.1557/PROC-19-295.
[322] Zi-Kui L, Ågren J. On the transition from local equilibrium to paraequilibrium during the growth of ferrite in Fe-Mn-C austenite. Acta Metall [Internet]. 1989;37:3157–3163. Available from: https://linkinghub.elsevier.com/retrieve/pii/0001616089901879.
[323] Liu ZK, Agren J. Thermodynamics of constrained and unconstrained equilibrium systems and their phase rules. J Phase Equilibria. 1995;16:30–35.
[324] Bhadeshia HKDH. Some difficulties in the theory of diffusion-controlled growth in substitutionally alloyed steels. Curr Opin Solid State Mater Sci [Internet]. 2016 [cited 2021 Dec 22];20:396–400. Available from: https://linkinghub.elsevier.com/retrieve/pii/S1359028616301103.
[325] Hackl K, Fischer FD, Svoboda J. A study on the principle of maximum dissipation for coupled and non-coupled non-isothermal processes in materials. Proc R Soc A Math Phys Eng Sci [Internet]. 2011 [cited 2021 Feb 14];467:1186–1196. Available from: https://royalsocietypublishing.org/doi/10.1098/rspa.2010.0179.
[326] Ågren J. A simplified treatment of the transition from diffusion controlled to diffusion-less growth. Acta Metall [Internet]. 1989 [cited 2022 Jan 9];37:181–189. Available from: https://linkinghub.elsevier.com/retrieve/pii/0001616089902770.
[327] Hillert M, Sundman B. A treatment of the solute drag on moving grain boundaries and phase interfaces in binary alloys. Acta Metall [Internet]. 1976;24:731–743. Available from: https://linkinghub.elsevier.com/retrieve/pii/0001616076901085.
[328] Hillert M, Sundman B. A Solute-Drag Treatment of the Transition from Diffusion-Controlled to Diffusionless Solidification. Acta Metall. 1977;25:11–18.
[329] Suehiro M, Liu ZK, Agren J. Effect of niobium on massive transformation in ultra low carbon steels: A solute drag treatment. Acta Mater. 1996;44:4241–4251.
[330] Liu ZK, Agren J, Suehiro M. Thermodynamics of interfacial segregation in solute drag. Mater Sci Eng a-Structural Mater Prop Microstruct Process. 1998;247:222–228.
[331] Fan D, Chen SP, Chen L-Q. Computer simulation of grain growth kinetics with solute drag. J Mater Res [Internet]. 1999 [cited 2022 Jan 9];14:1113–1123. Available from: https://link.springer.com/article/10.1557/JMR.1999.0147.
[332] Hillert M. Solute drag in grain boundary migration and phase transformations. Acta Mater [Internet]. 2004 [cited 2022 Jan 9];52:5289–5293. Available from: https://linkinghub.elsevier.com/retrieve/pii/S1359645404004409.
[333] Wu Z, Bei H, Otto F, et al. Recovery, recrystallization, grain growth and phase stability of a family of FCC-structured multi-component equiatomic solid solution alloys. Intermetallics [Internet]. 2014 [cited 2022 Jan 9];46:131–140. Available from: https://linkinghub.elsevier.com/retrieve/pii/S0966979513002872.
[334] Liu Z-K, Ågren J. Morphology of cementite decomposition in an fe-cr-c alloy. Metall Trans A [Internet]. 1991;22:1753–1759. Available from: http://link.springer.com/10.1007/BF02646499.
[335] Christian JW. The Theory of Transformations in Metals and Alloys. Oxford: Pergamon Press; 1965.
[336] Agren J. On the classification of phase transformations. Scr Mater. 2002;46:893–898.
[337] Chookajorn T, Murdoch HA, Schuh CA. Design of stable nanocrystalline alloys. Science. 2012;337:951–954.

[338] Dervichian DG. Changes of Phase and Transformations of Higher Order in Monolayers. J Chem Phys [Internet]. 1939 [cited 2022 Jan 9];7:931–948. Available from: http://aip.scitation.org/doi/10.1063/1.1750347.
[339] Borrmann P, Mülken O, Harting J. Classification of Phase Transitions in Small Systems. Phys Rev Lett [Internet]. 2000 [cited 2022 Jan 9];84:3511–3514. Available from: https://link.aps.org/doi/10.1103/PhysRevLett.84.3511.
[340] Cejnar P, Jolie J, Casten RF. Quantum phase transitions in the shapes of atomic nuclei. Rev Mod Phys [Internet]. 2010 [cited 2022 Jan 9];82:2155–2212. Available from: https://link.aps.org/doi/10.1103/RevModPhys.82.2155.
[341] Deger A, Brange F, Flindt C. Lee-Yang theory, high cumulants, and large-deviation statistics of the magnetization in the Ising model. Phys Rev B [Internet]. 2020 [cited 2022 Jan 9];102:174418. Available from: https://link.aps.org/doi/10.1103/PhysRevB.102.174418.
[342] Hillert M. No Title. In: Zackay VF, Aaronson HI, editors. Decompos Austenite by Diffus Process. Interscience Publishers; 1962. p. 197.
[343] De Graef M, Kral M V., Hillert M. A modern 3-D view of an “old” pearlite colony. JOM [Internet]. 2006 [cited 2022 Jan 9];58:25–28. Available from: https://link.springer.com/article/10.1007/BF02748491.
[344] Moelans N, Blanpain B, Wollants P. An introduction to phase-field modeling of microstructure evolution. CALPHAD [Internet]. 2008 [cited 2022 Jan 9];32:268–294. Available from: https://linkinghub.elsevier.com/retrieve/pii/S0364591607000880.
[345] Liu Z-K. Ocean of Data: Integrating first-principles calculations and CALPHAD modeling with machine learning. J Phase Equilibria Diffus [Internet]. 2018;39:635–649. Available from: https://doi.org/10.1007/s11669-018-0654-z.

*Table 1: Physical quantities related to the first directives of molar quantities (first column) to potentials (first row), symmetric due to the Maxwell relations* [1,11,345]

| | $T$, Temperature | $\sigma$, Stress | $E$, Electrical field | $\mathcal{H}$, Magnetic field | $\mu_i$, Chemical potential |
|---|---|---|---|---|---|
| $S$, Entropy | Heat capacity | Piezocaloric effect | Electrocaloric effect | Magnetocaloric effect | $\frac{\partial S}{\partial \mu_k}$ |
| $\varepsilon$, Strain | Thermal expansion | Elastic compliance | Converse piezoelectricity | Piezomagnetic moduli | $\frac{\partial \varepsilon_{ij}}{\partial \mu_k}$ |
| $\theta$, Electrical displacement | Pyroelectric coefficients | Piezoelectric moduli | Permittivity | Magnetoelectric coefficient | $\frac{\partial D_i}{\partial \mu_k}$ |
| $B$, Magnetic induction | Pyromagnetic coefficient | Piezomagnetic moduli | Magnetoelectric coefficient | Permeability | $\frac{\partial B_i}{\partial \mu_k}$ |
| $N_j$, Moles | *Thermoreactivity* | *Stressoreactivity* | *Electroreactivity* | *Magnetoreactivity* | $\frac{\partial N_i}{\partial \mu_k}$, *Thermodynamic factor* |

*Table 2: Cross phenomenon coefficients represented by derivatives between potentials* [23].

| | $\boldsymbol{T}$, **Temperature** | $\boldsymbol{\sigma}$, **Stress** | $\boldsymbol{E}$, **Electrical field** | $\boldsymbol{\mathcal{H}}$, **Magnetic field** | $\boldsymbol{\mu_i}$, **Chemical potential** |
|---|---|---|---|---|---|
| $\boldsymbol{T}$ | 1 | $-\frac{\partial S}{\partial \varepsilon}$ | $-\frac{\partial S}{\partial \theta}$ | $-\frac{\partial S}{\partial B}$ | $-\frac{\partial S}{\partial c_i}$ Partial entropy |
| $\boldsymbol{\sigma}$ | $\frac{\partial \sigma}{\partial T}$ | 1 | $-\frac{\partial \varepsilon}{\partial \theta}$ | $-\frac{\partial \varepsilon}{\partial B}$ | $-\frac{\partial \varepsilon}{\partial c_i}$ Partial strain |
| $\boldsymbol{E}$ | $\frac{\partial E}{\partial T}$ | $\frac{\partial E}{\partial \sigma}$ | 1 | $-\frac{\partial \theta}{\partial B}$ | $-\frac{\partial \theta}{\partial c_i}$ Partial electrical displacement |
| $\boldsymbol{\mathcal{H}}$ | $\frac{\partial \mathcal{H}}{\partial T}$ | $\frac{\partial \mathcal{H}}{\partial \sigma}$ | $\frac{\partial \mathcal{H}}{\partial E}$ | 1 | $-\frac{\partial B}{\partial c_i}$ Partial magnetic induction |
| $\boldsymbol{\mu_i}$ | $\frac{\partial \mu_i}{\partial T}$ Thermodiffusion | $\frac{\partial \mu_i}{\partial \sigma}$ Stressmigration | $\frac{\partial \mu_i}{\partial E}$ Electromigration | $\frac{\partial \mu_i}{\partial \mathcal{H}}$ Magnetomigration | $\frac{\partial \mu_i}{\partial \mu_j} = -\frac{\partial c_j}{\partial c_i} = \frac{\Phi_{ii}}{\Phi_{ji}}$ Crossdiffusion |

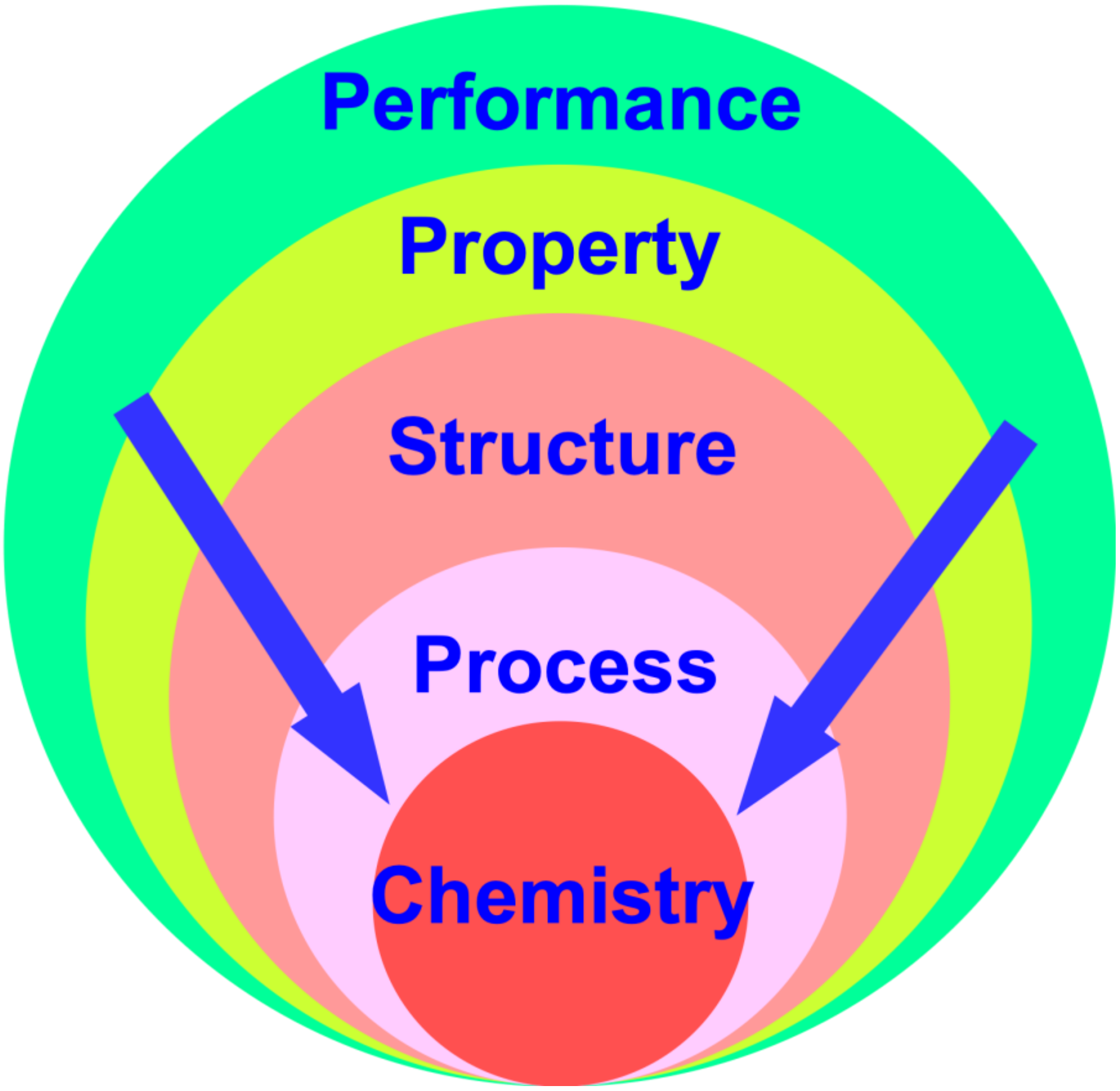

(i)

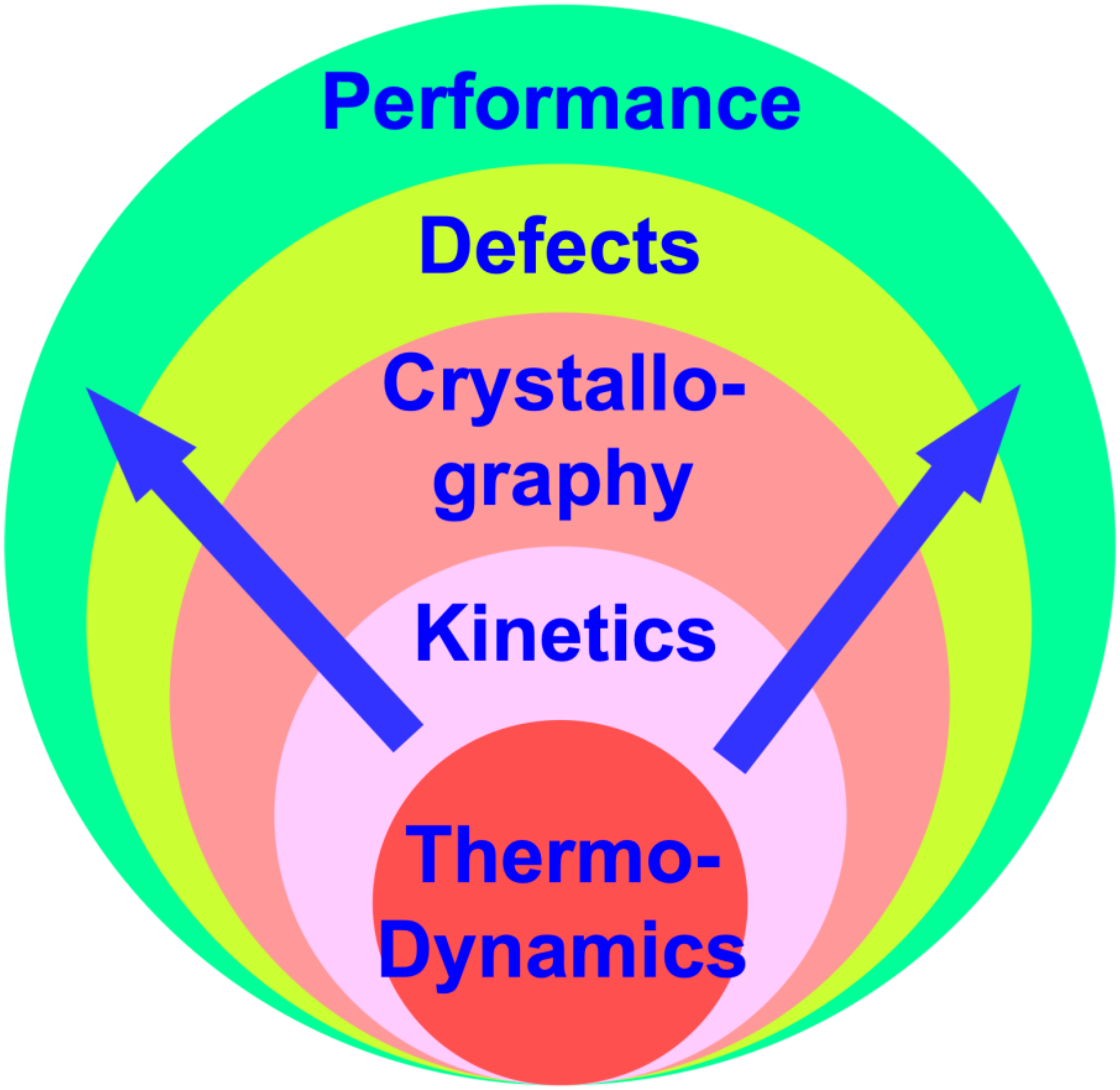

(ii)

Figure 1

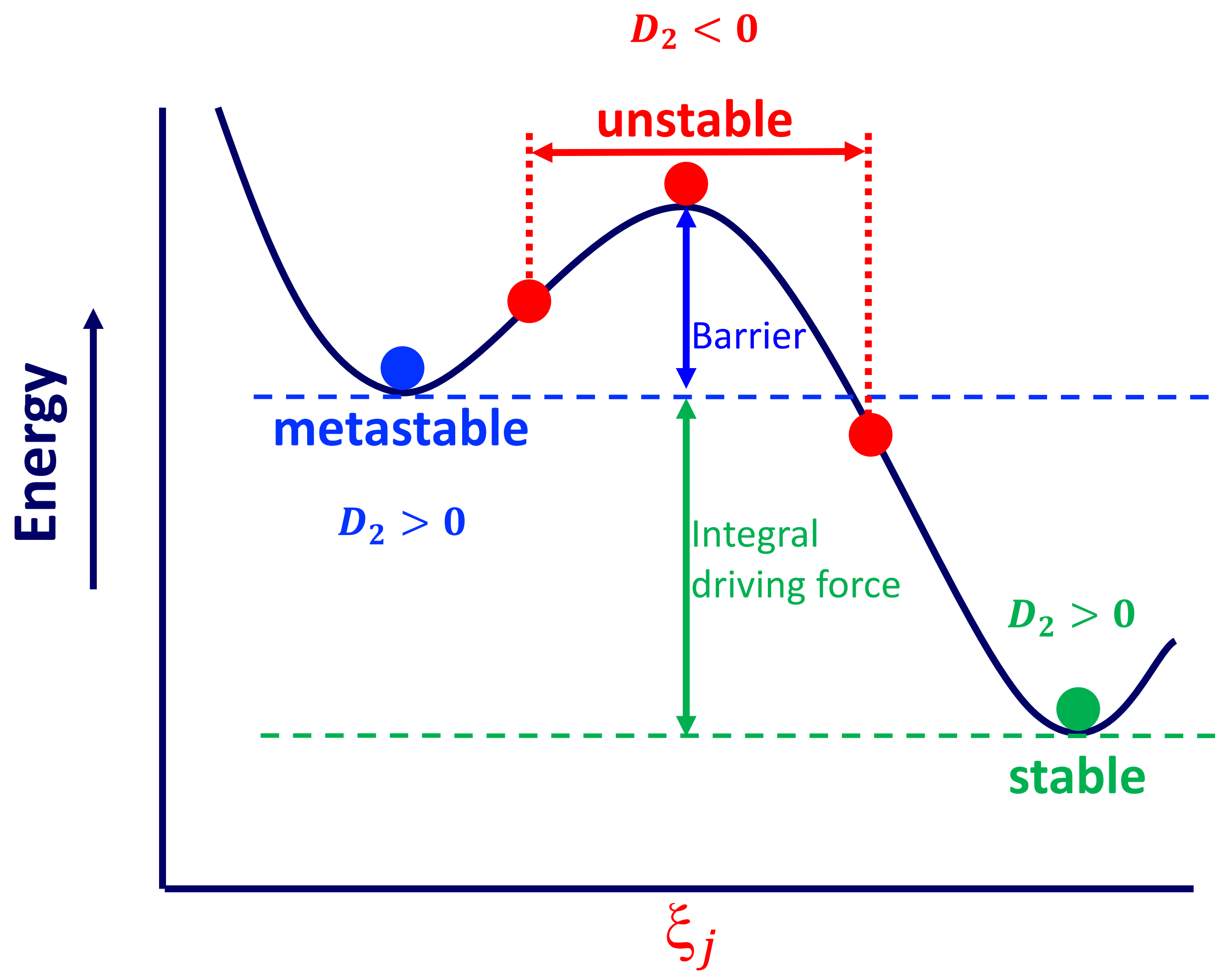

Figure 2

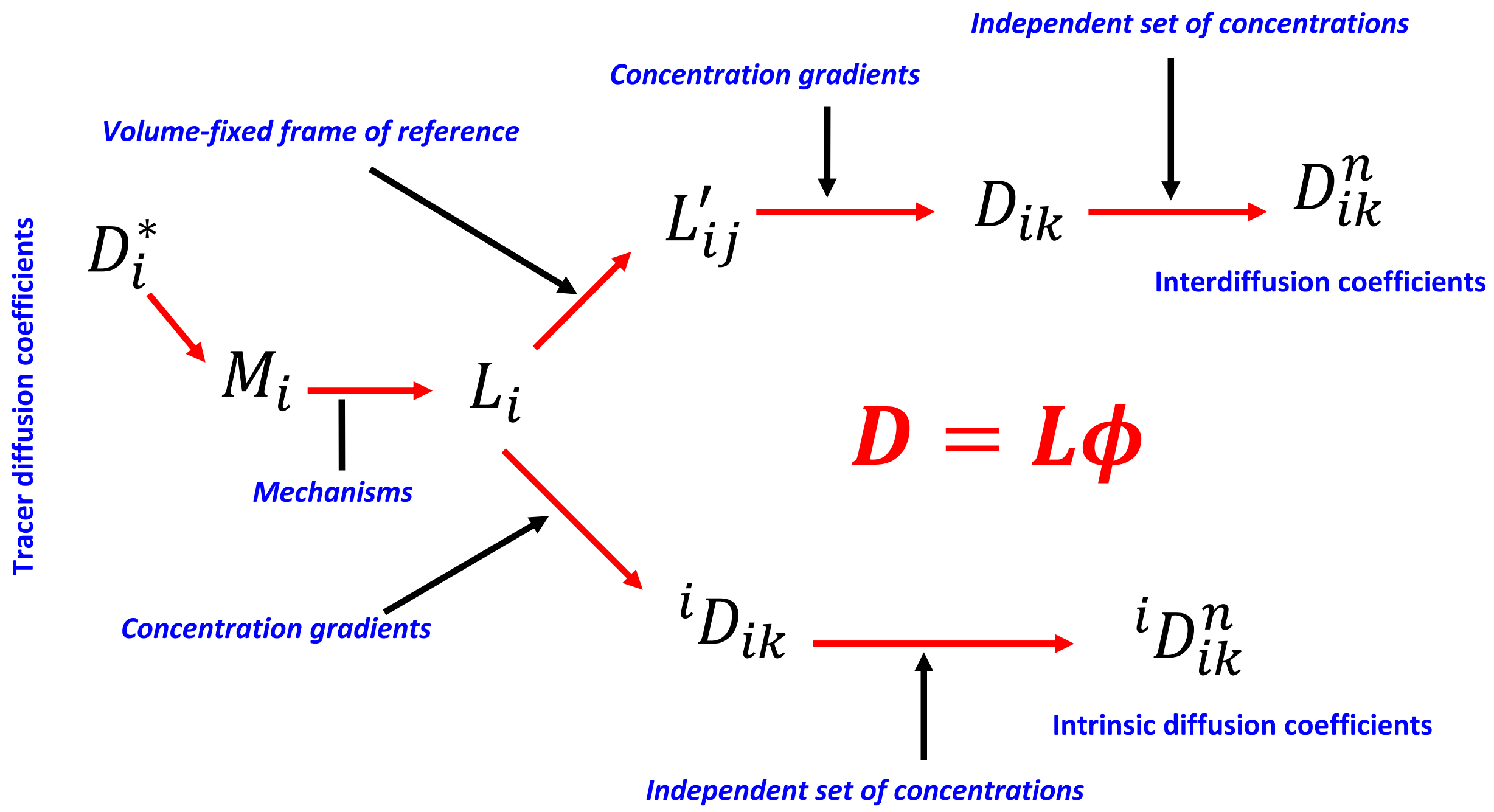

Figure 3

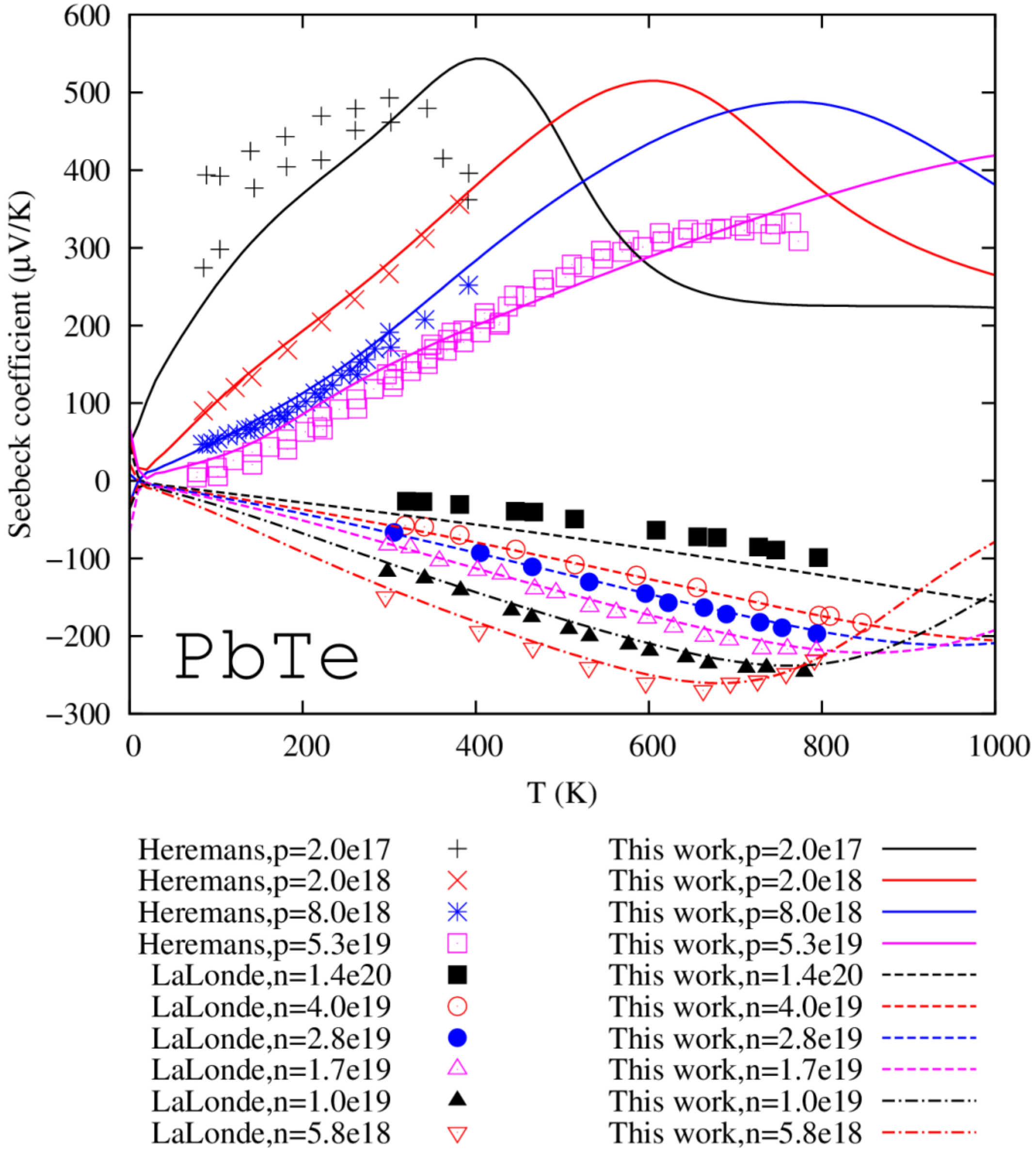

(i)

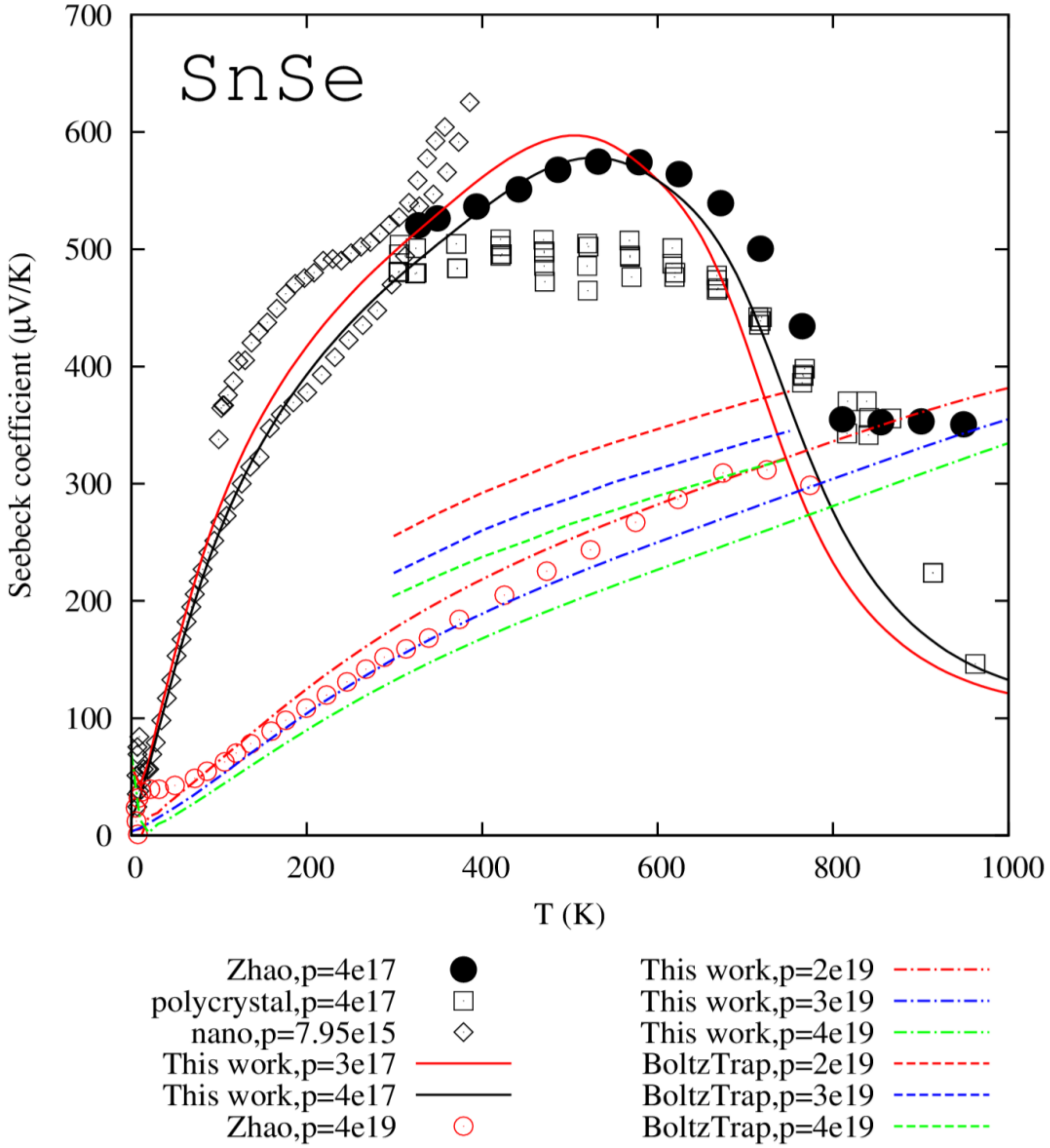

(ii)

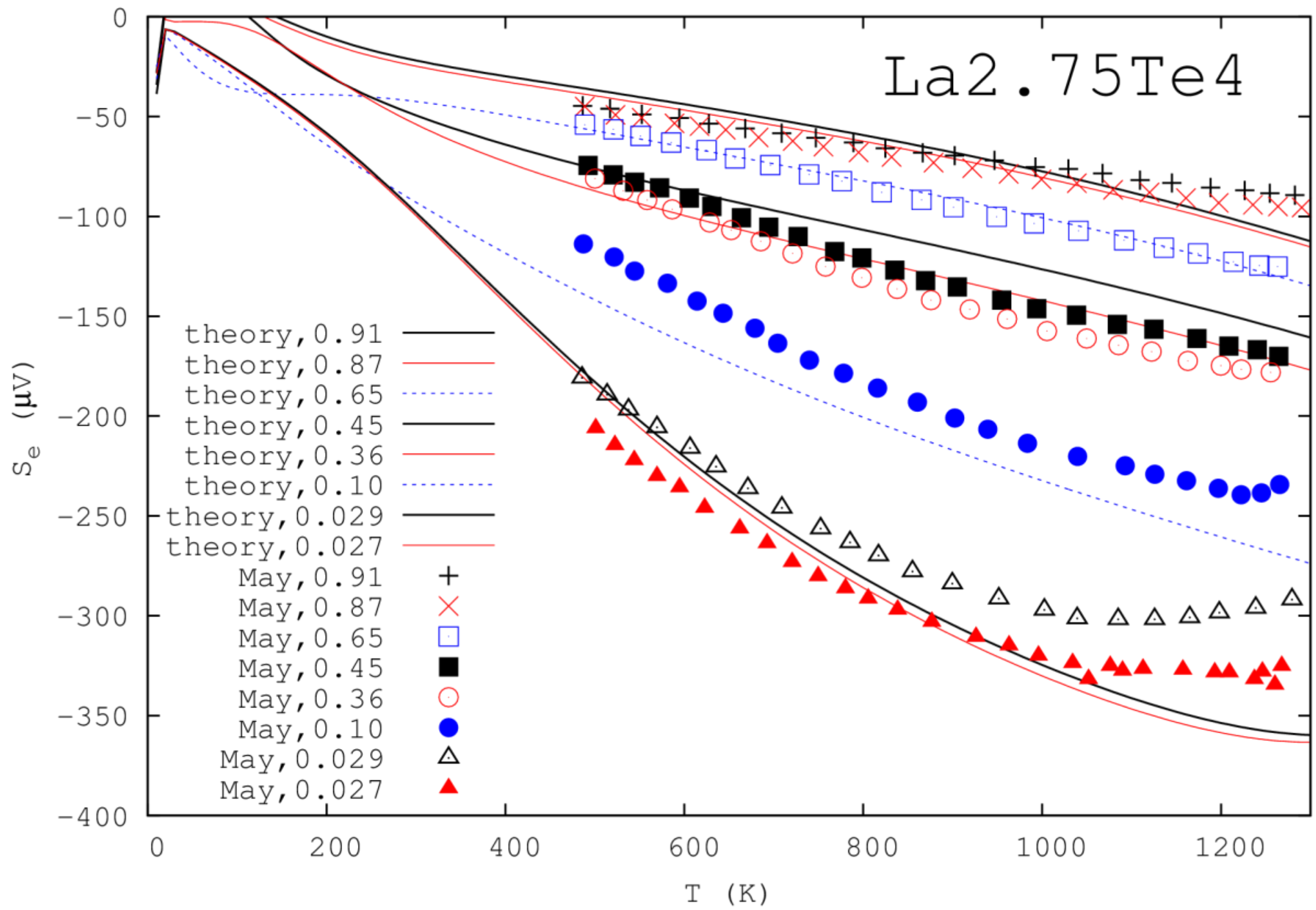

(iii)

Figure 4

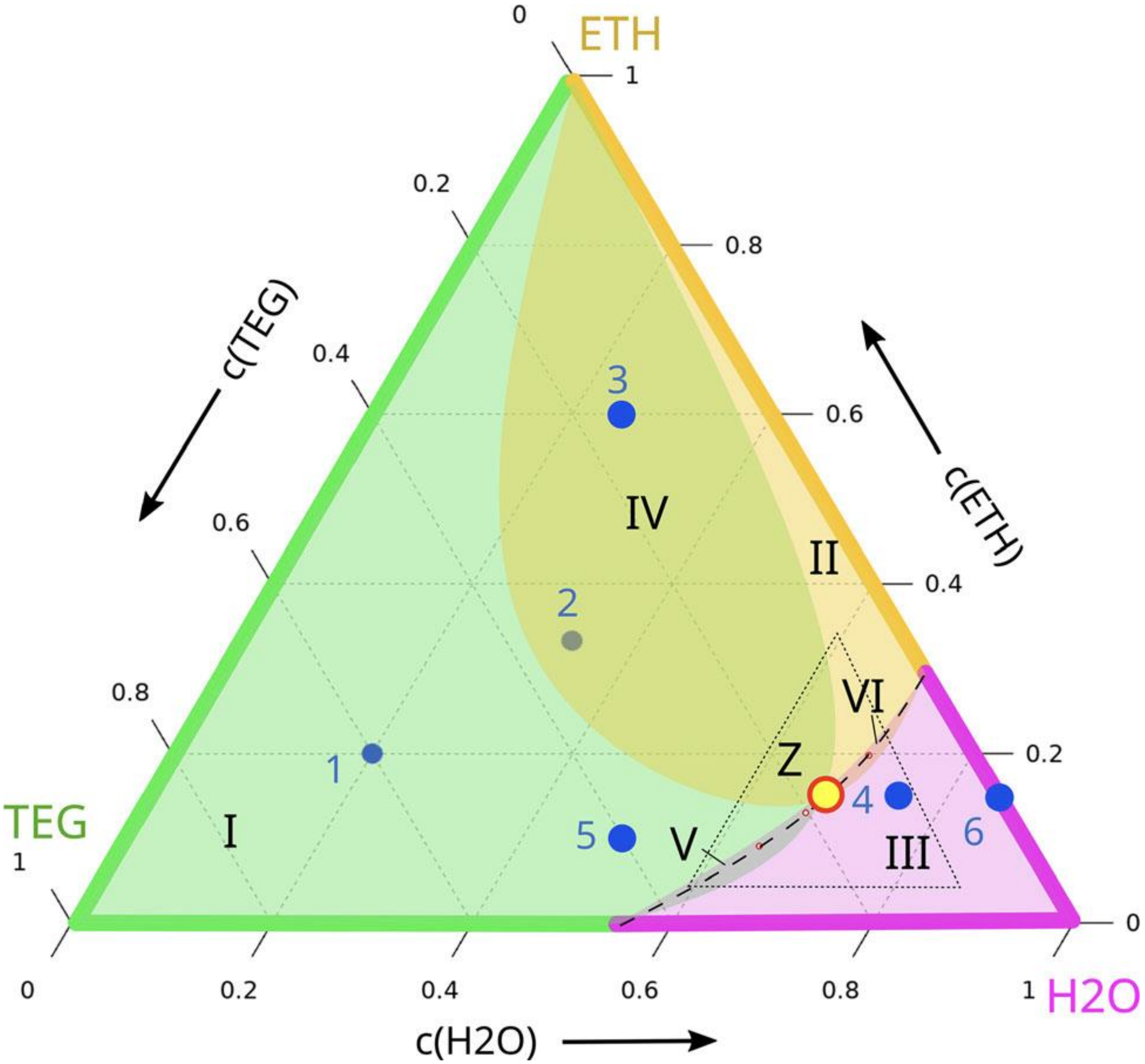

Figure 5

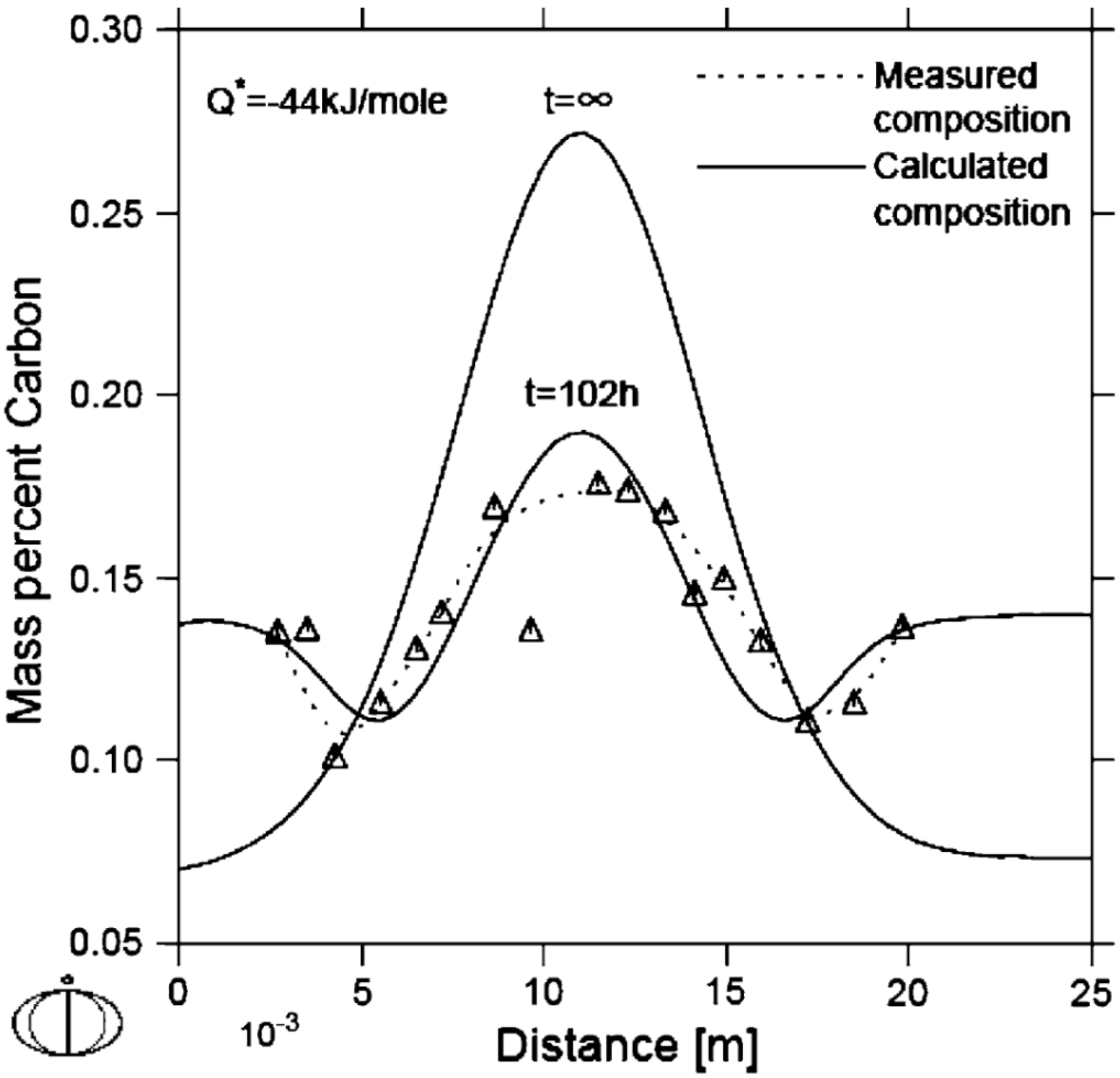

Figure 6

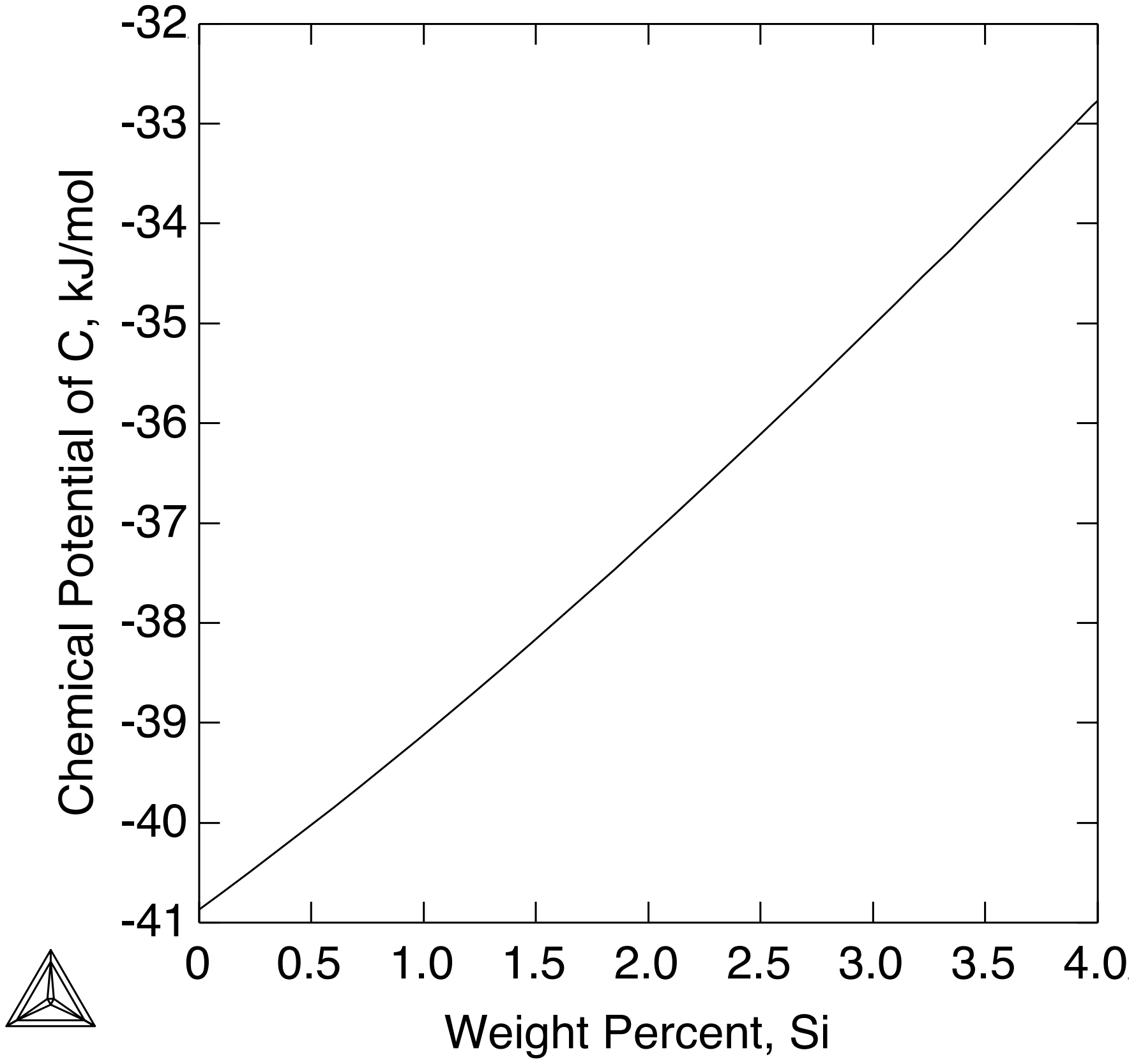

(i)

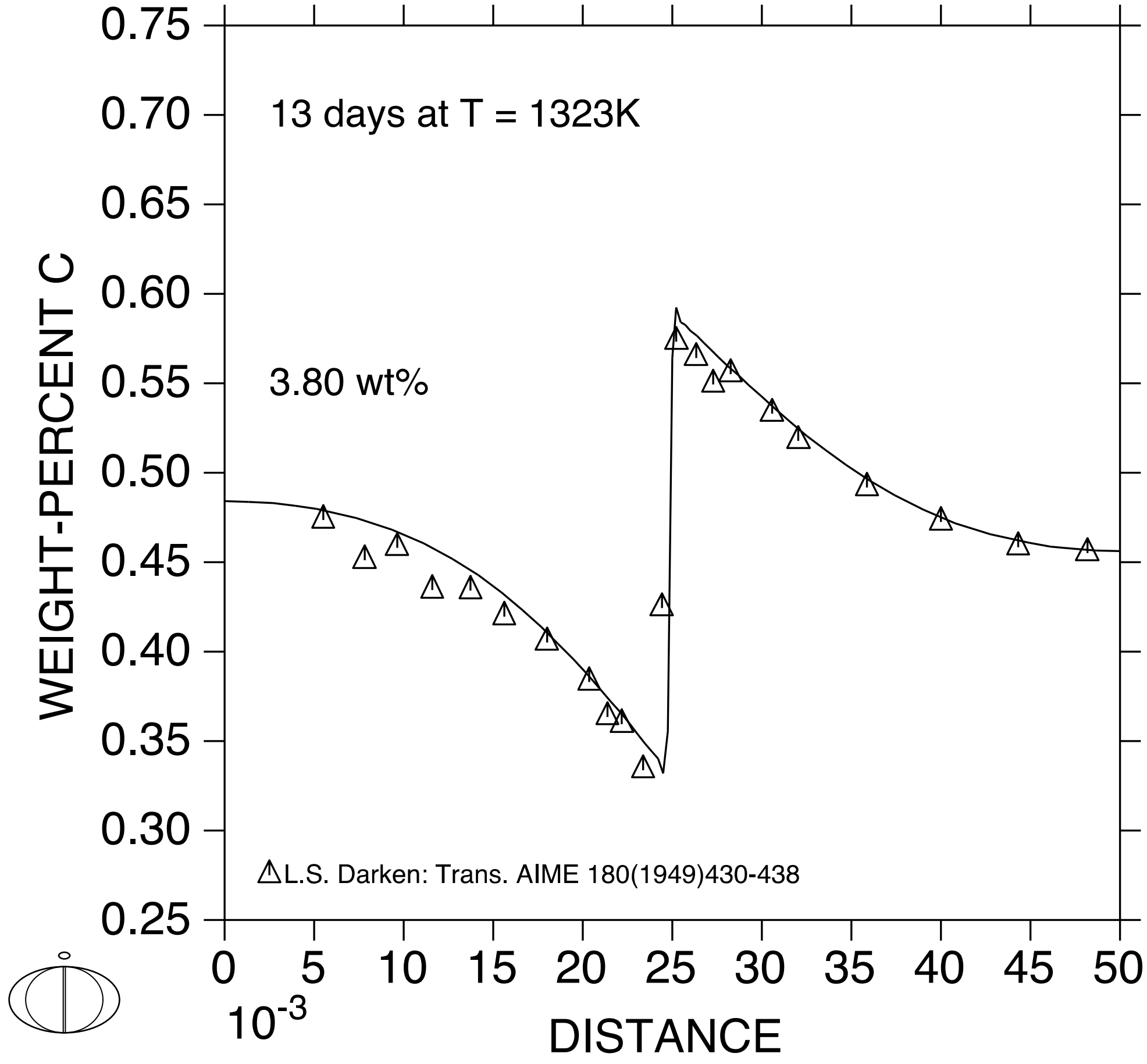

(ii)

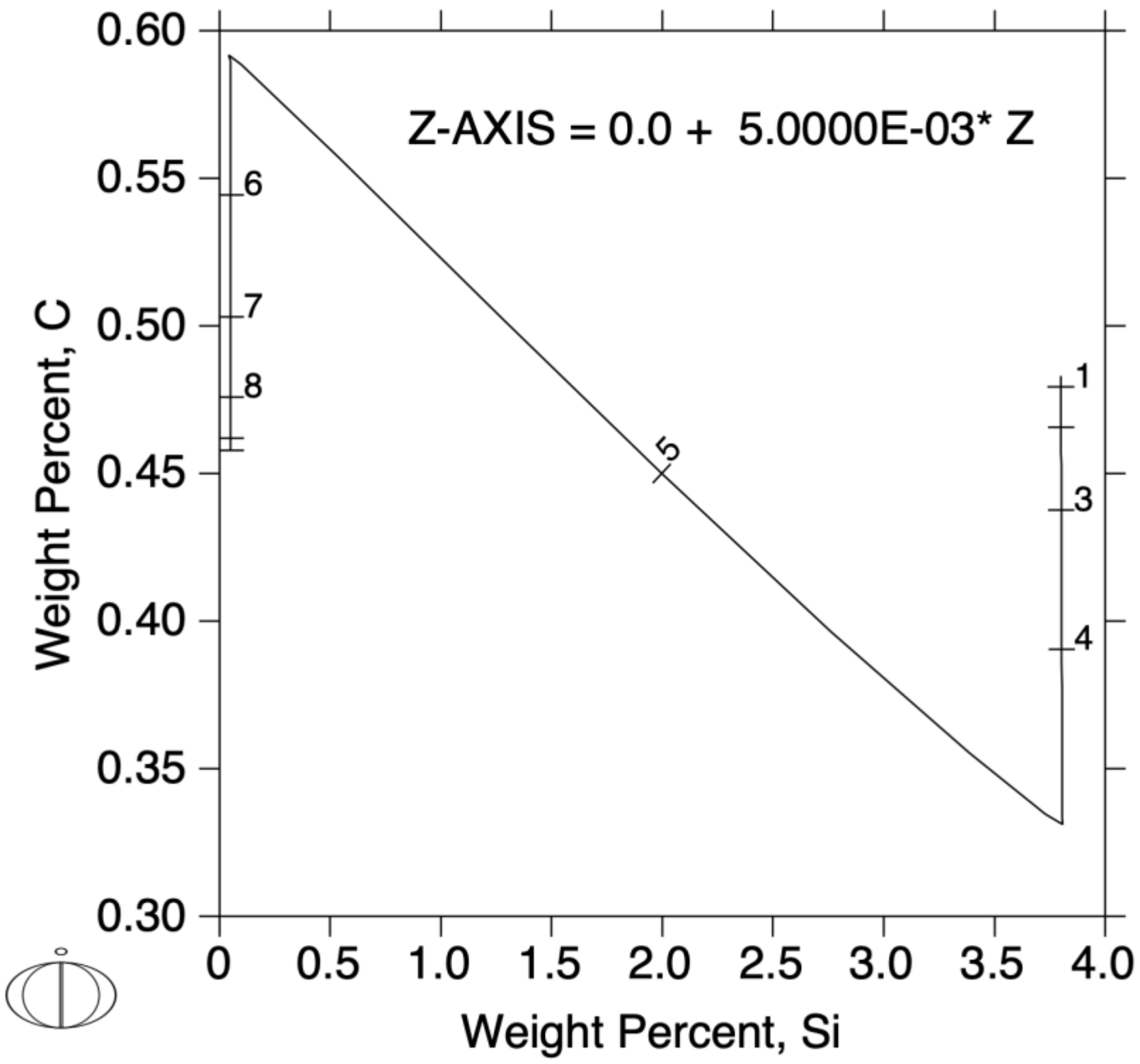

(iii)

Figure 7

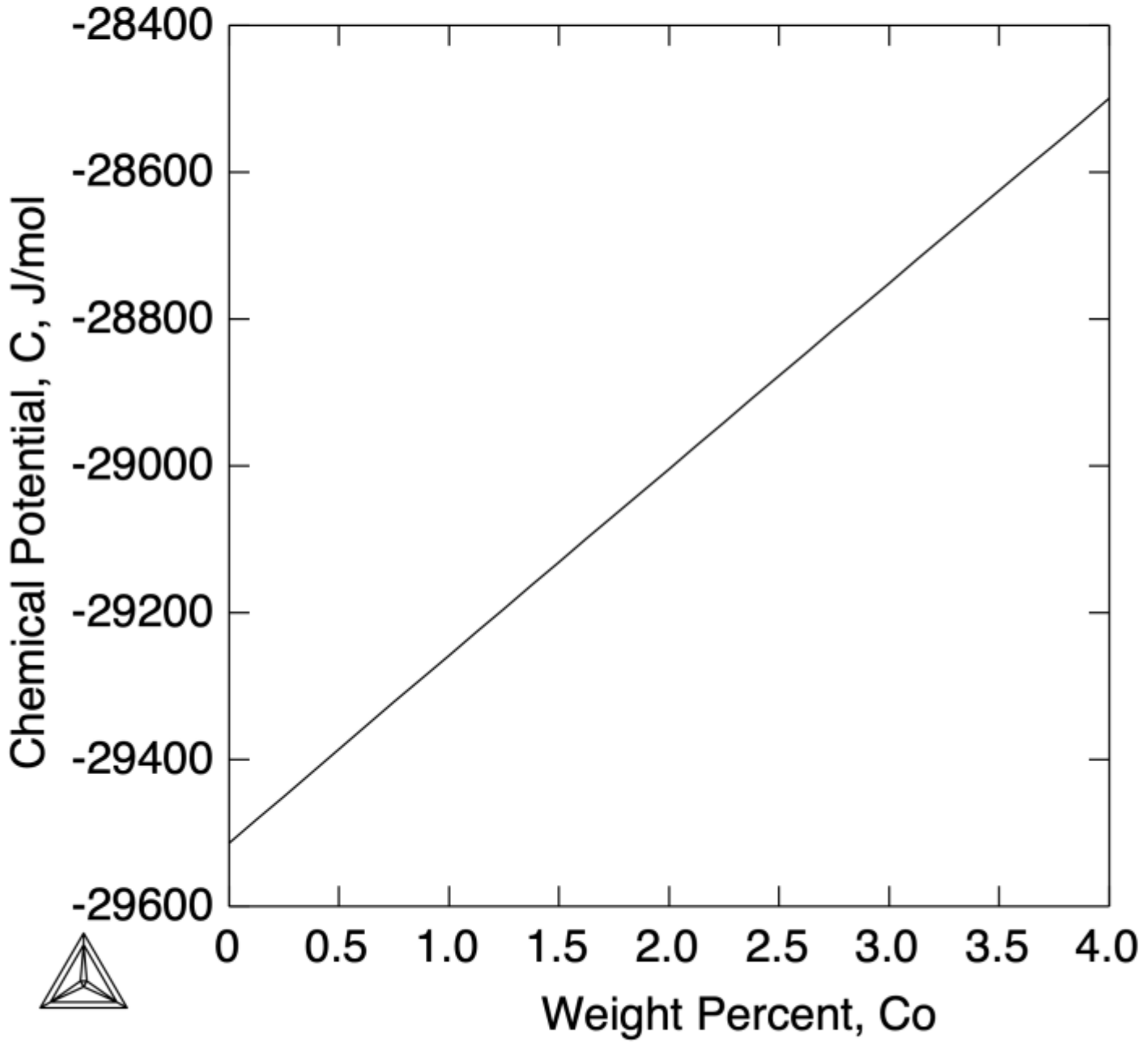

Figure 8

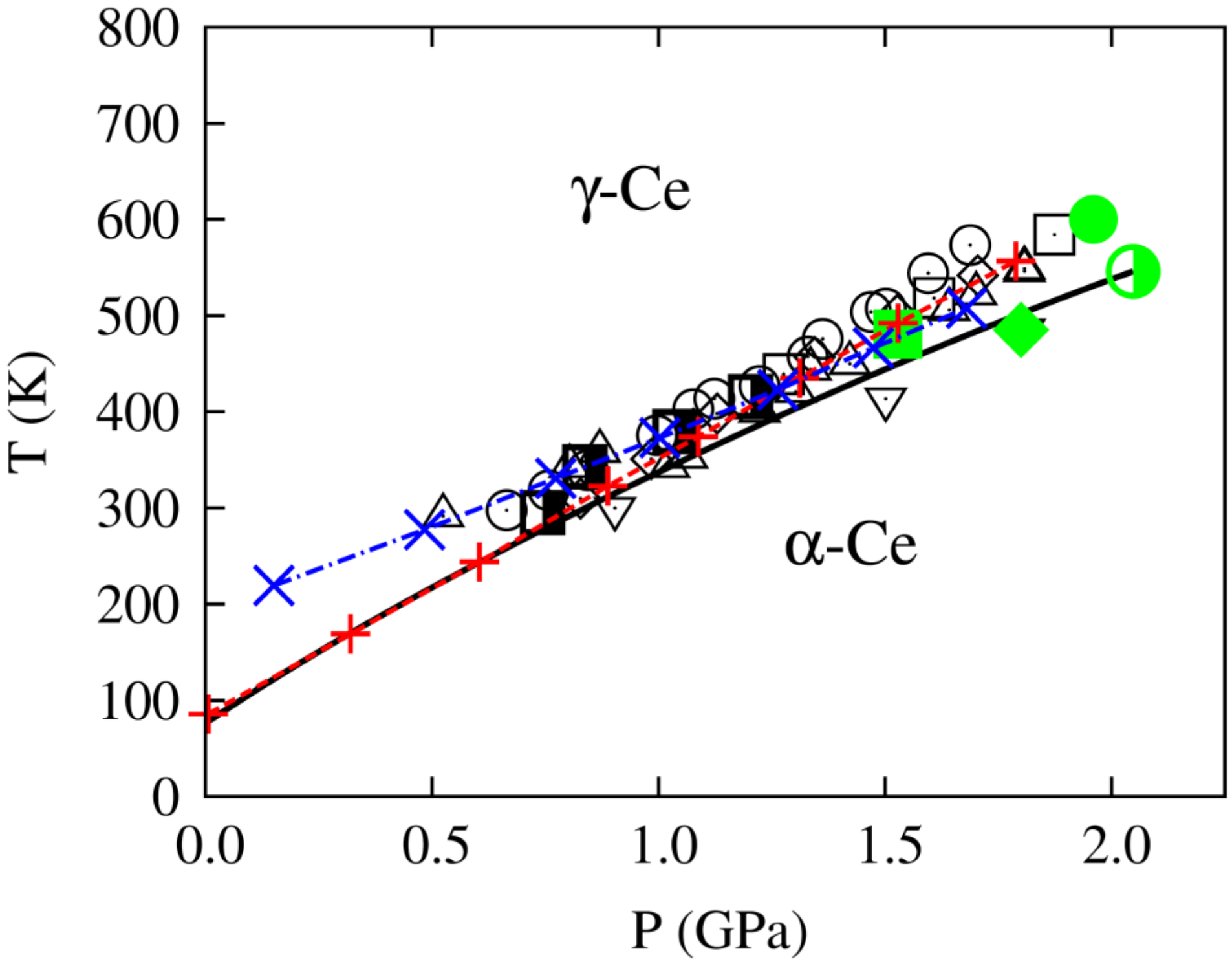

(i)

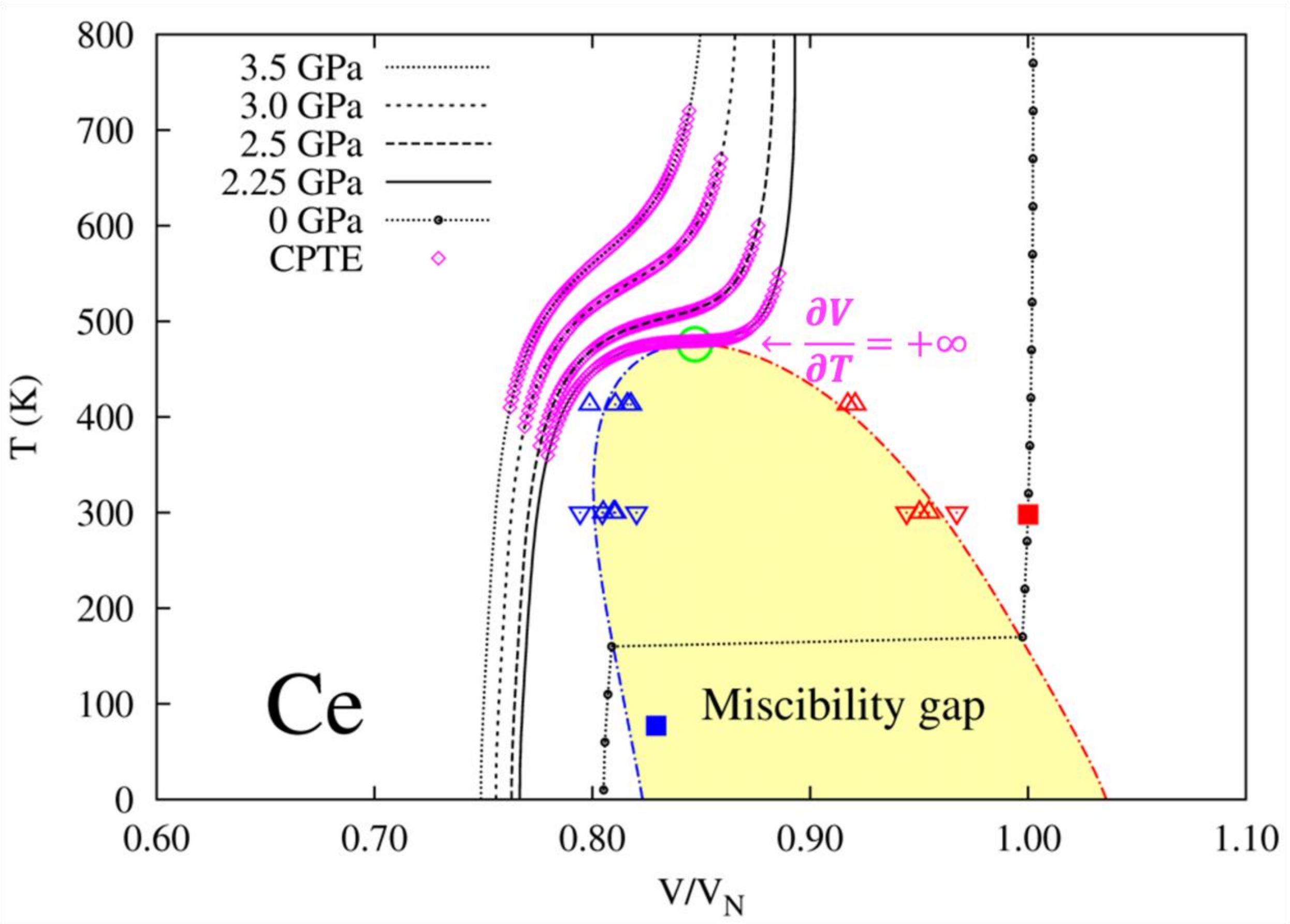

(ii)

Figure 9

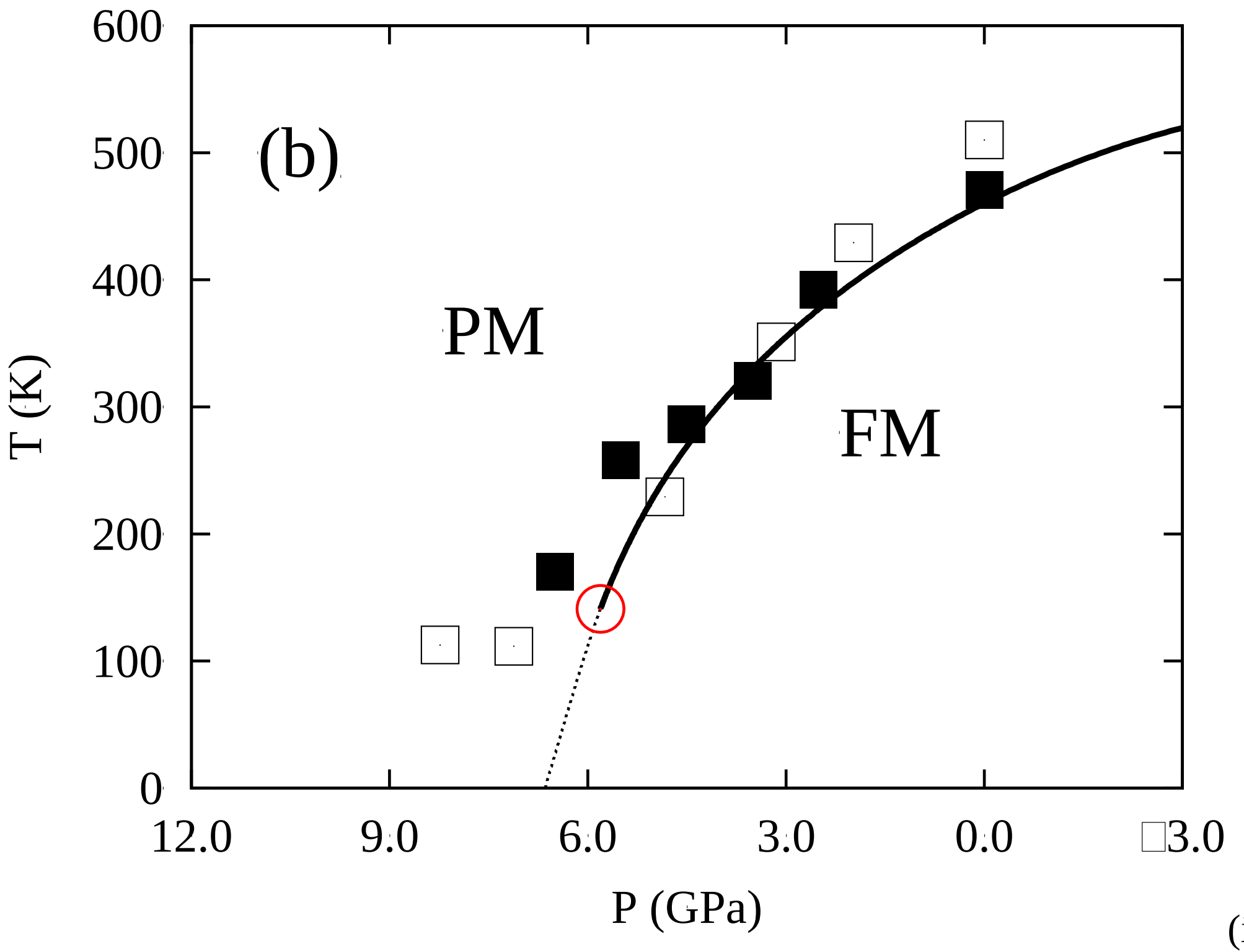

(i)

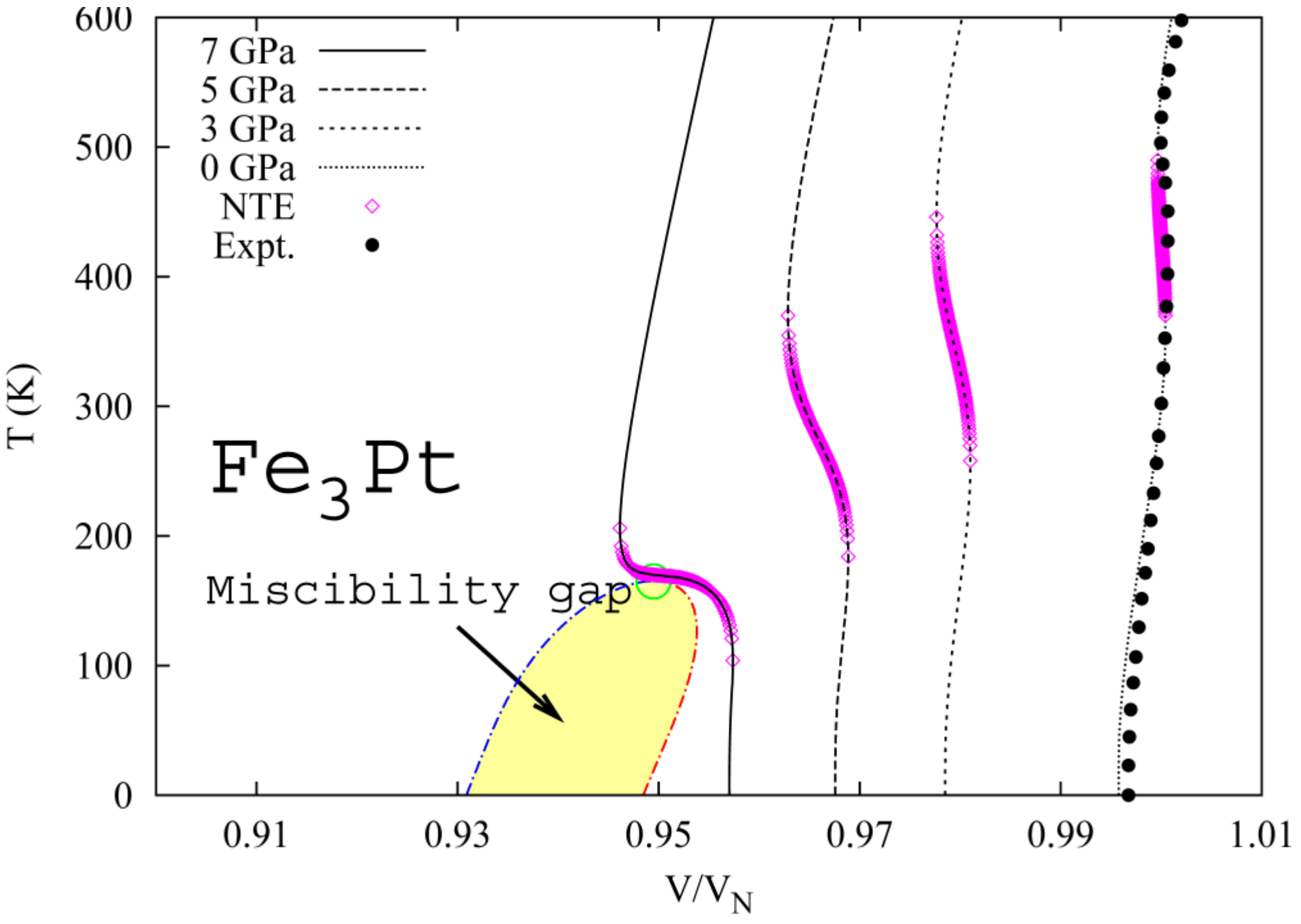

(ii)

Figure 10

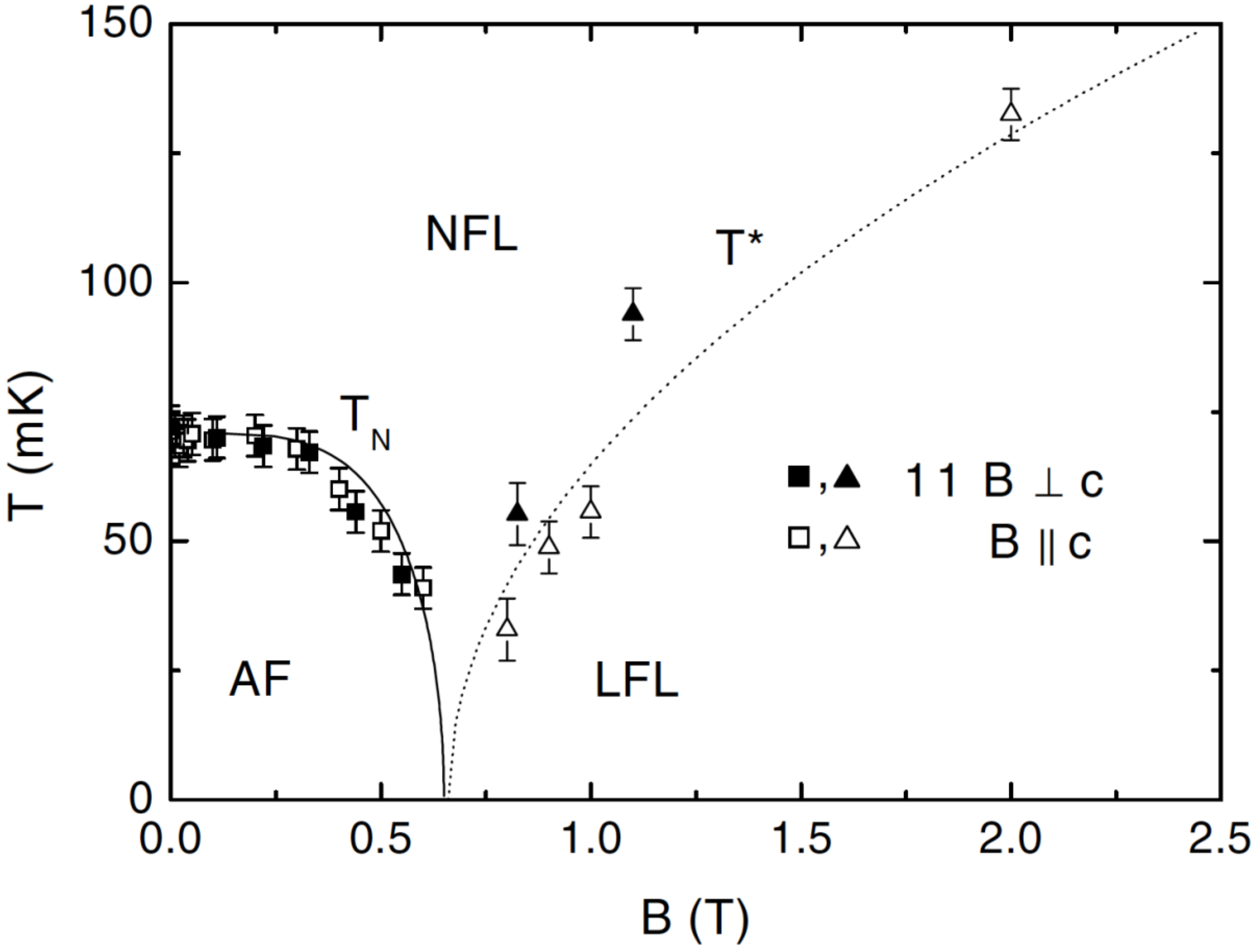

(i)

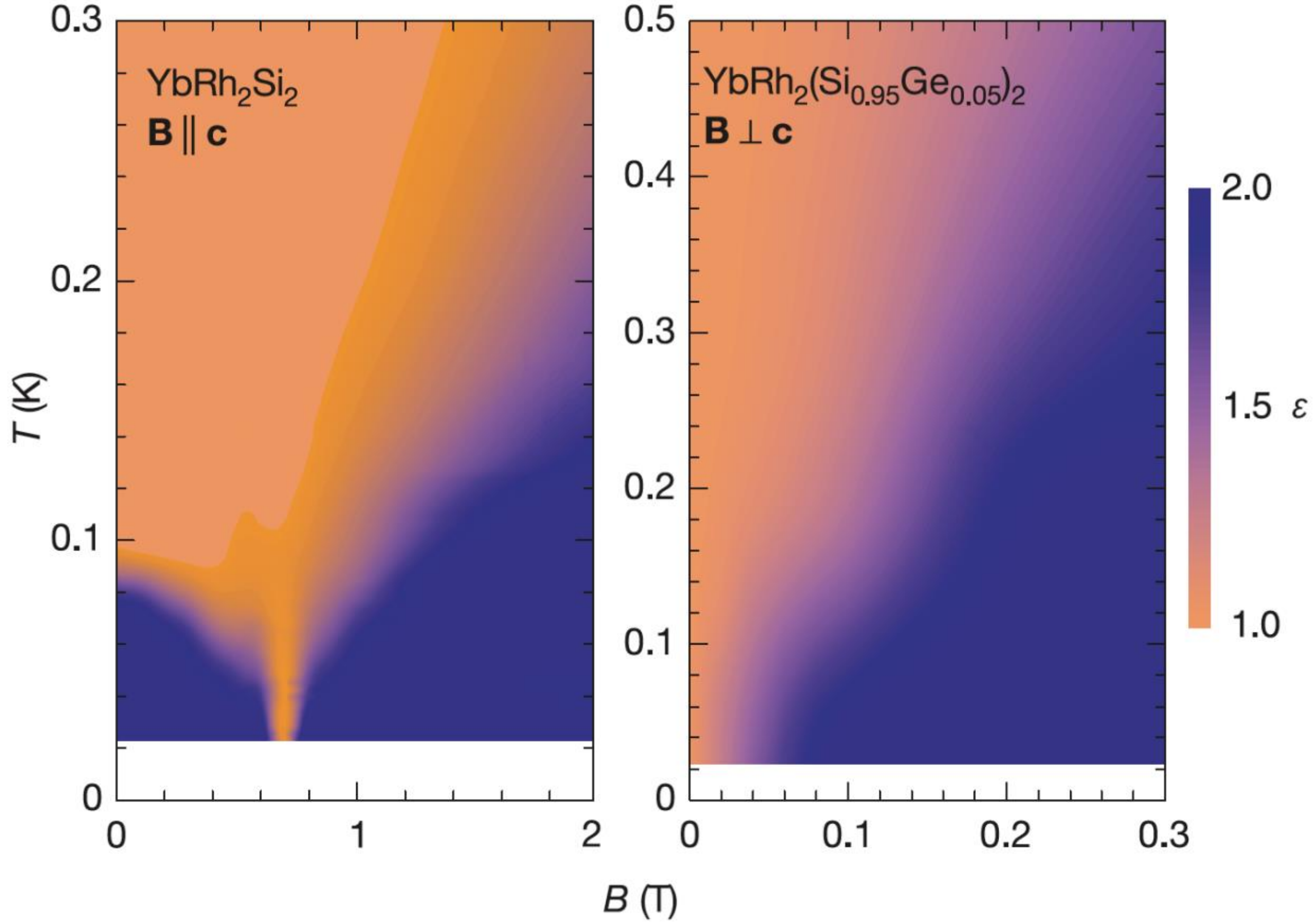

(ii)

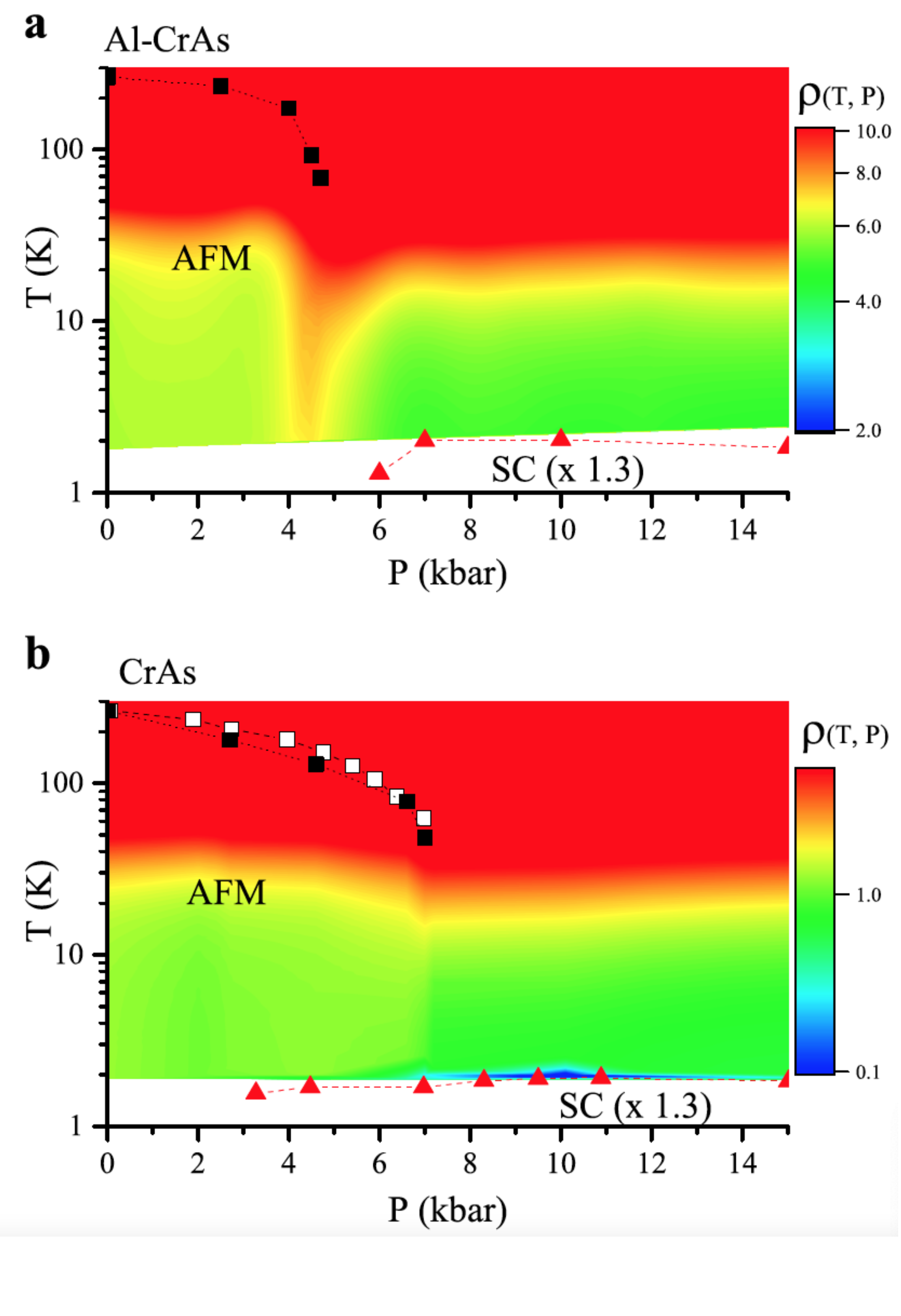

(iii)

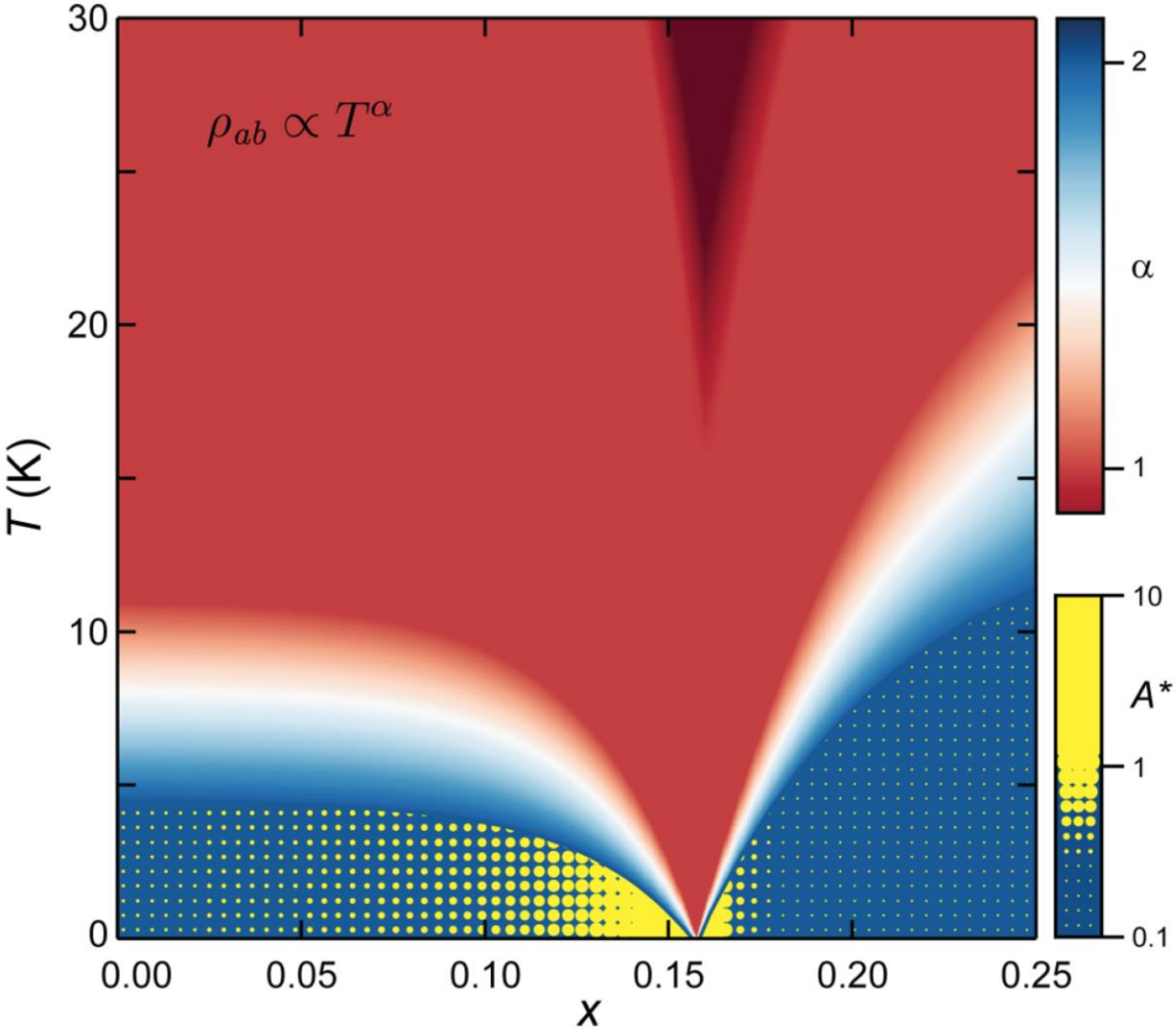

(iv)
Figure 11

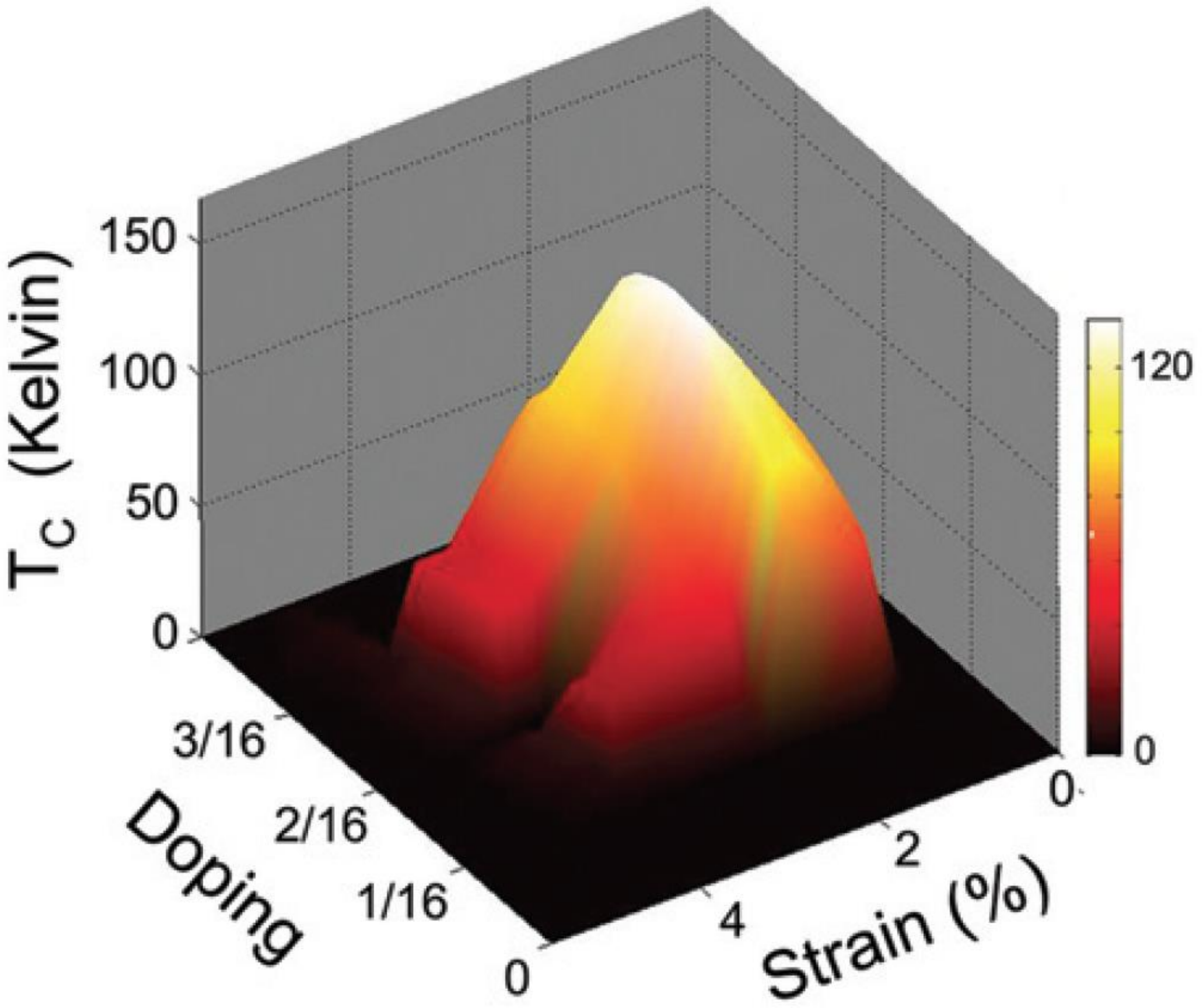

(i)

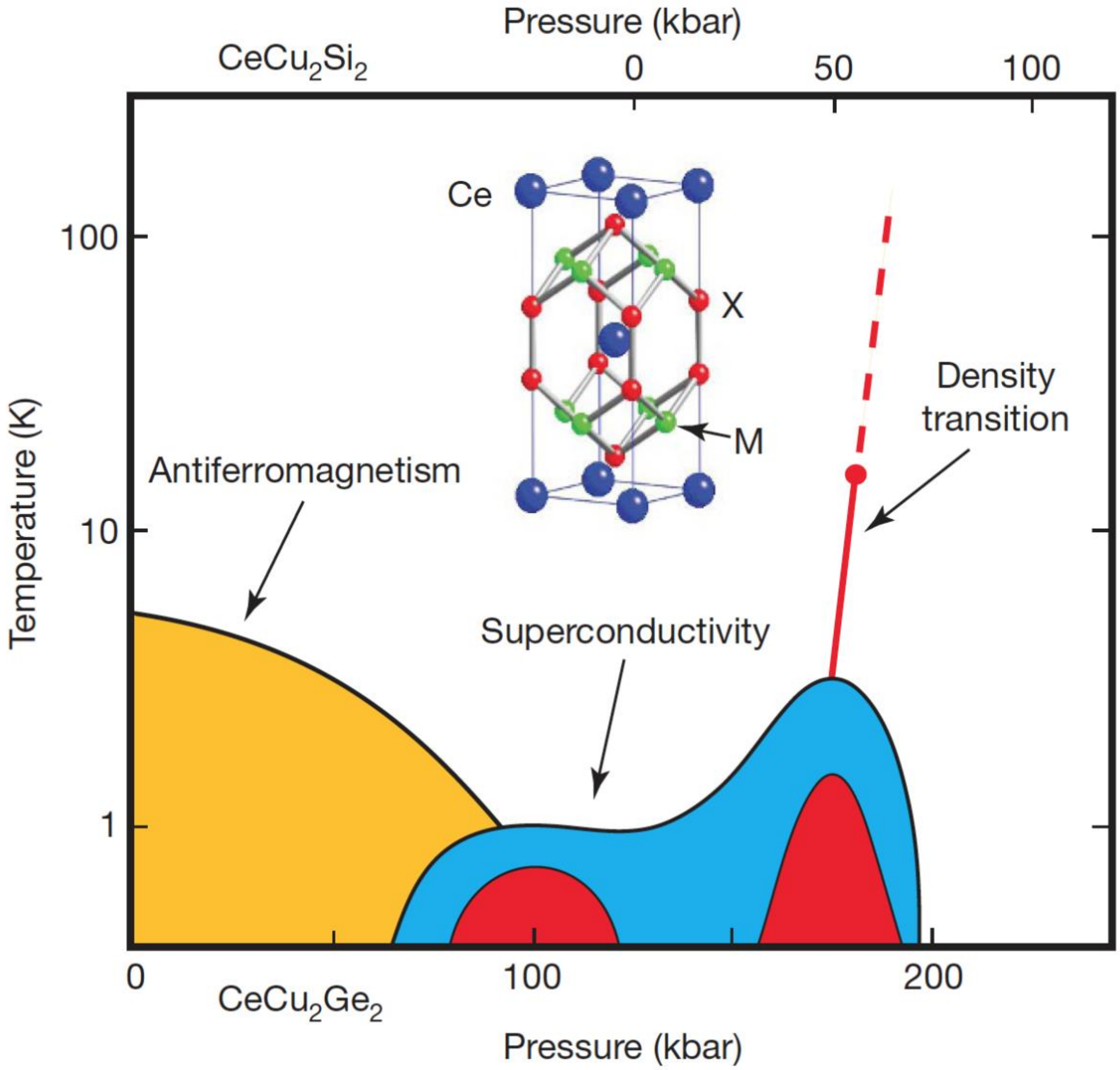

(ii)

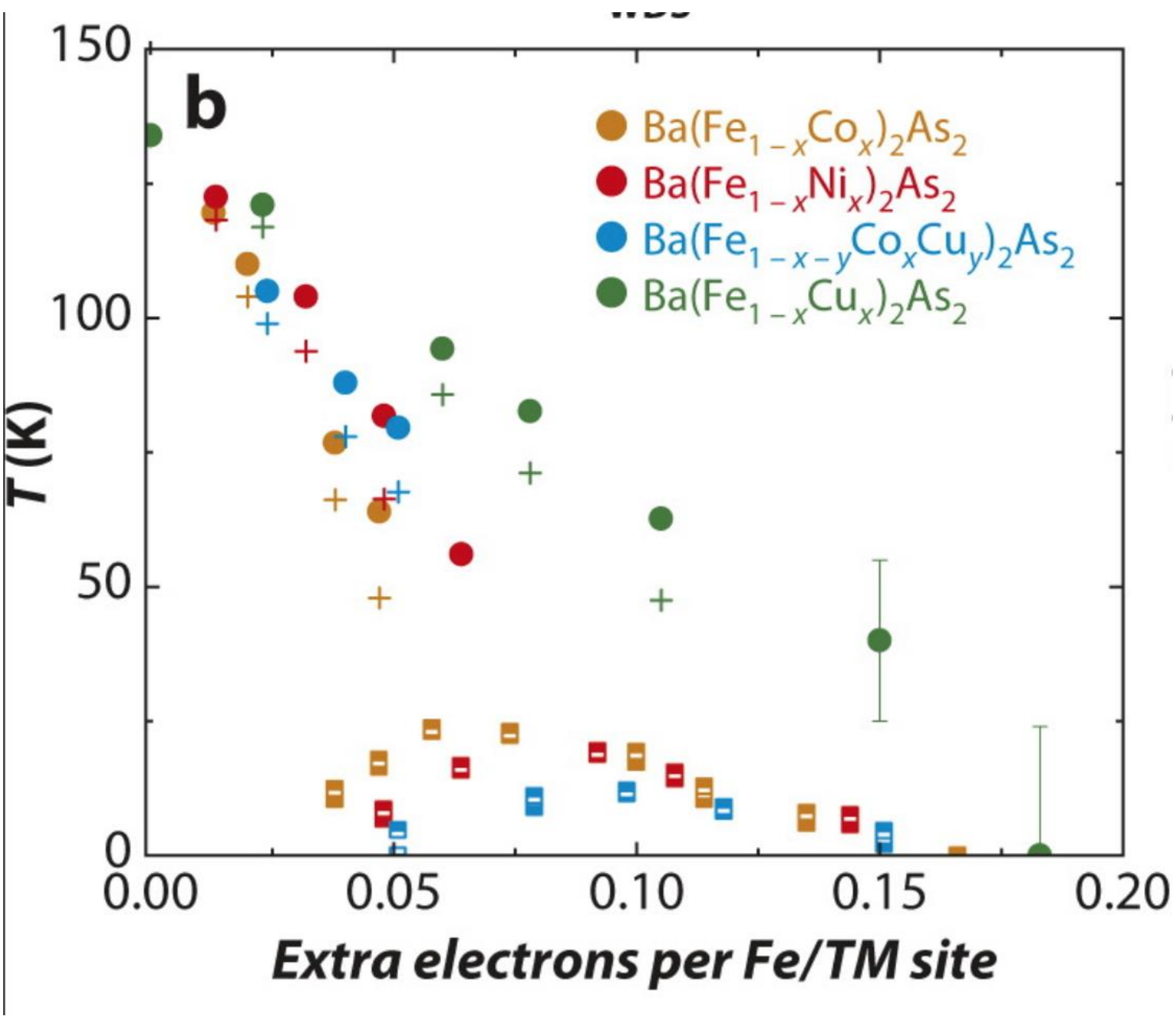

(iii)

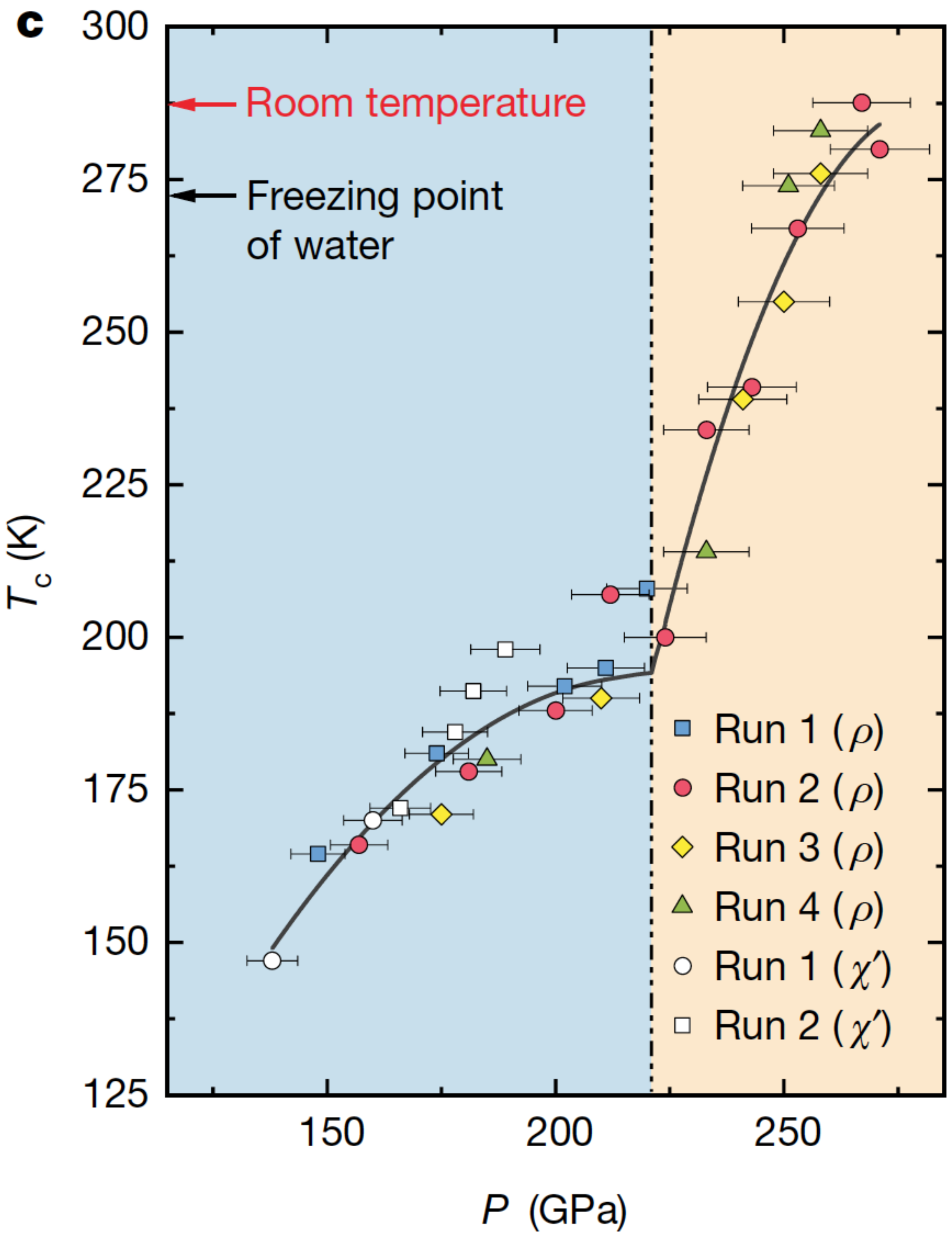

(iv)

Figure 12

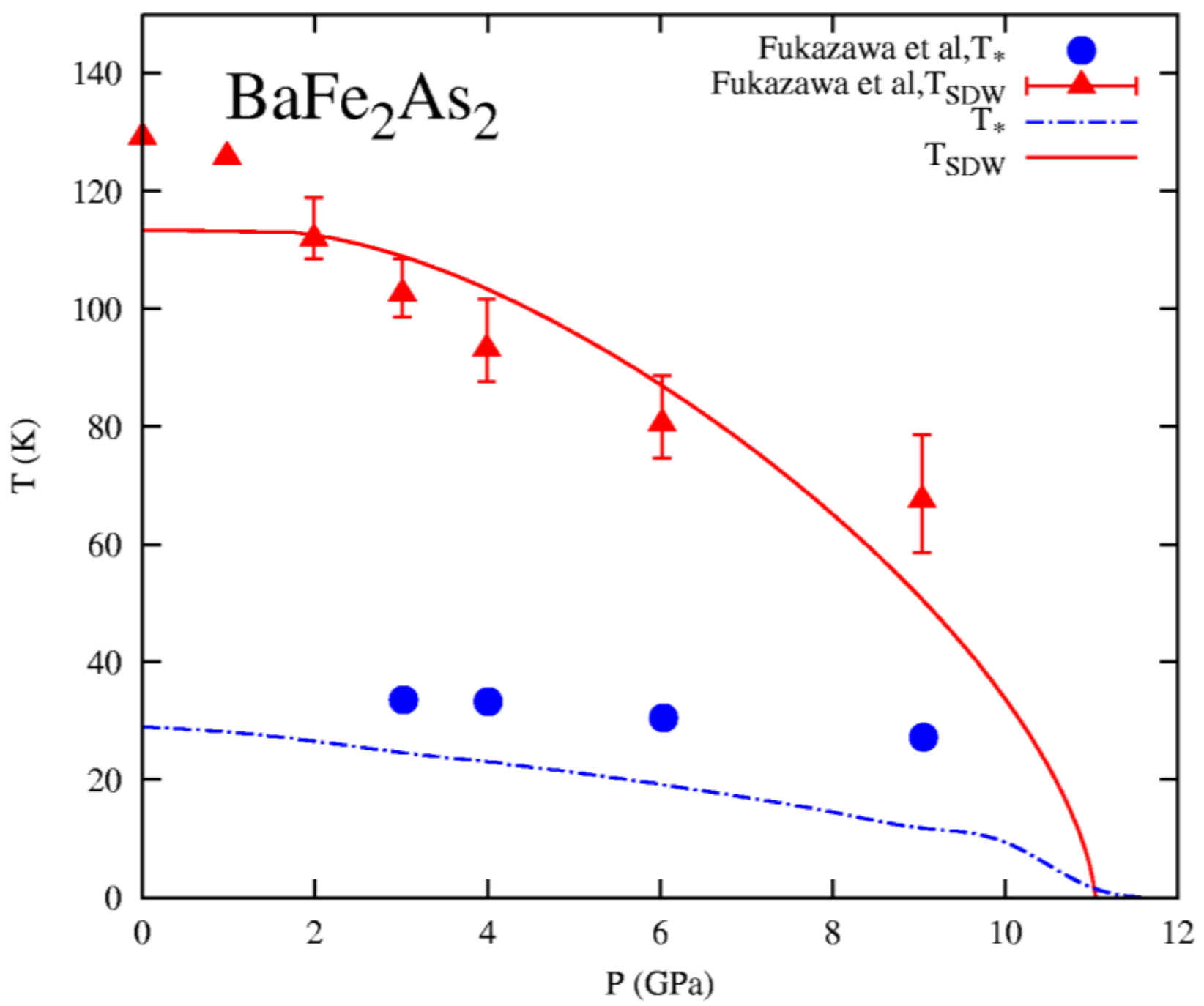

(i)

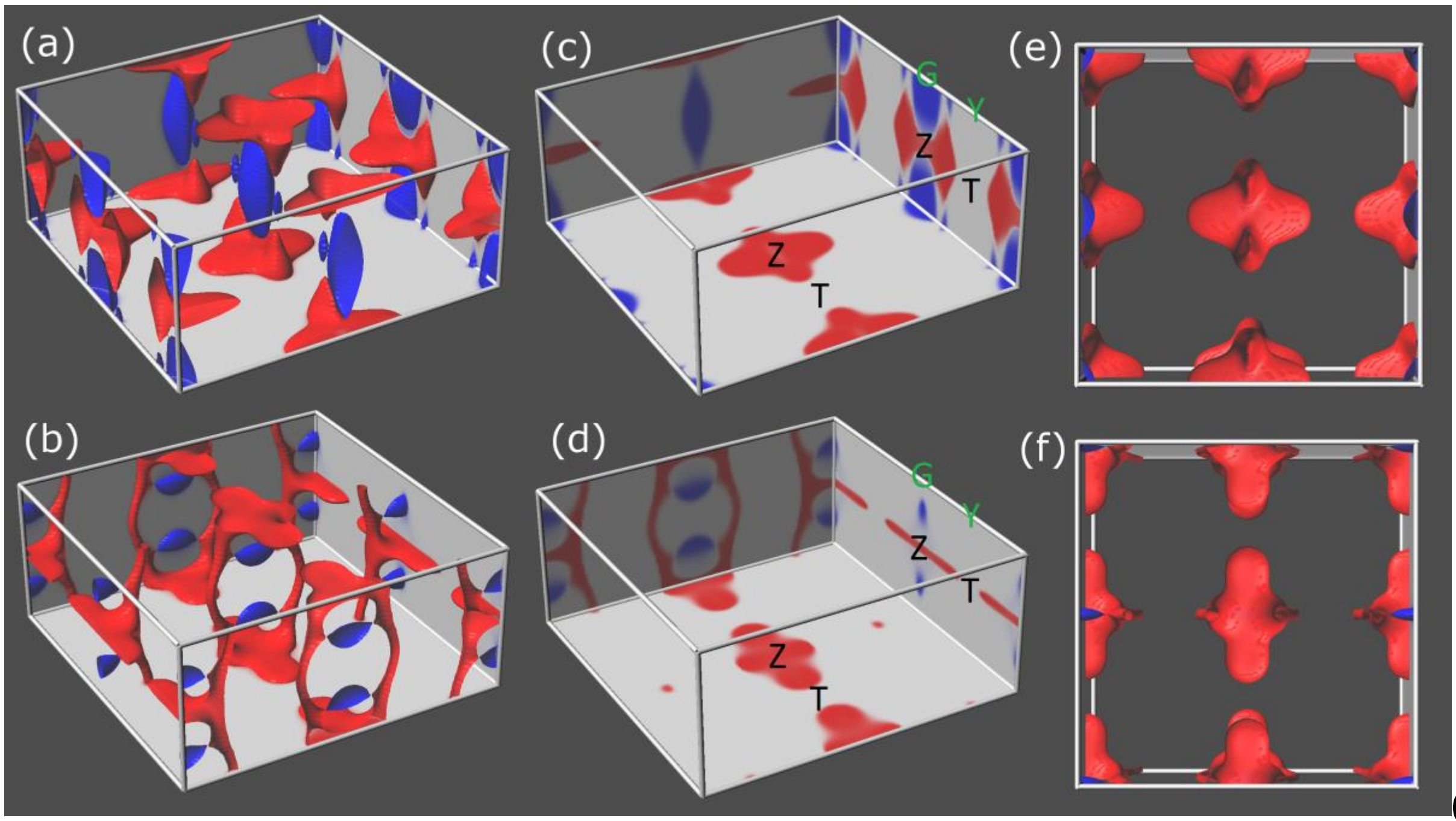

(ii)

Figure 13

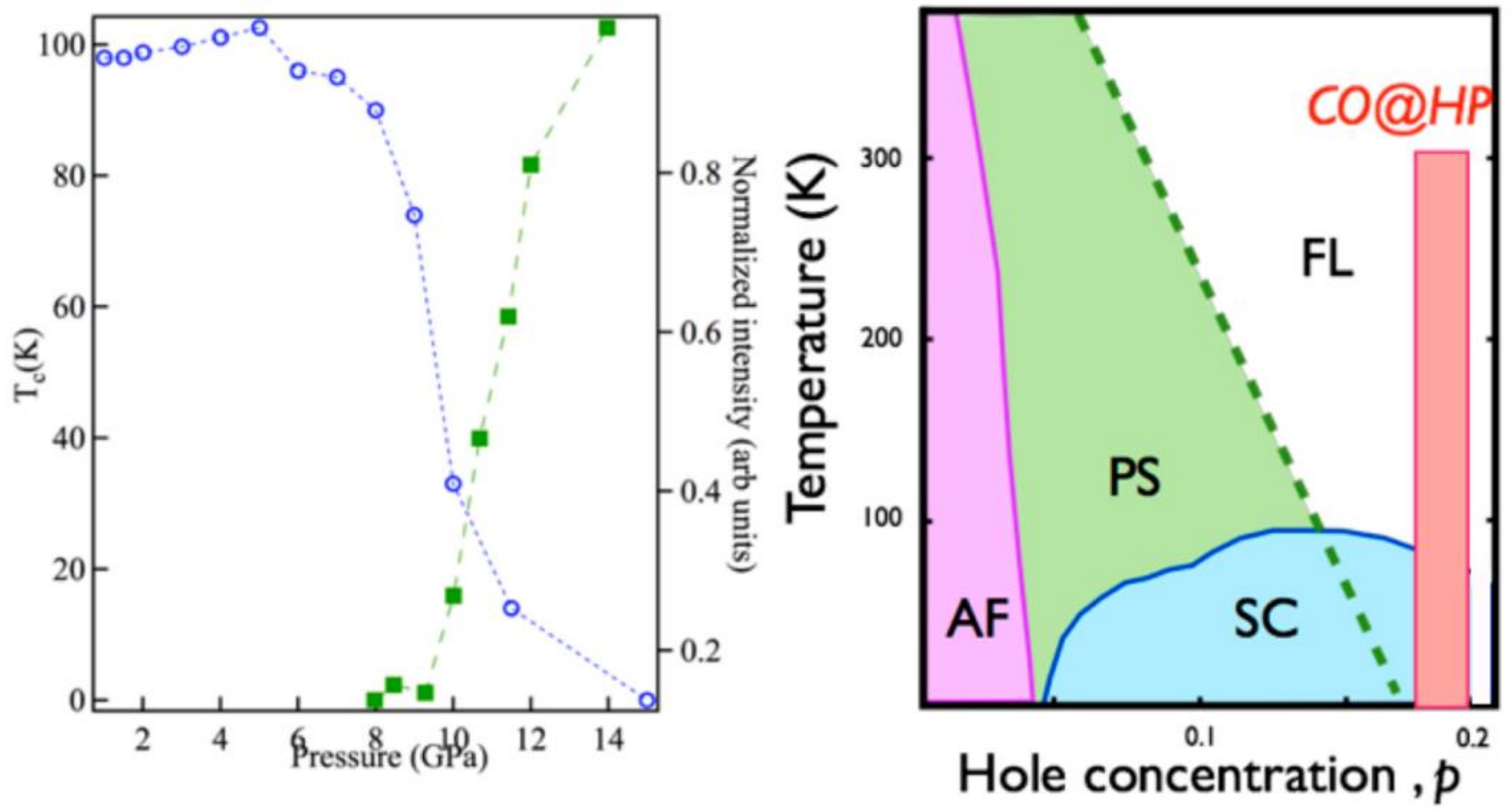

(i)

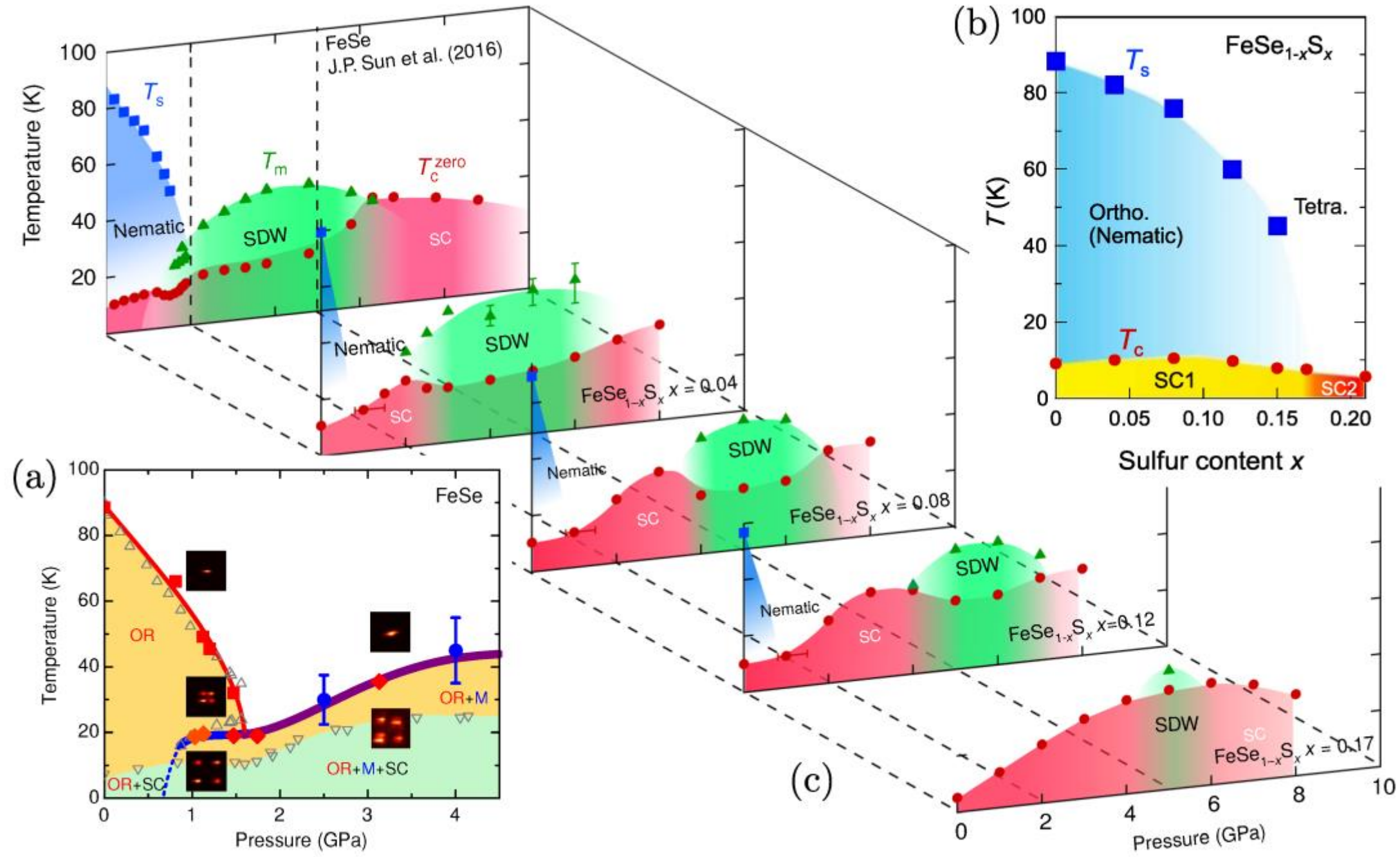

(ii)

Figure 14

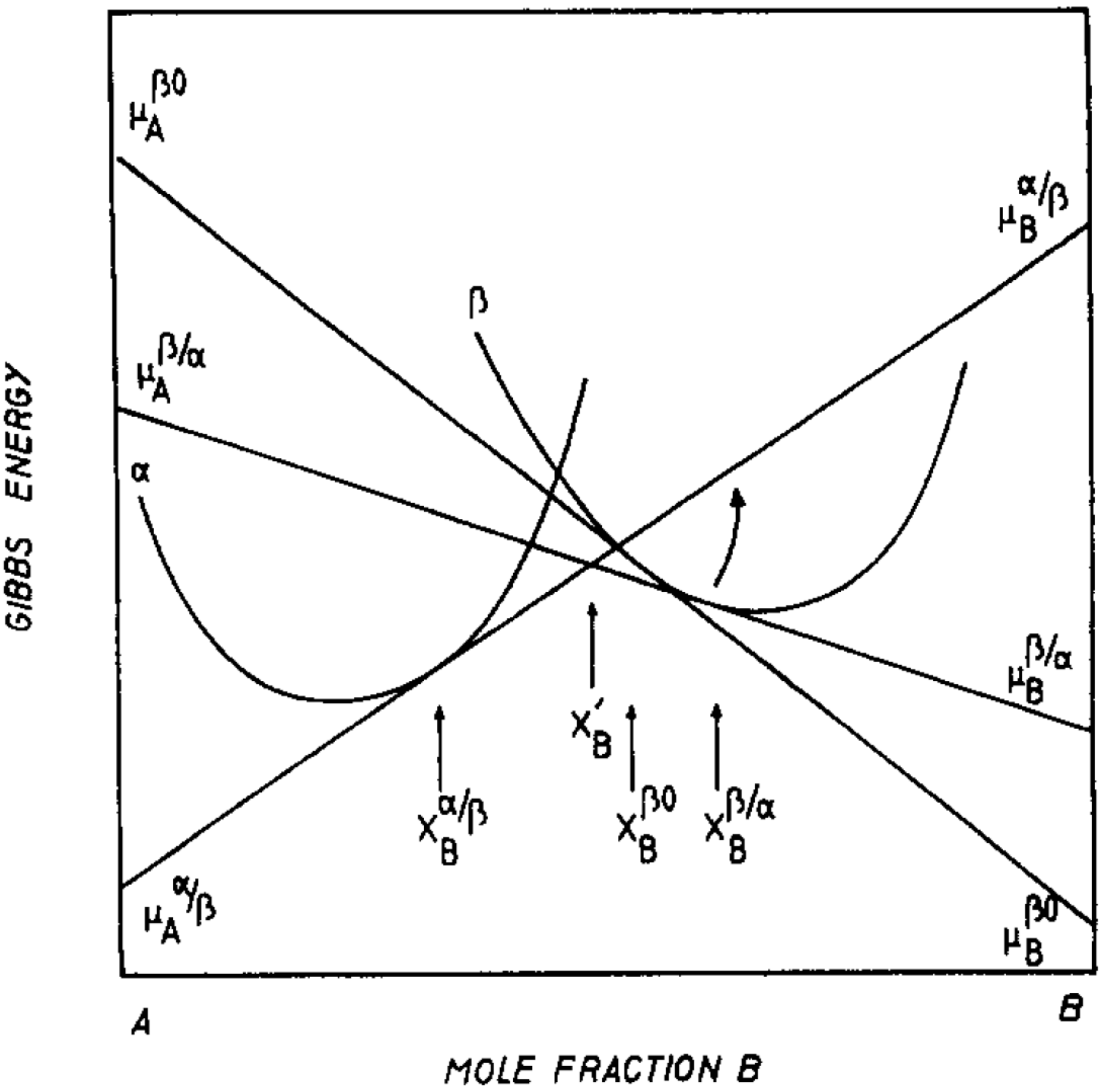

Figure 15

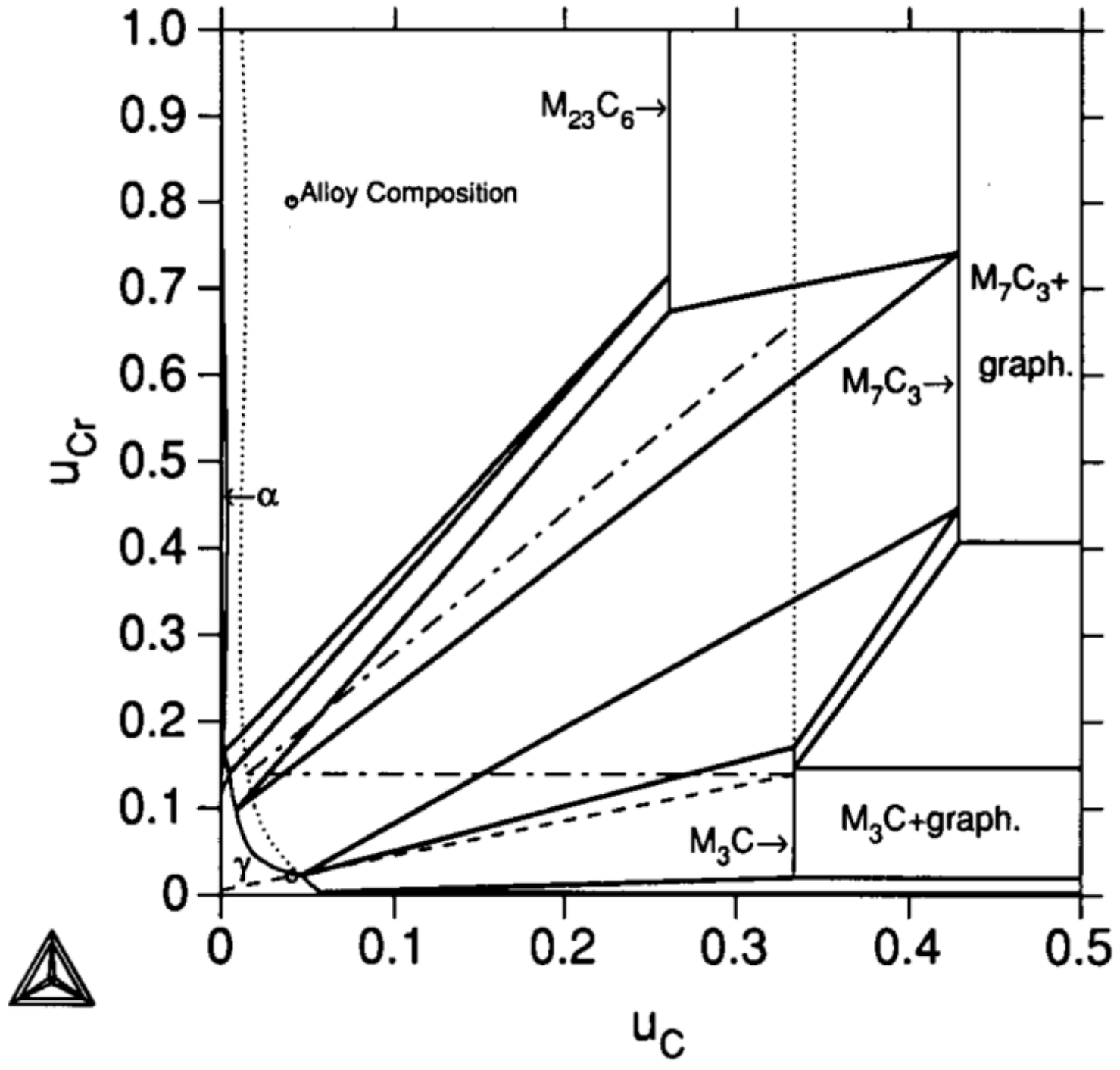

(i)

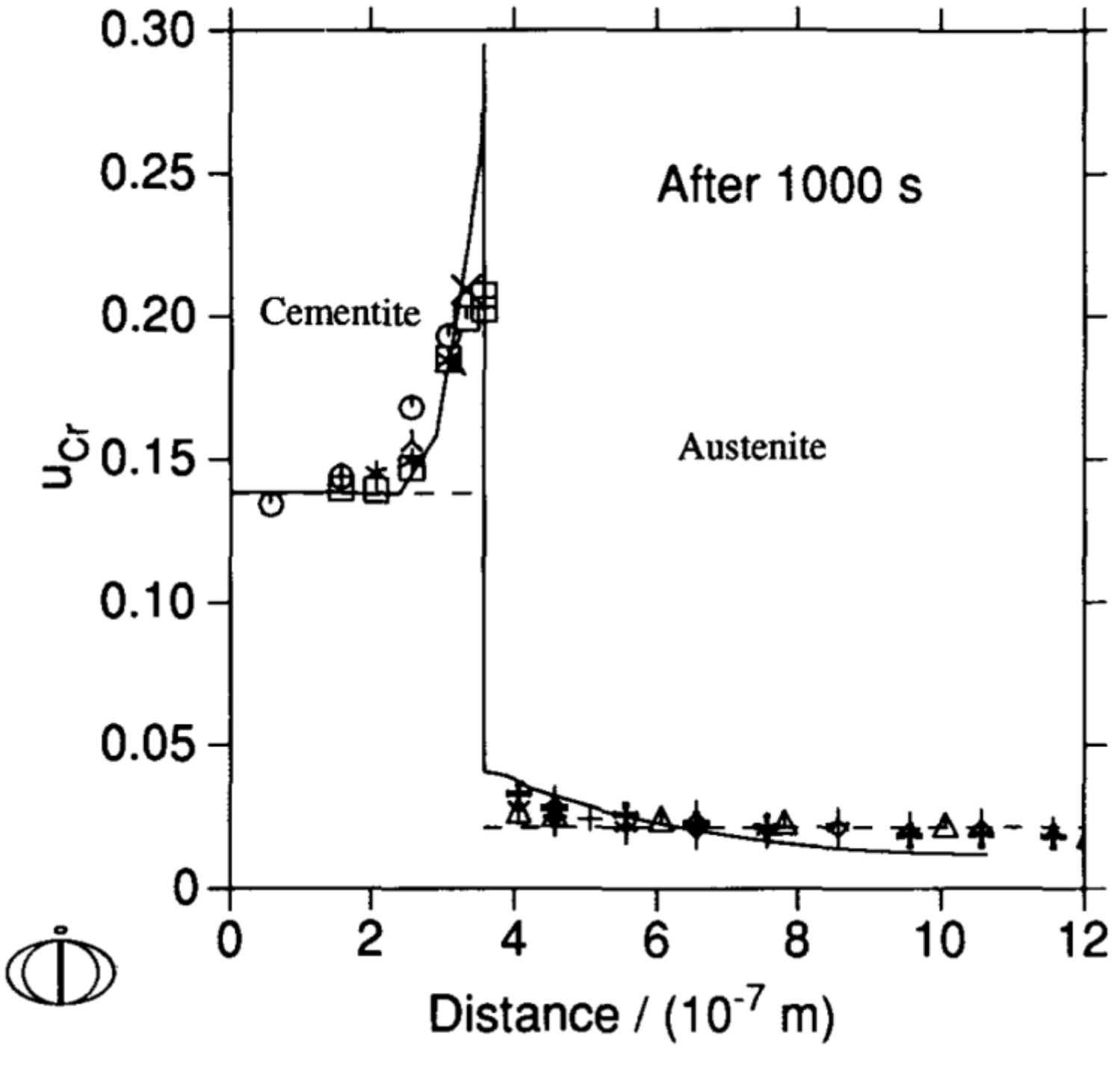

After 1000 s
Cementite
Austenite
uCr
Distance / (10-7 m)

(ii)

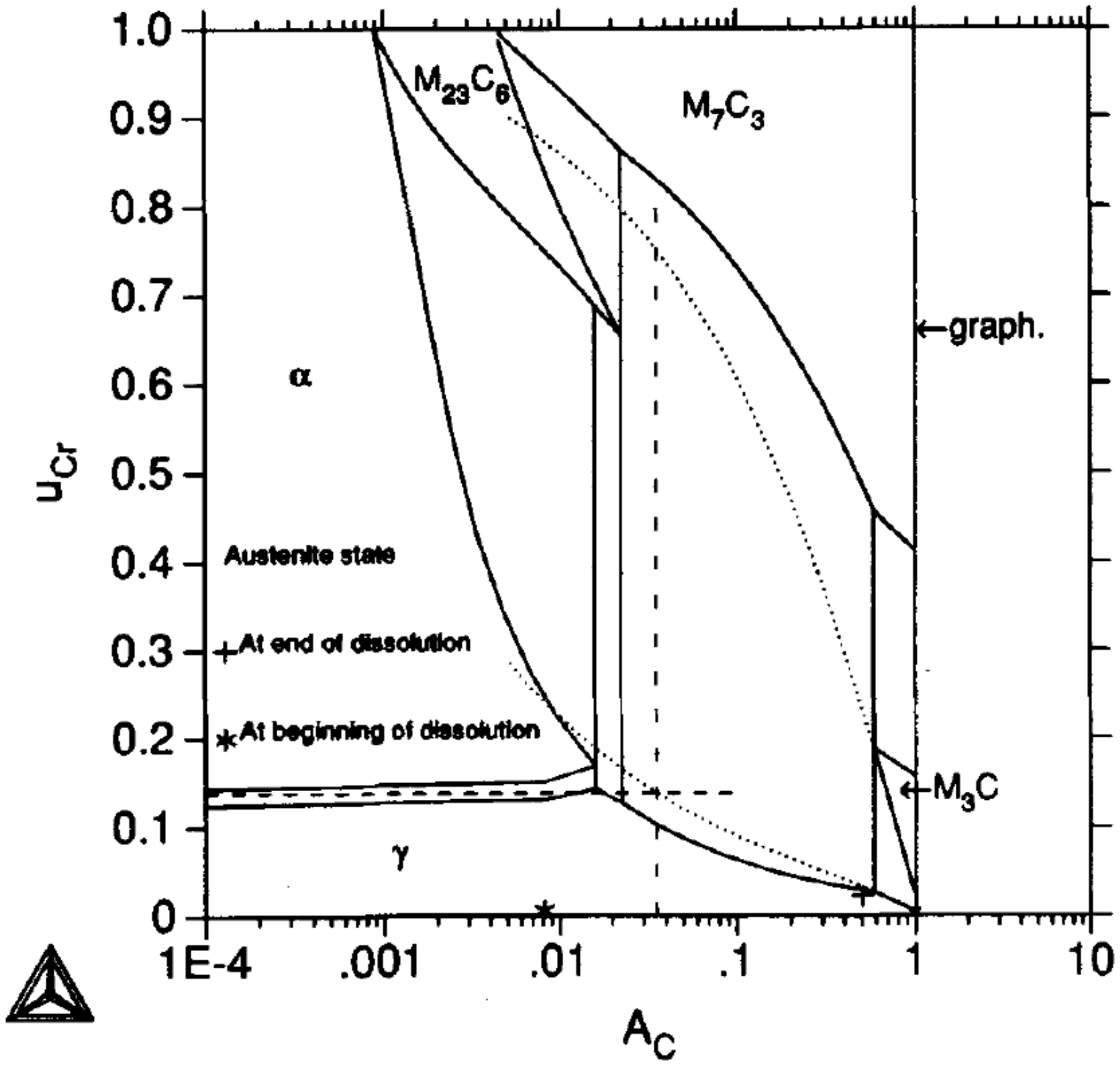
(iii)

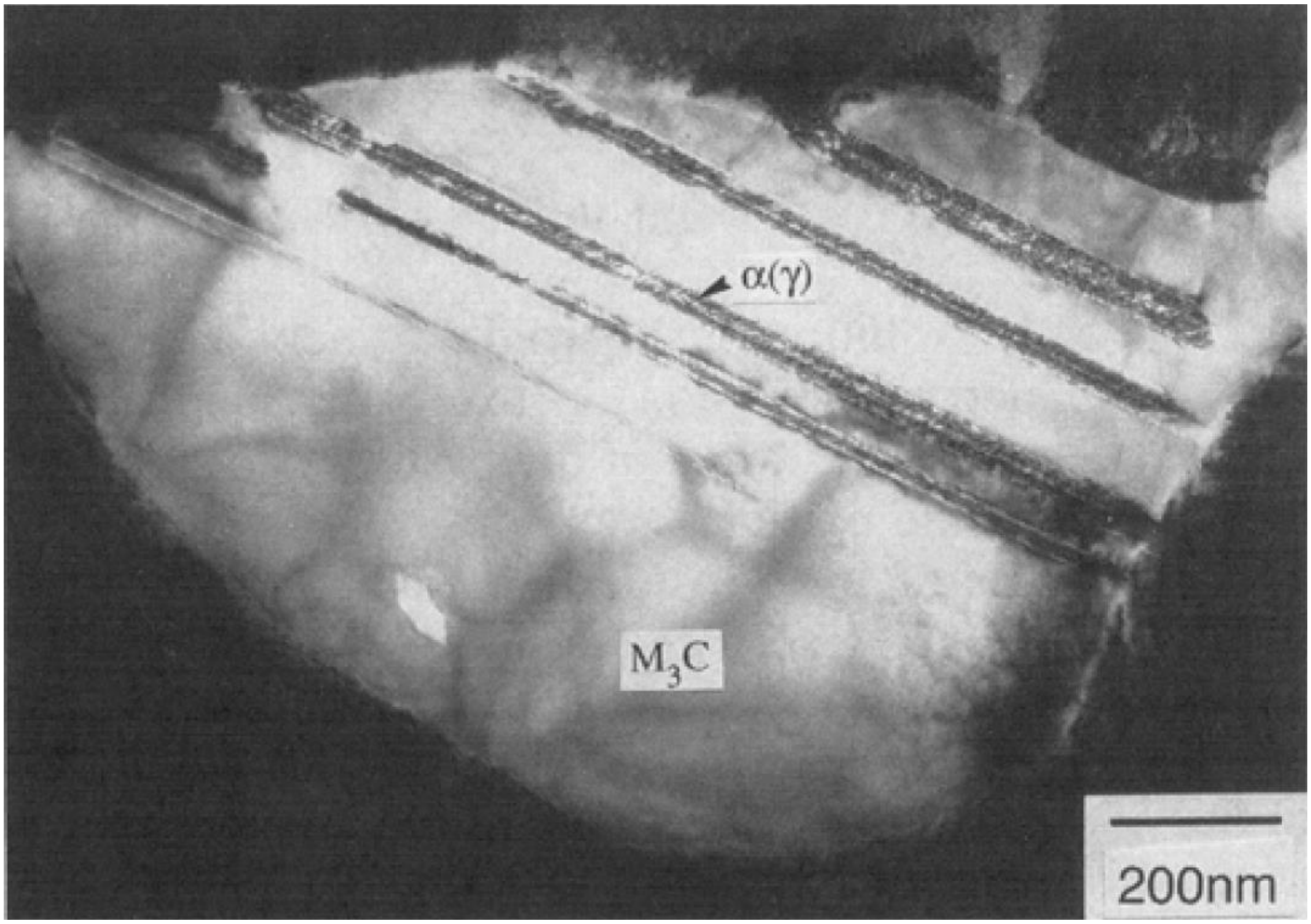

(iv)

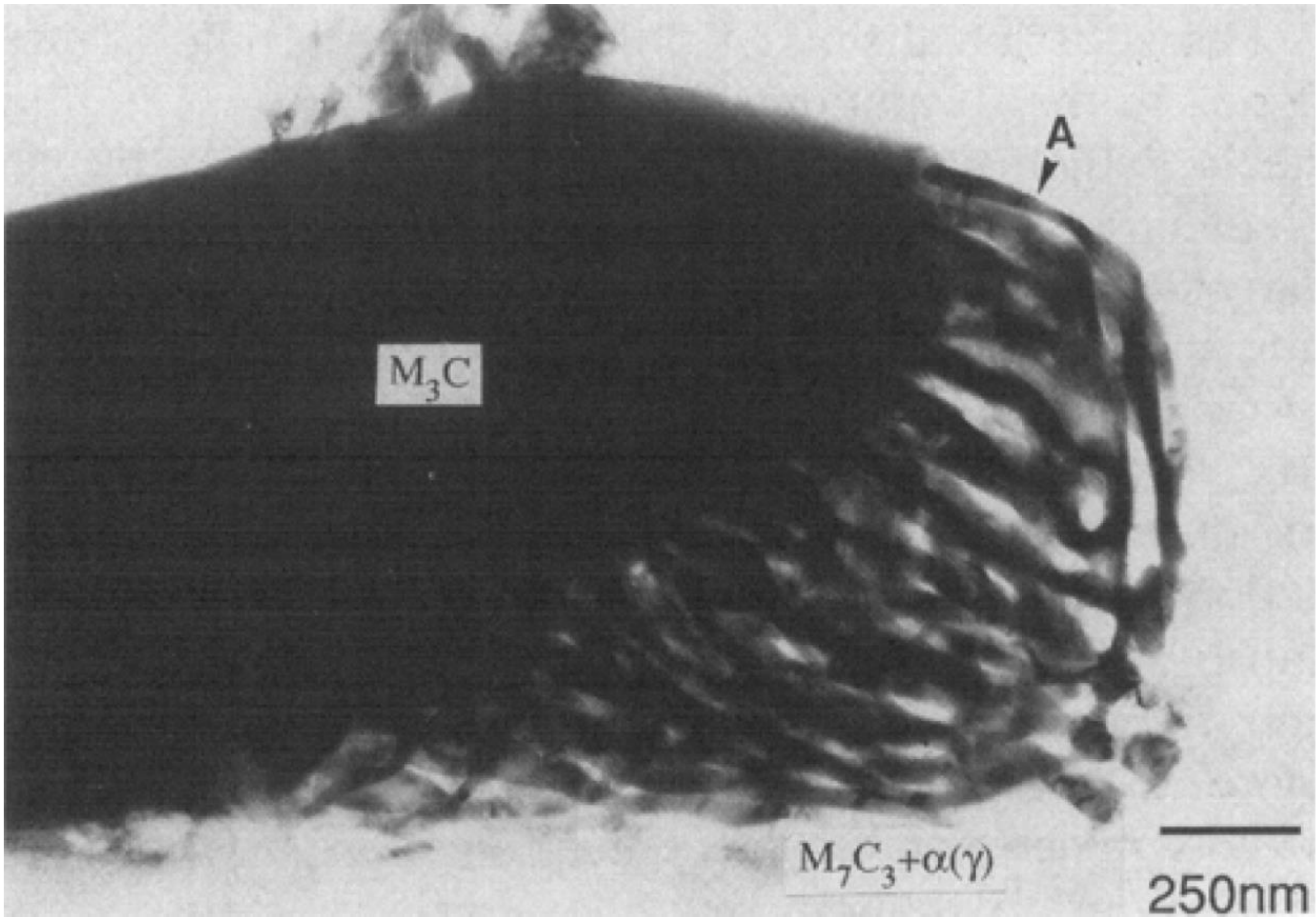

(v)

Figure 16

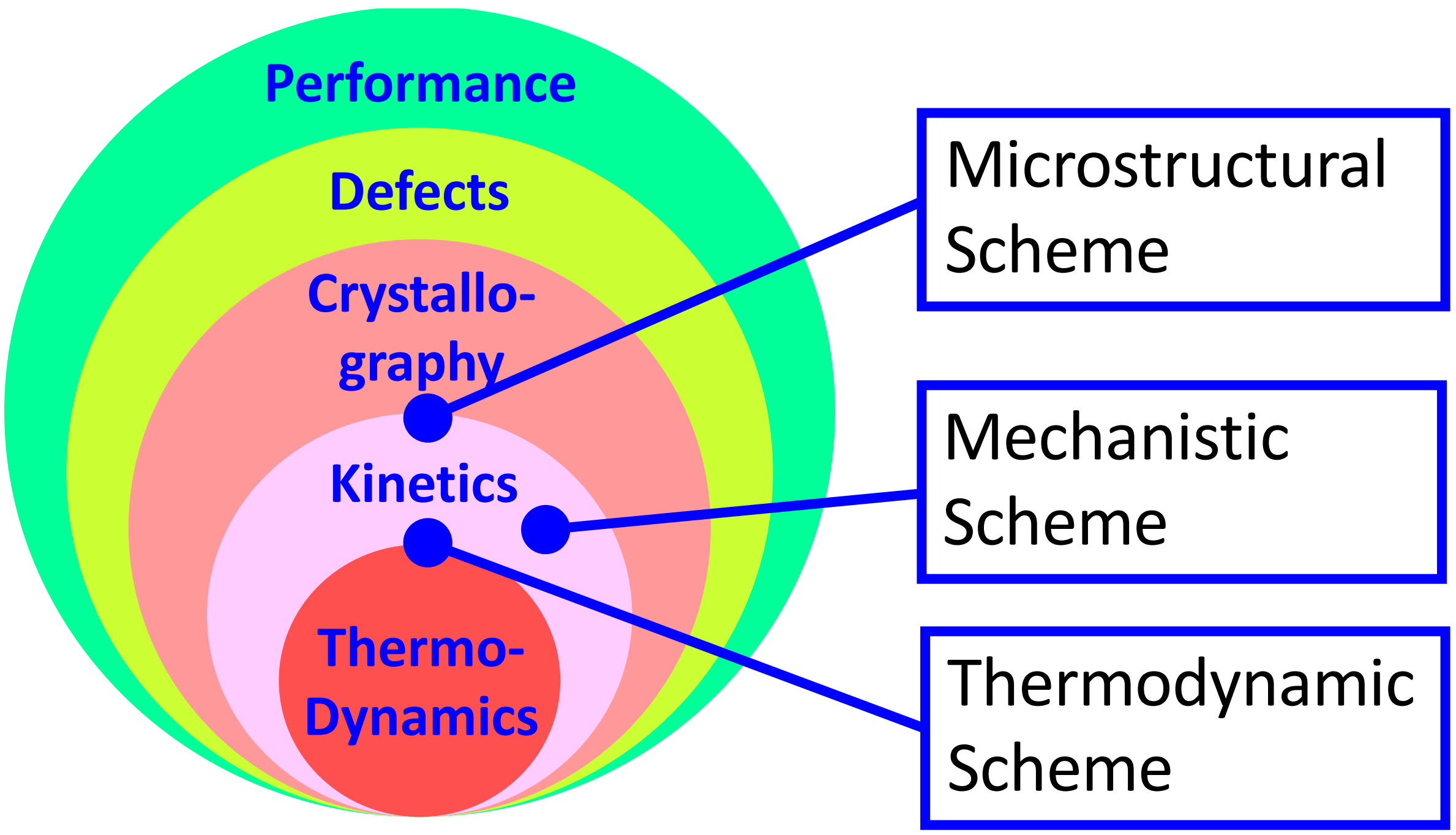

Figure 17